\documentclass[aps,pra,reprint, superscriptaddress,twocolumn, floatfix,showpacs,longbibliography]{revtex4-2}
\usepackage[utf8]{inputenc}
\usepackage[T1]{fontenc}
\usepackage{subcaption}
\usepackage{graphicx,color}
\usepackage[utf8]{inputenc}
\usepackage{amsmath,amssymb,bm,amsfonts,dsfont,mathrsfs,amsthm}
\usepackage{braket}
\usepackage{txfonts,comment}
\usepackage{enumitem}
\usepackage[colorlinks=true,linkcolor=blue,citecolor=blue,urlcolor=blue]{hyperref}
\usepackage{soul}
\usepackage[justification=raggedright]{caption}
\usepackage{multirow}
\usepackage{tabularx} 
\usepackage{diagbox}
\usepackage{graphicx} 
\usepackage{natbib}
\usepackage{colortbl}
\usepackage{pgf}  
\usepackage{xcolor}

\newcommand{\colornumber}[3]{
    \pgfmathsetmacro{\value}{#1}
    \pgfmathsetmacro{\min}{#2}
    \pgfmathsetmacro{\max}{#3}
    
    \pgfmathsetmacro{\ratio}{(\value-\min)/(\max-\min)}
    
    \ifdim \ratio pt < 0pt
        \pgfmathsetmacro{\ratio}{0}
    \fi
    \ifdim \ratio pt > 1pt
        \pgfmathsetmacro{\ratio}{1}
    \fi
    
    \xdef\cellcolor{\ratio*100!red!white!30}%
    
    \cellcolor{\cellcolor}#1%
}

\begin{document}

\title{Quantum Reservoir Computing for Realized Volatility Forecasting}

\author{Qingyu Li} 
\affiliation{Institute of Fundamental and Frontier Sciences, University of Electronic Sciences and Technology of China, Chengdu 611731, China}
\affiliation{Key Laboratory of Quantum Physics and Photonic Quantum Information, Ministry of Education, University of Electronic Science and Technology of China, Chengdu 611731, China}
\author{Chiranjib Mukhopadhyay} 
\affiliation{Institute of Fundamental and Frontier Sciences, University of Electronic Sciences and Technology of China, Chengdu 611731, China}
\affiliation{Key Laboratory of Quantum Physics and Photonic Quantum Information, Ministry of Education, University of Electronic Science and Technology of China, Chengdu 611731, China}
\author{Abolfazl Bayat} 
\affiliation{Institute of Fundamental and Frontier Sciences, University of Electronic Sciences and Technology of China, Chengdu 611731, China}
\affiliation{Key Laboratory of Quantum Physics and Photonic Quantum Information, Ministry of Education, University of Electronic Science and Technology of China, Chengdu 611731, China}
\affiliation{Shimmer Center, Tianfu Jiangxi Laboratory, Chengdu 641419, China}
\author{Ali Habibnia}
\affiliation{Department of Economics, Virginia Tech, U.S.A. }
\affiliation{Dataism Laboratory for Quantitative Finance, Virginia Tech, U.S.A.}

\begin{abstract}
Recent advances in quantum computing have demonstrated its potential to significantly enhance the analysis and forecasting of complex classical data. Among these, quantum reservoir computing has emerged as a particularly powerful approach, combining quantum computation with machine learning for modeling nonlinear temporal dependencies in high-dimensional time series. As with many data-driven disciplines, quantitative finance and econometrics can hugely benefit from emerging quantum technologies. In this work, we investigate the application of quantum reservoir computing for realized volatility forecasting. Our model employs a fully connected transverse-field Ising Hamiltonian as the reservoir with distinct input and memory qubits to capture temporal dependencies. The quantum reservoir computing approach is benchmarked against several econometric models and standard machine learning algorithms. The models are evaluated using multiple error metrics and the model confidence set procedures. To enhance interpretability and mitigate current quantum hardware limitations, we utilize wrapper-based forward selection for feature selection, identifying optimal subsets, and quantifying feature importance via Shapley values. Our results indicate that the proposed quantum reservoir approach consistently outperforms benchmark models across various metrics, highlighting its potential for financial forecasting despite existing quantum hardware constraints. This work serves as a proof-of-concept for the applicability of quantum computing in econometrics and financial analysis, paving the way for further research into quantum-enhanced predictive modeling as quantum hardware capabilities continue to advance.


\end{abstract}

\maketitle

\section{Introduction}
Quantum mechanics promises to enhance computing capacity in a fundamental way ~\cite{shor1994algorithms,steane1998quantum,kitaev2002classical}. In recent years, quantum computers, albeit with severe limitations, are rapidly emerging in various physical platforms including superconducting qubits \cite{arute2019quantum,wu2021strong,zhikun2024multilevel,ren2022experimental}, ion traps \cite{kielpinski2002architecture,zhang2017observation,ringbauer2022universal,monroe2021Programmable}, Rydberg atoms~\cite{bernien2017probing,ebadi2021quantum}, photonic setups~\cite{zhong2021phase,xiao2020nonHermitian}, nitrogen vacancy centers in diamond \cite{nemoto2014photonic} and topological qubits \cite{microsoft2025interferometric}. 
While near-term quantum computers have demonstrated quantum supremacy over their classical counterparts, they suffer from various imperfections such as finite coherence time and a limited number of qubits~\cite{preskill2018quantum}. 
Therefore, many algorithms with proven quantum advantages cannot be implemented on noisy near-term quantum computers. 
A natural emergent question is - \emph{can near-term quantum computers be used for machine learning?}
The advent of variational quantum algorithms~\cite{cerezo2021variational} and related quantum approximate optimization algorithms~\cite{farhi2014quantum} are attempts to answer this question where they have  
been adapted to solving problems in a truly diverse array of disciplines, including molecular simulations \cite{cao2019quantum,li2023fermionic}, biology \cite{baiardi2023quantum}, or particle physics \cite{paulson2021simulating}.  
In parallel, methods of classical time series analysis like Crutchfield computational mechanics framework \cite{crutchfield1989inferring, shalizi2001computational}, have also been shown to be improved by quantum encodings which reduce model complexity \cite{gu2012quantum}. Quantum computing methods are also starting to become important for modern quantitative finance problems like asset pricing \cite{da2023quantum}, portfolio optimization \cite{mugel2021hybrid,mugel2022dynamic}, credit sales classification \cite{wisniewska2023variational}, risk management \cite{leclerc2023financial}, loan eligibility prediction \cite{innan2024lep} among others. See Refs ~\cite{orus2019quantum,herman2022survey,naik2025portfolio} for recent surveys. 
The key approach behind these algorithms is to optimize a parametrized quantum circuit to optimize the loss function in an iterative way like a classical neural network~\cite{cerezo2021variational}. However, there is a different paradigm of classical machine learning, namely \textit{reservoir computing}~\cite{jaeger2004harnessing}, which does not seek to optimize the parameters of a neural network, and training takes place only at the final output layer.While this approach to machine learning has transformed time series modeling, it still operates in the realm of classical computing~\cite{LiDeep2024,kimComprehensiveSurveyDeep2025}.
Quantum versions of this approach are particularly promising for near-term quantum computers and have been proposed with various underlying platforms like disordered spin chains \cite{fujii2017harnessing}, quantum optical energy levels \cite{govia2021quantum}, spin-boson platforms \cite{das2025quantum}, or superconducting devices \cite{yasuda2023quantum} for some mathematical and physical problems. They are exceptionally useful for time-series forecasting, as demonstrated with canonical models like Mackey-Glass \cite{mackey1977oscillation} and autoregressive moving-average (\texttt{ARMA})~\cite{whittle1951hypothesis}. Thus, it is natural to ask whether quantum reservoir computing can be useful for accurate forecasting of real-world financial time series data. 

Volatility modeling is a fundamental aspect of financial econometrics, essential to understand and manage the uncertainty or risks associated with financial markets. Accurate modeling and forecasting of volatility are crucial for various applications, including risk management, portfolio optimization, and derivative pricing \citep{black1973pricing, poon2003forecasting}. Given its critical role, developing robust volatility forecasting models has been a major focus of financial research. While traditional models such as Generalized Autoregressive Conditional Heteroskedasticity (\texttt{GARCH})~\citep{bollerslev1986generalized} and its standard extensions are widely used, their reliance on low-dimensional parametric recursions for conditional variance, typically linear in squared returns, imposes strong functional restrictions. These constraints may limit their ability to flexibly represent complex nonlinear and multiscale volatility dynamics governing realized volatility at medium and long horizons. Recognizing these limitations, subsequent research has emphasized the importance of understanding the distribution of realized stock return volatility for more effective financial analysis, as highlighted in \citet{andersen2001distribution}. Furthermore, econometric analysis of realized volatility, particularly its application in estimating stochastic volatility models, has provided crucial insights into the behavior of financial time series, as demonstrated by \citet{barndorff2002econometric}. To address some of the shortcomings of traditional models, the Heterogeneous Autoregressive (\texttt{HAR}) model \citep{corsi2009simple} was developed, offering a more nuanced approach by incorporating realized volatility over different time horizons, thus providing a more accurate and comprehensive measure of market risk. Its variations, including \texttt{HAR-J} (\texttt{HAR} with jumps) and \texttt{CHAR} (continuous \texttt{HAR}) \citep{andersen2007roughing}, \texttt{SHAR} (semivariance-\texttt{HAR}) \citep{patton2015good}, and \texttt{HARQ} (\texttt{HAR} with realized quarticity) \citep{bollerslev2016exploiting}, further enhanced its ability to capture different aspects of market volatility. The nonlinear and often non-Gaussian nature of financial data has driven researchers to explore more sophisticated approaches, such as machine learning techniques to capture complex relationships that traditional econometric models may miss \citep{kuan1994artificial, habibnia2016essays, gu2020empirical, bucci2020realized, 
gu2021autoencoder, habibnia2021forecasting, zhu2023forecasting, jiang2023reimaging, chen2024deep}.
In the realm of realized volatility forecasting, machine learning models have shown promising results. While some studies found machine learning models to perform similarly to or slightly worse than traditional \texttt{HAR}-family models \citep{hillebrand2010bagging, fernandes2014modeling, audrino2016lassoing, branco2022forecasting}, others reported significant improvements using machine learning approaches \citep{bucci2020realized, christensen2023machine, zhu2023forecasting}. For instance, \citet{christensen2023machine} applied machine learning to volatility forecasting, demonstrating that these models can significantly improve prediction accuracy, especially when combined with macroeconomic variables. Similarly, \citet{zhang2024volatility} explored the integration of machine learning with intraday commonality, further enhancing the precision of volatility forecasts. \cite{bucci2020realized} also demonstrated the effectiveness of Long Short-Term Memory (\texttt{LSTM}) neural networks in forecasting monthly S\&P 500 realized volatility, and have found strong links between volatility and macroeconomic factors paving the way for more advanced machine learning applications in this field. The importance of capturing the multifaceted nature of volatility is underscored by the work of \citet{ghysels2006predicting}, who emphasized the value of high-frequency data for more accurate volatility estimation. By leveraging data sampled at different frequencies, their approach offers a richer understanding of market dynamics, which is crucial for effective risk management and portfolio allocation. Recent reviews, such as the one by \citet{gunnarsson2024prediction}, have highlighted the growing role of machine learning in volatility modeling, particularly in the prediction of realized and implied volatility indices. These advancements underscore the trend toward more data-driven approaches in finance, which offer superior adaptability to the complexities of modern financial markets. Among machine learning algorithms, recurrent neural networks and their variant, \texttt{LSTM} networks, have shown particular success in modeling time series data due to their inherent ability to capture temporal dependencies. Recurrent neural networks are designed to recognize patterns in sequences of data by maintaining a hidden state that is influenced by previous inputs, making them ideal for time-dependent data such as financial time series~\citep{hochreiter1997long,goodfellow2016deep,bucci2020realized}. 


This work introduces a novel approach using quantum reservoir computing for realized volatility forecasting of the S\&P 500 index. Quantum reservoir computing, an extension of classical reservoir computing, has been chosen for this study due to its ability to efficiently process temporal data on near-term quantum computers~\citep{fujii2017harnessing}, while maintaining a low training cost and  providing a state space that grows exponentially with the number of qubits to capture features. Quantum reservoir computing models harness quantum features to enhance the prediction power of time series data. They utilize the dynamics of fixed quantum systems known as the ``reservoir". Unlike conventional neural networks, in reservoir computing, the parameters of the reservoir are not trained, and the learning procedure takes place at the final output layer after performing measurements~\cite {lukovsevivcius2009reservoir}. This makes reservoir computing particularly advantageous for tasks requiring the capture of temporal dynamics without the heavy computational burden associated with training traditional neural networks. Quantum reservoir computing extends this concept into the quantum domain, leveraging quantum states to enhance computational power and efficiency. This approach can be promising for volatility forecasting, where capturing complex, time-dependent relationships is critical \citep{fujii2017harnessing, mujal2023time, garcia2023scalable,llodra2025quantum, garcia2024quantum, thakkar2023improved, rivera2022time}. Such quantum reservoir platforms have already been conceptualized in various atomic and spin lattices \cite{fujii2017harnessing,settino2024memory,llodra2025quantum} as well as photonic platforms \cite{garcia2023scalable,nerenberg2024photon, suprano2024experimental}, It is of particular interest to note that very recent results on reservoir computing already hint that quantum properties may lead to fast and reliable forecasts with smaller resources \citep{abbas2024reservoir}. We aim to showcase the proof of concept and capability of quantum machine learning in real-world time series analysis, particularly in financial forecasting. We benchmark our quantum model against several classical models, including the \texttt{HAR} and \texttt{HARX} models, as well as traditional neural network algorithms. By leveraging a comprehensive set of features, including market microstructure and macroeconomic variables, we explore whether quantum nonlinear models can more effectively capture the intricate relationships governing market volatility. Prior research, such as that by~\citet{alaminos2022forecasting} and~\citet{thakkar2023improved}, has demonstrated the potential of quantum machine learning in financial forecasting, suggesting that quantum models may eventually outperform classical machine learning algorithms, particularly when dealing with large datasets and complex patterns. \\

The paper is organized as follows. Section~\ref{sec: technical_background} outlines the classical models, including traditional regression methods and machine learning models, for realized volatility forecasting. 
Sec.~\ref{sec: Proposed QRC} demonstrates our method of volatility forecasting with quantum reservoir computing. If you are not familiar with quantum computing, Sect. Appendix~\ref{sec: quantum computing} gives a brief discussion of the principle of quantum computing, bridging concepts from both financial econometrics and quantum computing to facilitate interdisciplinary understanding. Sec.~\ref{sec: Empirical Results} compares our results with other volatility forecasting strategies, before concluding discussions in Sec.~\ref{sec: conclusions and outlook}.

\section{Technical Background}
\label{sec: technical_background}

This paper lies at the intersection of three different subjects, namely econometrics, machine learning, and quantum computation.  In this section, we provide a brief overview of the former two subjects for physicists, introducing the concepts and methodologies that we will use later. For non-physicists, we include an appendix on quantum computation preliminaries as well as a notation table. Depending on their expertise, the readers can skip the following subsections.

\subsection{Classical Models for Realized Volatility Forecasting}

Stock market volatility is an important economic factor that reflects the risk of investment at a given time. The concept of realized volatility, formally introduced by \citet{andersen1998answering}, marked a significant advancement in volatility measurement by providing an accurate, model-free estimate utilizing high-frequency financial data. Realized volatility, \(RV_t\), at time \(t\), is defined as the square root of the sum of squared returns within a given time interval:

\begin{equation}
RV_t = \sqrt{ \sum_{i=1}^{N_t} r_{i,t}^2 }, \quad \text{and we model} \; \log(RV_t)
\end{equation}

Where \(r_{i,t}\) represents the return on the day \(i\) within period \(t\), and \(N_t\) indicates the number of observations (e.g., trading days) in that period. Autoregressive (\texttt{AR}) models and their various extensions have since become essential tools for modeling and forecasting realized volatility due to their computational simplicity and effectiveness as baseline methodologies in time series analysis. Although traditional \texttt{AR} models effectively capture short-term volatility dependencies, they often fail to reflect the long-memory characteristics typically present in volatility dynamics. To overcome this limitation, the Heterogeneous Autoregressive (\texttt{HAR}) model proposed by \citet{corsi2009simple} incorporates realized volatility at multiple time horizons, thus effectively capturing both persistence and scaling features. The general form of the \texttt{HAR} model is expressed as the following:

\begin{equation}
    RV_t = \beta_0 + \beta_d RV_{t-1}^{(d)} + \beta_w RV_{t-1}^{(w)} + \beta_m RV_{t-1}^{(m)} + \varepsilon_t,
\end{equation}

where \(RV_{t-1}^{(d)}\), \(RV_{t-1}^{(w)}\), and \(RV_{t-1}^{(m)}\) represent realized volatility averages computed over daily, weekly, and monthly horizons. The coefficients \(\beta_d\), \(\beta_w\), and \(\beta_m\) measure the contributions of these short-term, intermediate-term, and long-term volatility components to the current volatility level. The parameter $\varepsilon_t$ is the residual term assumed to follow a conditionally heteroskedastic process, with $\mathbb{E}[\varepsilon_t|\mathcal{F}_{t-1}] = 0$ and $\operatorname{Var}(\varepsilon_t|\mathcal{F}_{t-1}) = \sigma_t^2$, where \(\mathcal{F}_{t-1}\) denotes the information set available at time \(t-1\). This specification allows the \texttt{HAR} model to capture volatility at multiple frequencies, accommodating heterogeneity in market participants' investment horizons. The model assumes weak stationarity of the log-volatility series and serially uncorrelated errors. Estimation is typically performed via ordinary least squares on the log-transformed realized volatility series to stabilize variance and reduce the impact of heteroskedasticity. For statistical inference, heteroskedasticity and autocorrelation-consistent (HAC) standard errors are employed following \citet{newey1986simple}, ensuring robustness to serial correlation and time-varying error variance.

To incorporate exogenous macroeconomic and financial variables, we adopt a monthly version of the \texttt{HAR} model, resulting in a \texttt{HARX} specification that captures realized volatility dynamics at multiple time scales and permits additional predictors. Specifically, we define the monthly \texttt{HARX} model as:

\begin{equation}
\resizebox{\columnwidth}{!}{$
    \begin{split}
        RV_t &= \beta_0 + \beta_1 RV_{t-1} + \beta_2 \left( \frac{1}{3} \sum_{i=1}^{3} RV_{t-i} \right) + \beta_3 \left( \frac{1}{12} \sum_{i=1}^{12} RV_{t-i} \right) \\
        &\quad + \sum_{k=1}^k\gamma_k^TX_{t-k} + \varepsilon_t,
    \end{split}
$}
\end{equation}

where \(RV_{t-1}\) represents the monthly realized volatility lagged by one month, capturing short-term dynamics; \(\frac{1}{3}\sum_{i=1}^{3} RV_{t-i}\) and \(\frac{1}{12}\sum_{i=1}^{12} RV_{t-i}\) represent the quarterly and annual realized volatility averages, respectively; and \(X_{t-k}\) denotes the vector of macroeconomic and financial features available at lag \(k\), with corresponding coefficient vector \(\gamma_k\). Residual diagnostics and model stability tests are performed to ensure robustness. The \texttt{HARX} model serves as a high-performing linear benchmark, capturing persistence at different time horizons while allowing flexible enhancement through $X_t$. \\

However, linear models such as the \texttt{HAR} may still fall short in capturing the non-linear dynamics inherent in financial time series. This limitation has motivated the development of more advanced models, such as the Realized \texttt{GARCH} model by \citet{hansen2012realized}, which integrates realized measures into the GARCH framework, and extensions of the standard \texttt{HAR} framework designed to accommodate non-linearities and regime-switching behaviors \citep{mcaleer2008realized, hillebrand2010bagging}. More recently, hybrid approaches combining \texttt{HAR} structures with advanced methodologies, such as neural networks and regularization-based procedures (e.g., LASSO or elastic net), have further enhanced forecast performance and adaptability in high-dimensional settings \citep{audrino2016lassoing}. \\

Machine learning techniques, briefly discussed later in this section, offer promising alternatives due to their ability to effectively capture complex nonlinear relationships without requiring explicit parametric functional forms. Models like Long Short-Term Memory (\texttt{LSTM}) networks and Reservoir Computing (\texttt{RC}) have demonstrated effectiveness in financial forecasting tasks \citep{fischer2018deep, butcher2013reservoir}. Incorporating exogenous variables leads to extensions such as Long Short-Term Memory with Exogenous features (\texttt{LSTMX})  and Reservoir Computing with Exogenous features (\texttt{RCX}), enhancing forecast performance by utilizing additional information.

\begin{figure*}
    \centering   \includegraphics[width=0.8\textwidth]{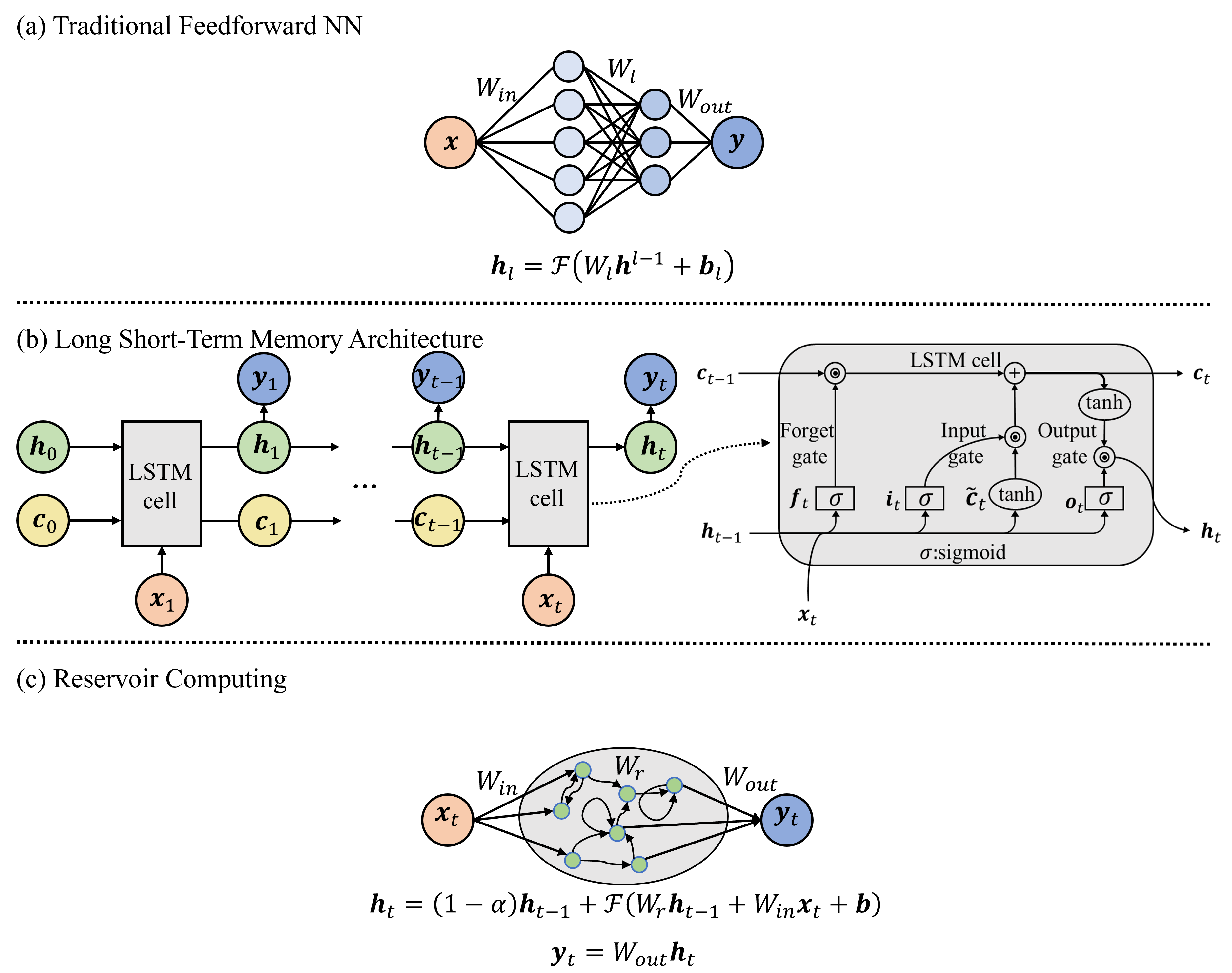}
    \caption{\textbf{Schematic of various machine-learning paradigms.} (a) Traditional feedforward neural networks, where $W_{in}$, $\{W_l\}$ and $W_{out}$ are the trainable weight matrices for the input layer, hidden layer(s), and output layer, respectively. Noticeably, the information flows unidirectionally layer by layer from the input to the output.
    (b) \texttt{LSTM} networks are a special form of the recurrent neural networks architecture. \texttt{LSTM} introduces two key states as cell state~$\boldsymbol{c}_i$ for long-term memory and hidden state~$\boldsymbol{h}_i$-for short-term memory. In addition, \texttt{LSTM} layers utilize additional gates to control which information in the hidden state is output and passed to the next hidden state. These additional gates overcome the common issues recurrent neural networks face in learning long-term dependencies.
    (c) Reservoir Computing is a computational framework that inputs $\boldsymbol{x}_t$ at each step are mapped to a high-dimensional space by 
    a fixed random $W_{in}$. The hidden states $\boldsymbol{h}_t$ evolve dynamically through a fixed random reservoir $W_r$, which models the system's temporal denpendencies. Notably, both $W_{in}$ and $W_r$ are randomly initialized and reamin fixed throughout the process, while only the output weights matrix $W_{out}$ is trainable. This feature make reservoir computing efficient for handling time series and dynamic system tasks, requiring minimal training effort.}
    \label{Fig: schematic_different_computational_models}
\end{figure*}

\subsection{Machine Learning Preliminaries}
Machine learning involves the development of algorithms and statistical models that allow systems to perform specific tasks effectively by analyzing data, identifying patterns, and making predictions.
The artificial neural network~\cite{zou2009overview}, as one of the most powerful algorithms in machine learning, is widely used in many domains, including image recognition, natural language processing, recommendation systems, predictive analytics, and time series processes. 
\subsubsection{Feedforward Neural Networks}
Feedforward neural networks are the simplest type of artificial neural network, consisting of an input layer, an output layer, and one or more hidden layers that connect the input layer to output layer~(see Fig.~\ref{Fig: schematic_different_computational_models}(a)). 
The term "feedforward`` indicates that the architecture of these networks relies on transforming inputs into outputs through a series of operations, where each operation involves multiplying the input by a weight matrix and applying an activation function.
Specifically, consider the output of the $l$-th layer, denoted as $\boldsymbol{h}_l$, such that the input of the network is represented as $\boldsymbol{h}_0=\boldsymbol{x}$, and the final output is $\boldsymbol{h}_L=\boldsymbol{\hat{y}}$.
For the $l$-th layer in the Feedforward neural networks, the output takes the form
\begin{equation}
        \boldsymbol{h}_l = \mathcal{F}(W_l\boldsymbol{h}_{l-1}+\boldsymbol{b}_l),
\end{equation}
where $W_l$ is the weight matrix, $\boldsymbol{b}_l$ is the bias vector, and $\mathcal{F}(\cdot)$ is a certain activation function.
The overall mathematical model of an feedforward neural networks is a nested composition of these computations, where the output $\boldsymbol{h}_l$ of each layer serves as the input of the next layer. 
Because there are no cycles or loops in the networks, the information flows strictly in a single direction: from the input node, through the hidden nodes, and to the output nodes~(see Fig.~\ref{Fig: schematic_different_computational_models}(a)). 
Training for a neural network is based on a set of labeled data $\{ (\boldsymbol{x}_i,\boldsymbol{y}_i) \}$, where $\boldsymbol{x}_i$ and $\boldsymbol{y}_i$ represent input and output, respectively. 
We train the weight matrices $\{W_l\}_l$ as well as the bias vectors $\{\boldsymbol{b}_l\}_l$ so that for any input data $\boldsymbol{x}_i$, the corresponding output of the neural network $\boldsymbol{\hat{y}}_i$ closely approximates the real label $\boldsymbol{y}_i$. This is accomplished by minimizing a loss function such as
\begin{equation}
    L=\min_{\{W_l,\boldsymbol{b}_l\}} \frac{1}{n}\sum_i^n (\boldsymbol{y}_i-\boldsymbol{\hat{y}}_i)^2, 
    \label{cost_fun}
\end{equation}
where $n$ is the size of the training set. 
Feedforward neural networks are well-suited for tasks where each input is independent of the others, such as image classification or regression. However, they are not ideal for tasks that require capturing temporal or sequential dependencies.

\subsubsection{Long Short-Term Memory Neural Networks}
For addressing tasks with spatio-temporal dependencies, it is often necessary to utilize information from past data to make accurate predictions about future results. For example, when reading an article, the meaning of each word is interpreted based on the understanding of the preceding words.
Long Short-Term Memory (\texttt{LSTM}) networks~\cite{hochreiter1997long} address this problem by introducing a specialized architecture designed to capture both short-term and long-term dependencies in sequential data. Unlike standard recurrent neural networks, which rely solely on a hidden state $\boldsymbol{h}_t$ to carry forward information, \texttt{LSTM} introduce an additional cell state $\boldsymbol{c}_t$, which serves as long-term memory. The cell state enables \texttt{LSTM} to selectively retain or discard information over time. 
At each time step, as Fig.~\ref{Fig: schematic_different_computational_models}(b) shows, the \texttt{LSTM} cell receives the following inputs:
\begin{enumerate}
    \item Hidden state $\boldsymbol{h}_{t-1}$: The short-term memory from the previous time step.
    \item Cell state $\boldsymbol{c}_{t-1}$: The long-term memory from the previous time step.
    \item Current input $\boldsymbol{x}_t$: The input in the current time step.
\end{enumerate}
The operation of the \texttt{LSTM} cell can be described in terms of three different gates: (i) forget gate; (ii) input gate; and (iii) output gate (see Fig.~\ref{Fig: schematic_different_computational_models}(b)).
The forget gate determines which parts of the previous cell state $\boldsymbol{c}_{t-1}$ should be ``forgotten" or retained. It uses a sigmoid activation function to produce values between 0~(completely forget) and 1~(completely keep) for each element, defined as 
\begin{equation}
    \boldsymbol{f}_t = \text{sigmoid}(W_f[\boldsymbol{h}_{t-1},\boldsymbol{x}_t]+\boldsymbol{b}_f),
\end{equation}
where $W_f$ is the weight matrix for the forget gate, $\boldsymbol{b}_f$ is the bias term, and $\boldsymbol{f}_t$ is a vector whose elements take values between $0$ and $1$. 
The input gate determines which parts of the current input $\boldsymbol{x}_t$ should be added to the cell state $\boldsymbol{c}_t$. 
It has two components, a sigmoid layer to decide which values to update, and a tanh layer to create new candidate values to potentially add to the cell state $\boldsymbol{c}_t$, which formula are defined as
\begin{equation}
    \begin{split}
        \boldsymbol{i}_t &= \text{sigmoid}(W_i[\boldsymbol{h}_{t-1},\boldsymbol{x}_t]+\boldsymbol{b}_i)\\
        \tilde{\boldsymbol{c}}_t &= \text{tanh}(W_c[\boldsymbol{h}_{t-1},\boldsymbol{x}_t]+\boldsymbol{b}_c) 
    \end{split}
\end{equation}
where $\boldsymbol{i}_t$ is the input gate output, a vector of values between $0$ and $1$, and $\tilde{\boldsymbol{c}}_t$ is the candidate cell state.
The information from the forget gate and the input gate are used to update the cell state $\boldsymbol{c}_t$, as
\begin{equation}
    \boldsymbol{c}_t = \boldsymbol{f}_t\odot \boldsymbol{c}_{t-1}+\boldsymbol{i}_t\odot \tilde{\boldsymbol{c}}_t,
\end{equation}
where $\odot$ is the element-wise multiplication operator. 
As a result, $\boldsymbol{c}_t$ contains a nonlinear combination of the input state $\boldsymbol{x}_t$ and the previous input data. 
The output gate determines the vector $\boldsymbol{o}_t$ which is a nonlinear function of the the hidden state $\boldsymbol{h}_{t-1}$ as well as the input $\boldsymbol{x}_t$ as
\begin{equation}
     \boldsymbol{o}_t =\text{sigmoid}(W_o[\boldsymbol{h}_{t-1},\boldsymbol{x}_t]+\boldsymbol{b}_o).
\end{equation}
The next hidden state $\boldsymbol{h}_t$ is then determined through combination of $\boldsymbol{o}_t$ and the cell state $\boldsymbol{c}_t$ as 
\begin{equation}
        \boldsymbol{h}_t =\boldsymbol{o}_t \odot \text{tanh}(\boldsymbol{c}_t). 
\end{equation}
In general, the output $\hat{\boldsymbol{y}}_t$ is a function of $h_t$ as $\hat{\boldsymbol{y}}_t=g(h_t)$, where $g(\cdot)$ depends on the specific problem.
Training is carried out by optimizing a loss function, of the form of Eq.~(\ref{cost_fun}), which aims to minimize the difference between $\hat{\boldsymbol{y}}_t$ and the real output $\boldsymbol{y}_t$. 
During the training procedure, the weight matrices $\{W_f,W_i,W_c,W_o\}$ and bias vectors $\{\boldsymbol{b}_f,\boldsymbol{b}_i,\boldsymbol{b}_c,\boldsymbol{b}_o\}$ are all trained and updated iteratively, see Fig.~\ref{Fig: schematic_different_computational_models}(b). 
The \texttt{LSTM} architecture uses gating mechanisms, see Fig.\ref{Fig: schematic_different_computational_models}(b), to regulate information flow and gradient propagation~\cite{hochreiter1997long}, effectively mitigating vanishing and exploding gradients common in standard recurrent neural networks and enabling the retention of long-term information.
These characteristics make \texttt{LSTM} widely used in fields such as natural language processing and time-series prediction.

\subsubsection{Classical Reservoir Computing}

In the early 2000s, echo state networks~\cite{jaeger2004harnessing} and liquid state machine~\cite{maass2002real} were independently proposed as the seminal approach of the time series model, which are grouped into the framework of classical reservoir computing. 
Reservoir computing models share the common principle of using a ``reservoir'', comprising $W_r$~(the recurrent weight matrix) and $W_{in}$ (the input-to-reservoir weight matrix), to project inputs into a high-dimensional feature space.
This transformation enables the effective capture of complex patterns, relationships, and temporal dynamics in the data.
Unlike conventional neural networks, reservoir computing does not train weight matrices $W_r$ and $W_{in}$ and bias vectors $\boldsymbol{b}$. 
Instead, these parameters are randomly initialized and remain fixed during training. 
A simple and trainable read-out mechanism, i.e. linear regression, is then used to generate information for this high-dimensional representation.
In the whole reservoir computing process, only the weights $W_{out}$ in the linear regression layer are trained, significantly reducing training cost.
The mathematical representations of these models can be expressed as: 
\begin{equation}
    \begin{split}
        h_t&=(1-\alpha)\boldsymbol{h}_{t-1}+ \mathcal{F}(W_r\boldsymbol{h}_{t-1}+W_{in}\boldsymbol{x}_t+\boldsymbol{b})\\
        y_t&=W_{out}\boldsymbol{h}_t,
    \end{split}\label{eq:classical_reservoir_learning}
\end{equation}
where $\boldsymbol{h}_t$ represents the hidden state at time $t$, and $\alpha$ is the leak rate, a hyperparameter controlling the update speed of hidden state, see Fig.~\ref{Fig: schematic_different_computational_models}(c).

Since only $W_{out}$ needs to be trained, reservoir computing is highly suitable for deployment in artificial or natural physical systems characterized by high-dimensionality and nonlinear transformations.
To date, reservoir computing has been widely implemented in various systems, such as cellular automata~\cite{yilmazSymbolicComputationUsing2015,mcdonaldReservoirComputingExtreme2017}, coupled oscillators~\cite{yamaneWaveBasedReservoirComputing2015}, analog circuits~\cite{royLiquidStateMachine2014,katumbaLowLossPhotonicReservoir2018}, optical node arrays~\cite{duportAllopticalReservoirComputing2012,mesaritakisMicroRingResonators2013,dejonckheereAllopticalReservoirComputer2014} and biological organization~\cite{goudarziDNAReservoirComputing2013}.

\section{Quantum Reservoir Computing for Realized Volatility Forecasting}
\label{sec: Proposed QRC}

\begin{figure*}
    \centering
    \includegraphics[width=0.8\textwidth]{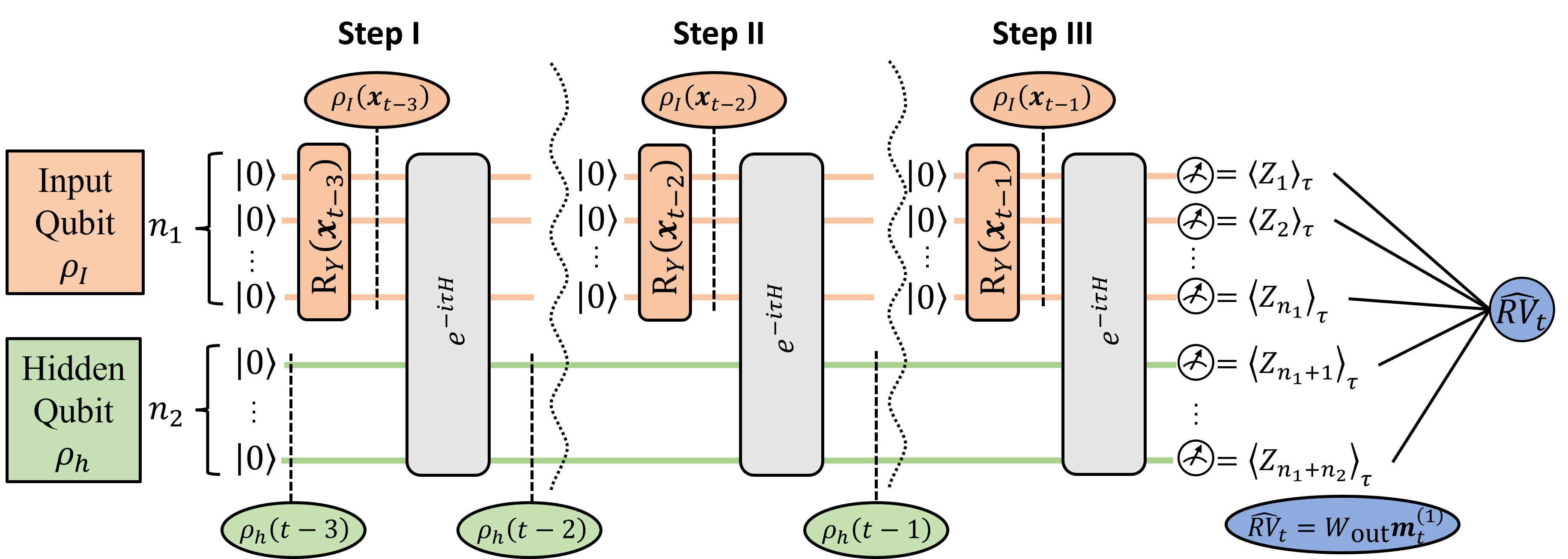}
    \caption{\textbf{Quantum reservoir computing schematic.} The task is to forecast $\widehat{RV}_t$ based on the past information of features $\boldsymbol{x}$ at discrete time from $t{-}3$ to $t{-}1$. The Quantum Reservoir consists of two subsystems, the input qubits $\rho_I$ for inputting features and the hidden qubits $\rho_h$ for hidden state, composed of $n_1$ and $n_2$ qubits respectively. In the first, the feature variables $\boldsymbol{x}_{t-3}$ are encoded into the input qubits via single-qubit gates $R_Y(\boldsymbol{x}_{t-3})$ applied on $|0\rangle^{\otimes n_1}$. Therefore the input density matrix is given as $\rho_I(\boldsymbol{x}_{t-3})$. The hidden state at first is initialized as  $\rho_h(t{-}3)=|0\rangle\langle 0|^{\otimes n_2}$.
    Thus the quantum reservoir state is given by $\rho_I (\boldsymbol{x}_{t-3})  \otimes \rho_{h}(t{-}3)$. The state evolves under the action of the Hamiltonian $H$ for time $\tau$, described by $e^{-iH\tau}$. To update the hidden state for the next step, the input $n_1$ qubits are discarded such that the new hidden state is given by a partial trace as $\rho_h(t{-}2)$.
    This process continues until all features from $t{-}3$ to $t{-}1$ have been encoded within the quantum reservoir. After processing all features, measurements on Pauli-Z basis of all qubits are performed to read out useful information from the quantum reservoir, yielding the expectation value $\langle Z_j\rangle_\tau$ for the $j$-th qubit. These expectation values collectively form the readout vector $\boldsymbol{m}_t^{(1)}=[\langle Z_1\rangle_\tau,\cdots,\langle Z_{n_1+n_2}\rangle_\tau]$. A linear regression model is used to predict $\widehat{RV}_t$ based on this readout vector as $\widehat{RV}_t=W_{out}\boldsymbol{m}_t^{(1)}$, where $W_{out}$ is a weight matrix trained using the ridge regression method.}\label{Fig: schematic_qrl}
\end{figure*}

By harnessing the unique properties of quantum mechanics, such as superposition and entanglement, quantum computing demonstrates a quantum advantage in solving specific problems, including integer factorization~\cite{shor1999polynomial}, random circuit sampling~\cite{arute2019quantum} and quantum simulation~\cite{sethUniversal1996}. 
These achievements identify quantum computing as a powerful method for information processing and thus drive researchers to explore more fields that might benefit from quantum computing.  
A natural extension of the quantum computing application is the investigation of reservoir computing with quantum systems. 
Fujii and Nakajima took the first theoretical step by proposing disordered quantum spin ensembles as reservoirs \cite{fujii2017harnessing}, which has since been extended to photonic \cite{ghosh2019quantum,garcia2023scalable,nerenberg2024photon}, non-linear oscillator~\cite{govia2021quantum}, and neutral atomic Rydberg array~\cite{bravo2022quantum} platforms. In terms of physical implementation, nuclear-spin-based reservoirs~\cite{negoro2018machine} as well as superconducting qubit platforms such as IBM \cite{chen2020temporal} have already been successful in demonstrating various aspects of reservoir computing. 
See Ref.~\cite{mujal2021opportunities} for a more detailed recent overview.

Here, we introduce a quantum reservoir computing approach, see Fig.~\ref{Fig: schematic_qrl}, based on a qubit system driven by a fully connected transverse-field Ising model as a reservoir.
The quantum reservoir consists of two subsystems: (i) input qubits, whose quantum state is shown as $\rho_I$, for encoding variables;  and (ii) hidden qubits, whose quantum state is shown as $\rho_h$, to store past information to predict future outcomes.
The whole reservoir is controlled by full connected Hamiltonian $H$ which is given by 
\begin{equation}\label{eq:Hamiltonian_Ising}
    H=\sum_{ij}J_{ij}X_iX_j + v\sum_iZ_i
\end{equation}
where $X_i$ and $Z_i$ are the Pauli operators acting on $i$-th qubits (see Table~\ref{table:notation_table_adjusted} for the definition of Pauli matrices), $v$ is the strength of the magnetic field which is used as the unit of energy and thus set to be $v=1$ and finally 
$J_{ij}$ are exchange couplings between qubits $i$ and $j$ which are randomly sampled from  $J_{ij}/v\in [0,1]$.
After being randomly initialized, the Hamiltonian $H$ is fixed throughout the process, similar to a classical reservoir.

As shown in the leftmost part of Fig.~\ref{Fig: schematic_qrl}, the state of the ensemble quantum system is initialized as $|0\rangle$ at the beginning of the quantum reservoir computing process: 
\begin{equation}
    \rho_I\otimes\rho_h=|0\rangle\langle0|^{\otimes n_1}\otimes|0\rangle\langle0|^{\otimes n_2},
\end{equation} 
where $n_1$ and $n_2$ are the respective sizes of the input and hidden subsystems, which can be adjusted for different tasks. In this paper, we consider $n_1+n_2=10$, which is more accessible size on current quantum computers.

The goal in this paper is to predict $RV_t$ by using past $k$ steps features, i.e., the time series $\boldsymbol{x}_{t-k}, \cdots,\boldsymbol{x}_{t-1}\}$. 
For convenience of notation, we denote $\boldsymbol{x}_{t-k} = [x_{t-k,1}, x_{t-k,2},\cdots,x_{t-k,n_1}]^T$, where each element $x_{t-k, i}$ represents the $i$-th economic feature at time step $t{-}k$. For example the input vector $\boldsymbol{x}_{t-k}$ may contain three input data at time $t-k$ such as realized volatility $RV_{t-k}$, Dividend Yield  ($DP_{t-k}$) ratio and Earning-Price ($EP_{t-k}$) ratio~(note that all external features used in this paper are defined in Table.~\ref{Table : Macro_features}). In this situation, $\boldsymbol{x}_{t-k}=[RV_{t-k},DP_{t-k}, EP_{t-k}]^T$ contains three input features. 
In this work, we limit $k{=}3$, that is, we consider the memory depth for learning to be only three steps. 
Although generalizing to other memory depths is straightforward and it may appear that considering large memories may yield better results, one must also consider the fact that larger memory depths require deeper quantum circuits. 
Given the finite coherence times available in near-term quantum devices, this trade-off quickly becomes considerable. 
As we shall see, our choice of $k{=}3$ already yields a satisfactory predictive accuracy.
Note that all features are scaled between $[-\pi,+\pi]$, fulfilling the requirement of the $R_Y$ gate. The input vectors $\{ \boldsymbol{x}_{t-3}, \boldsymbol{x}_{t-2}, \boldsymbol{x}_{t-1}\}$ are fed into the reservoir model sequentially. 
The reservoir performs an iterative loop consisting of three steps: \\

\noindent \textbf{Step I: Encoding input data $\boldsymbol{x}_{t-3}$}.  The classical variables $\boldsymbol{x}_{t-3}$, assumed to have $n_1$ features, are encoded to the input qubits of the quantum reservoir through phase encoding, i.e., single-qubit rotation gates around the $Y$ direction, such that the input state is given by
    \begin{eqnarray}
    |\psi_I (\boldsymbol{x}_{t-3}) \rangle &=&R_Y(\boldsymbol{x}_{t-3})  |0\rangle^{\otimes n_1},  \cr 
    \text{where } \quad R_Y(\boldsymbol{x}_{t-3})&=&\bigotimes_{j=1}^{n_1} e^{-i\frac{x_{t-3,j}}{2}Y_j}.
    \end{eqnarray}
    Rotation encoding is a common choice for near-term quantum hardware because it is easy to implement and embeds classical features into quantum states via parameterized single-qubit rotations, thereby effectively leveraging the continuous degrees of freedom of a qubit state.
    Therefore, the input density matrix is given by $\rho_I (\boldsymbol{x}_{t-3}) =  |\psi_I (\boldsymbol{x}_{t-3}) \rangle\langle  \psi_I (\boldsymbol{x}_{t-3})|$. The hidden $n_2$ qubits (the total reservoir thus consists of $n_1+n_2$ qubits) at this stage are initialized as 
    \begin{equation}
        \rho_{h}(t-3) = \big[|0\rangle\langle0|\big]^{\otimes n_2}.
    \end{equation} 
Thus, the collective input state of the quantum reservoir state is given by $\rho_I (\boldsymbol{x}_{t-3})  \otimes \rho_{h}(t{-}3)$. Then the whole reservoir evolves under the action of the Hamiltonian $H$ for a specific time $\tau$, described by $U = e^{-iH\tau}$. The evolution scrambles the input data $\boldsymbol{x}_{t-3}$ across the entire reservoir. Note that since the reservoir is a fully connected graph we set $\tau=1/v$ which is enough to guarantee that the information is scrambled across the whole system.
Following this, while the original $n_1$ qubits are discarded, the quantum state of the hidden $n_2$ qubits carry this information forward to the next step. The quantum state of the hidden qubits after discarding the input qubits now becomes 
    \begin{equation}
        \rho_h (t-2) = \text{Tr}_{I} \left[ U(\tau) \left[ \rho_I (\boldsymbol{x}_{t-3}) \otimes \rho_{h}(t-3) \right] U^{\dagger}(\tau) \right]
    \end{equation}
    where $\text{Tr}_I[\cdot]$ represents partial trace over all the $n_1$ input qubits. \\

\noindent \textbf{Step II: Encoding input data $\boldsymbol{x}_{t-2}$}. After discarding the input $n_1$ qubits at the end of the last round, we replace them with fresh $n_1$ qubits with the encoding 
    \begin{eqnarray}
    |\psi_I (\boldsymbol{x}_{t-2}) \rangle &=&R_Y(\boldsymbol{x}_{t-2})  |0\rangle^{\otimes n_1},  \cr 
    \text{where } \quad R_Y(\boldsymbol{x}_{t-2})&=&\bigotimes_{j=1}^{n_1} e^{-i\frac{x_{t-2,j}}{2}Y_j}.
    \end{eqnarray}
   Therefore the new state of the whole reservoir becomes  $\rho_I (\boldsymbol{x}_{t-2})  \otimes \rho_{h}(t{-}2)$. Similar to the previous step the whole system undergoes another evolution for time $\tau$ which scrambles the input data $\boldsymbol{x}_{t-2}$ across the reservoir. Once again the $n_1$ input qubits are discarded and the quantum state of the hidden qubits is naturally updated to 
        \begin{equation}
        \rho_h (t-1) = \text{Tr}_{I} \left[ U(\tau) \left[ \rho_I (\boldsymbol{x}_{t-2}) \otimes \rho_{h}(t-2) \right] U^{\dagger}(\tau) \right]
    \end{equation}\\

\noindent \textbf{Step III: Encoding input data $\boldsymbol{x}_{t-1}$ and final  measurement}. Similar to the previous step, the  $n_1$ input qubits are replaced by fresh qubits to encode the input data $\boldsymbol{x}_{t-1}$ as
    \begin{eqnarray}
    |\psi_I (\boldsymbol{x}_{t-1}) \rangle &=&R_Y(\boldsymbol{x}_{t-1})  |0\rangle^{\otimes n_1},  \cr 
    \text{where } \quad R_Y(\boldsymbol{x}_{t-1})&=&\bigotimes_{j=1}^{n_1} e^{-i\frac{x_{t-1,j}}{2}Y_j}.
    \end{eqnarray}
   The new state of the whole reservoir thus becomes  $\rho_I (\boldsymbol{x}_{t-1})  \otimes \rho_{h}(t{-}1)$. Again the whole system evolves freely for time $\tau$ which scrambles the input data $\boldsymbol{x}_{t-1}$ across the whole reservoir. At this stage, no qubit is discarded and all of them are measured in the Pauli $Z$ basis. The expectation of measurement outcomes of qubit $j$ is given by 
\begin{equation}
    \langle Z_j \rangle_\tau = \text{Tr} \left[ U(\tau) \left[ \rho_I (\boldsymbol{x}_{t-1}) \otimes \rho_{h}(t-1) \right] U^{\dagger}(\tau) Z_j \right] .
\end{equation}
At any given  step $t$, the measurement outcomes on all the qubits form a vector $\boldsymbol{m}_t^{(1)} = [{\langle Z_1\rangle_\tau,\ldots,\langle Z_{n_1+n_2}\rangle_\tau}]^T$. The training is performed on these measured data, just as the classical reservoir learning. This is accomplished  by a linear regression which is used to approximate $RV_t$ by
\begin{equation}
    \widehat{RV}_t=W_{out}\boldsymbol{m}_t^{(1)},
\end{equation}
where $W_{out}$ is a weight matrix which has to be trained. If we consider the loss function as mean squared error, defined as:
\begin{equation}
    \text{MSE}=\frac{1}{T}\sum_{t=1}^T(RV_t-\widehat{RV}_t)^2
    \label{MSE_loss}
\end{equation}
Then the weight matrix takes an analytical form  given  by the ridge regression as 
\begin{equation}\label{eq:training_Wout}
    W_{out}=RV^\top M_1^\top {(M_1M_1^\top+\delta I)}^{-1},
\end{equation}
where $M_1=[\cdots,\boldsymbol{m}_{t}^{(1)},\cdots,\boldsymbol{m}_T^{(1)}]$, $I$ is the identity matrix and $\delta{=}10^{-8}$ is a small number which is used to guarantee that the matrix is non-singular. 
We call this model as Quantum Reservoir 1 (\texttt{QR1}), since here we only use one quantum reservoir.{\color{black}~see Appendix~\ref{Numerical} for more information.}

\begin{figure}
    \centering
    \includegraphics[width=0.9\linewidth]{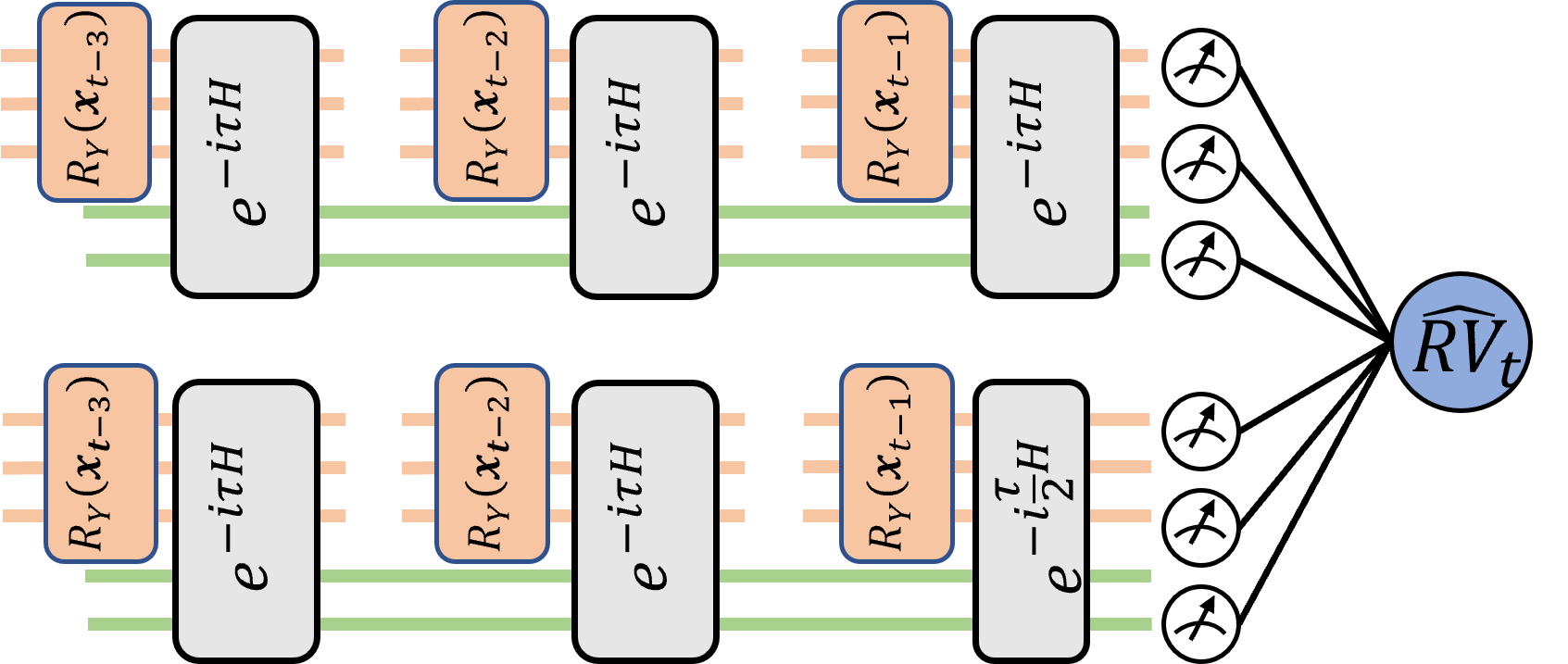}
    \caption{\textbf{Ensemble Reservoir}. The ensemble reservoir approach repeats the procedure twice, with the final step running the quantum dynamical evolution for times $\tau$ and $\tau/2$ in two separate ensembles respectively. This is referred to as \texttt{QR2} in this work, contrasted with \texttt{QR1} not employing ensemble learning. }
    \label{Fig: virtual_node}
\end{figure}

Furthermore, to obtain richer information from the quantum reservoir, we adopt the ensemble reservoir approach~\cite{fujii2017harnessing}.
In this approach, at any given step $t$ apart from the measurement outcomes $\boldsymbol{m}^{(1)}_t$, one can use another reservoir which is almost identical to the previous setup except that the evolution at the step III is run for a time duration of $\tau/2$, instead of $\tau$, see Fig.~\ref{Fig: virtual_node}.

The ensemble of these two reservoir are combined to make a larger vector $\boldsymbol{m}_t^{(2)} = [\langle Z_1\rangle_\tau, \ldots,\langle Z_{n_1+n_2}\rangle_\tau,\langle Z_1\rangle_{\tau/2},\ldots,\langle Z_{n_1+n_2}\rangle_{\tau/2}]^T$, where $\langle Z_j\rangle_{\tau/2}$ represents the measurement of the Pauli operator $Z_j$ in the second reservoir setup in which the last evolution is run for time $\tau/2$, as schematically shown in Fig.~\ref{Fig: virtual_node}. We call this model Quantum Reservoir 2 (\texttt{QR2}) which is expected to be more precise than the \texttt{QR1} as it uses twice as much resources.
{\color{black}It is worth noting that, in the simulation stage, we directly compute the measurement outcomes of the quantum system. In an actual experiment, however, one must perform many repeated measurements (shots) and use their average as the output.
Recent studies~\cite{xiongFundamentalAspectsQuantum2025} have shown that, as the system dimension increases, QRC may suffer from an exponential concentration of measurement results toward a value that is independent of the input. Concretely, the variance of measurement outcomes across different input variables can decay exponentially to zero. As a result, an exponential number of shots would be required to reliably distinguish the outputs corresponding to different inputs, which undermines the potential advantage of QRC.
In our approach, by contrast, the QRC system is kept at a fixed size~(10 qubits), so this issue does not arise~(see Appendix~\ref{Exponential} for more information). The measurement outputs corresponding to different inputs still exhibit sufficiently large variance, indicating that our scheme does not require an excessive number of measurements to obtain accurate output estimates.}

\section{Empirical Analysis}
\label{sec: Empirical Results}

In this section, we present the empirical findings of our study, with a focus on the performance of quantum algorithms in forecasting the realized volatility of the S\&P 500 index. We begin by detailing the data set and the variables used, which include monthly realized volatility, as well as a set of macroeconomic and financial features. Following this, we outline the fitting procedure for competing models, both classical and quantum, and describe the training and evaluation methodologies applied.

A key technical contribution of this study lies in our approach to feature selection within the quantum framework. To optimize the set of features for the quantum reservoir computing models, we employ a forward selection method, a wrapper-based approach that incrementally identifies the most impactful features. This process enables the model to systematically build an optimal feature set by evaluating the performance impact of each addition of variables. Furthermore, to assess the relative importance of each selected feature, we utilize the Shapley value, a method grounded in game theory that fairly distributes the contribution of each feature across different model configurations. The use of Shapley values provides valuable information on the explanatory power of individual features, enhancing the interpretability of the forecast results of the quantum reservoir computing model.

Finally, we analyze the predictive performance of the models, comparing them based on accuracy metrics and statistical tests. This comprehensive evaluation demonstrates the effectiveness of the quantum reservoir computing approach in capturing the complex and nonlinear dynamics of financial market volatility, underscoring its potential advantages over classical models in volatility forecasting.

\begin{table*}[hbt]
    \caption{\textbf{Economic variables used in this study.} Unless otherwise stated, equity data refers to S\&P 500 index.}\label{Table : Macro_features}
\centering
    \begin{tabularx}{\textwidth}{l c l}
    \hline
    Variable            & Symbol & Description \\ 
    \hline
    Realized Volatility & RV     & \begin{tabular}[c]{@{}c@{}}The natural log of the square root of the sum of squared daily returns for a month.\\  RVq and RVa represent the quarterly and annual realized volatility averages.\end{tabular} \\ 

    \begin{tabular}[c]{@{}c@{}}Dividend Yield Ratio \end{tabular} & DP  & \begin{tabular}[c]{@{}c@{}}Dividends over the past year relative to current market prices\end{tabular} \\ 

    \begin{tabular}[c]{@{}c@{}}Earning-Price Ratio \end{tabular} & EP     & \begin{tabular}[c]{@{}c@{}}Earning over the past year relative to current market prices index\end{tabular} \\ 
    Market Excess Return &  MKT  & Fama–French’s market factor: return of U.S. stock index minus the one-month T-bill rate\\ 

    Value Factor  &  HML  & Fama–French’s HML factor: average return on value stocks minus growth stocks\\ 
    
    Size Premium Factor & SMB  & Fama–French’s SMB factor: average return on small-cap stocks minus large-cap stocks\\
    \begin{tabular}[c]{@{}c@{}}Short-Term Reversal Factor\end{tabular} & STR  & \begin{tabular}[c]{@{}c@{}}Fama–French’s STR: average return on stocks with low prior minus high prior returns\\\end{tabular} \\ 

    T-bill Rate & TB & Three-month T-bill rate\\ 

    Monthly Inflation & INF & US monthly inflation rate\\
    Default Spread & DEF & Estimate the credit risk\\
    
    \begin{tabular}[c]{@{}c@{}}Monthly Industrial Production Growth Rate\end{tabular} & IP &  Monthly growth rate of U.S. industrial production\\ \hline
    \end{tabularx}
\end{table*}

\subsection{Data}
\label{Data_sec}
\begin{figure}[h]
    \centering
    \includegraphics[width=1\linewidth]{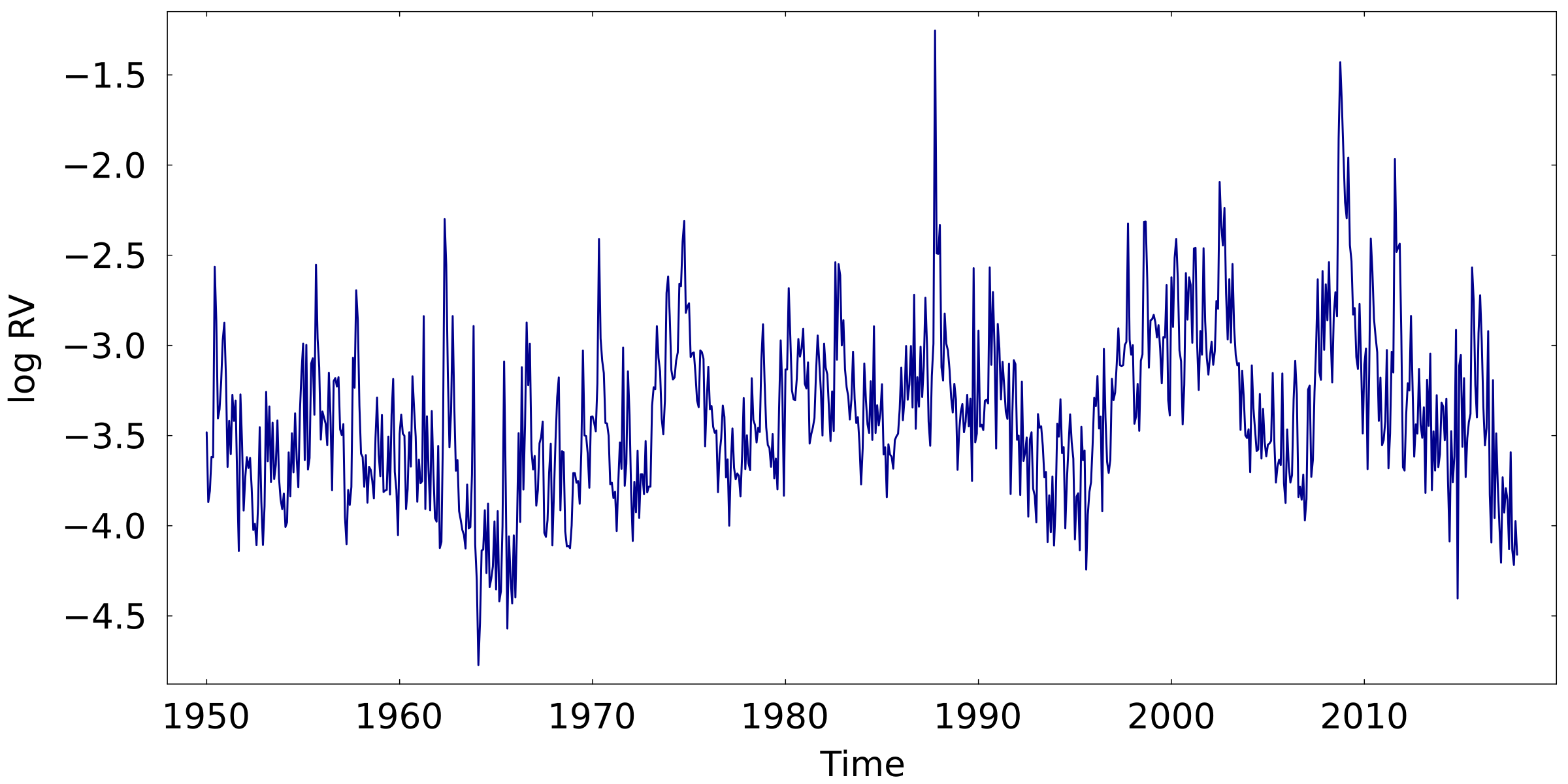}
    \caption{\textbf{Time series of the logarithm of monthly realized volatility data for the S\&P 500 Index.} Data is taken from Yahoo Finance's API~\cite{Yfinance} between January 1950 to December 2017.}
    \label{fig:RV}
\end{figure}

For this study, we utilize a dataset comprising monthly observations of Realized Volatility (RV) for the S\&P 500 index, covering the period from February 1950 to December 2017, resulting in a total of 815 data points. This approach aligns with that of \citet{bucci2020realized}, who employed a similar time frame to assess the performance of neural network models in forecasting realized volatility. Our primary variable of interest, realized volatility ($RV_t$), provides a foundational measure of market uncertainty. As illustrated in Figure~\ref{fig:RV}, realized volatility exhibits significant variability over time, reflecting alternating phases of market turbulence and stability. Capturing these dynamics accurately is critical for financial decision-making and risk management.

To facilitate robust comparison, we align our study with the data and sample period used by \citet{bucci2020realized}. This alignment enables a direct and fair comparison between our quantum reservoir computing approach and established neural network models while maintaining consistency in data characteristics and market environments.

Beyond realized volatility, we include a set of macroeconomic and financial variables known to capture important economic forces and market behavior. Research has shown that adding these features can significantly enhance the performance of volatility forecasting models. For example, \citet{schwert1989} and \citet{engle2009economic} demonstrate that indicators such as Inflation Rates (INF) and Industrial Production Growth (IP) provide an essential context about the state of the broader economy, improving a model’s reliability in forecasting financial volatility. Similarly, valuation measures such as the Dividend-Price Ratio (DP)  and the Earnings-Price ratio (EP) reflect market expectations and investor sentiment, providing valuable predictive power by signaling changes in expected returns and market risk \citep{welch2008comprehensive}. Additionally, market-based factors derived from the Fama-French model \citep{fama1993common,fama1996multifactor}, including Market Excess Returns (MKT), Value Factor (HML), Size Premium Factor (SMB) and Short-Term Reversal Factor (STR) capture critical aspects of equity market dynamics, offering insights into investor behavior and systematic market risk. The inclusion of financial indicators such as the Three-month Treasury Bill Rate (TB) and Default Spread (DEF) further enhances the predictive framework by incorporating measures of interest rate and credit risk conditions.

Including these diverse factors aligns with the findings of \citet{bucci2020realized}, \citet{christensen2023machine}, and \citet{zhang2024volatility}, who showed that combining macroeconomic and financial variables with machine learning models significantly enhances both explanatory power and forecasting accuracy. By incorporating this set of features into our quantum reservoir computing framework, we aim to capture the multifaceted drivers of market volatility. This approach is expected to not only improve forecast precision but also yield deeper insights into the complex interplay between financial markets and economic fundamentals.

Table~\ref{Table : Macro_features} provides a detailed summary of the features utilized in our study, consistent with the variables commonly employed in prior research.

\subsection{Training and Benchmarking}

We employ our quantum reservoir models \texttt{QR1} and \texttt{QR2} to predict realized volatility. To assess their performance, we conduct a comprehensive benchmarking study against established classical models. In particular, we compare quantum reservoir computing models \texttt{QR1} and \texttt{QR2} with the classical linear models \texttt{AR1} (which is \texttt{AR} with memory of one previous step), \texttt{AR3} (which is \texttt{AR} with memory of three previous steps), \texttt{ARMAX}, \texttt{HAR}, \texttt{HARX} as well as the non-linear machine learning based methods \texttt{LSTM}, \texttt{LSTMX}, \texttt{RC} and \texttt{RCX}, see the relevant parts of section \ref{sec: technical_background} for a short introduction to these models.  For all models, including classical models and quantum reservoir computing, we employ a rolling-window approach to estimate and train the models, optimize their parameters, and perform one-step-ahead and five-steps-ahead out-of-sample predictions. Rolling re-estimation at each monthly step is consistent with standard out-of-sample evaluation practices in the volatility forecasting literature, ensuring alignment with recent information and accommodating structural change \citep{Feng2024,PattonSheppard2015,bucci2020realized,PesaranTimmermann2007}. This design promotes comparability across models by isolating learning effects from differences in estimation windows and is particularly important for monthly realized volatility, which reflects aggregated market activity and is influenced by slowly evolving macroeconomic regimes. Without regular updating, parameter estimates may become obsolete in the presence of structural breaks. Our approach, therefore, maintains regime adaptiveness while avoiding excessive sensitivity to high-frequency noise. While our study uses monthly updates to maximize predictive accuracy for benchmarking, a lower re-estimation frequency could be adopted in industrial settings to enhance stability and ease of debugging.

The initial training window spans February 1950 to June 1997 (approximately 571 months). After training and optimizing these models on the initial window, we generate a forecast for the next out-of-sample month (August 1997). Subsequently, the rolling window advances by one month, using data from March 1950 to August 1997 to predict September 1997. This process continues iteratively, rolling through all 245 out-of-sample observations, spanning from August 1997 to December 2017.
During this process, all models are re-estimated at each step to ensure optimal performance for the new rolling window.

Among the classical linear models that we use, \texttt{AR1}, \texttt{AR3}, \texttt{ARMAX}, \texttt{HAR}, \texttt{HARX} are estimated using ordinary least squares or maximum likelihood methods, depending on the structure of the model.
In addition, the machine learning based \texttt{LSTMX} model is implemented with two layers of \texttt{LSTM} cells. The hidden state size is set to 60 for the \texttt{LSTM} model and 50 for the \texttt{LSTMX} variant. After that, a linear regression is used to transform the output of LSTM(X) into the prediction of $\widehat{RV}_t$. The model parameters are optimized using the ADAM optimizer, with a learning rate of $0.001$, the batch size of $64$, and $100$ epochs.
For the classical reservoir computing models (\texttt{RC} and \texttt{RCX}), the reservoir consists of 50 hidden neurons for \texttt{RC} and 20 hidden neurons for \texttt{RCX}. The leak rate, which controls the speed at which the state of the reservoir evolves, is set at 0.6. The spectral radius, which influences the stability and non-linearity of the reservoir, is fixed at 0.9. The input scaling, which is a coefficient applied on $W_{in}$, is set at $0.1$.  The readout matrix $W_{out}$, responsible for mapping the reservoir states to the output, is estimated using ridge regression to mitigate overfitting and improve generalization. 
All machine learning hyperparameters, including architecture choices and training settings, are selected using time-series-aware cross-validation procedures within the training window to ensure generalization and robustness.{\color{black}~(see Appendix~\ref{Hyperparameters} for more informaiotn)}

For quantum reservoir computing, taking into account the current limitations of quantum devices, we used a quantum reservoir with 10 qubits (comprising both input and hidden qubits such that $n_1+n_2=10$) and 3 layers, which means that features with lagged as $t-1$, $t-2$, and $t-3$ will be used to predict $RV_t$.
As mentioned before, we implement two quantum reservoir computing models, \texttt{QR1} and \texttt{QR2}. \texttt{QR1} does not utilize the ensemble reservoir approach, while \texttt{QR2} incorporates an additional reservoir to enhance computational capacity (see Fig.~\ref{Fig: virtual_node}). The quantum reservoir outputs, corresponding to evolved quantum states driven by lagged input encodings, are used as nonlinear regressors in a regularized linear model estimated via ridge regression. In both quantum reservoir models, the weight matrix $W_{out}$ is estimated using ridge regression, as given in Eq.~(\ref{eq:training_Wout}). Implementation-wise, we consider 100 different quantum reservoir instances and train and evaluate each of them separately. We then select the best-performing quantum reservoir and report the results. Similarly for \texttt{LSTM}, we test multiple hyperparameter configurations and report the best-performing results for presentation. Notice that we are reporting
the best-performing results for both \texttt{LSTM} and QRC, hence QRC does not specifically gain any unfair advantage.

\subsection{Closed-Loop Prediction of Multi-Step Future Outcomes }

So far, we have focused on the prediction of  outcomes which are only one step ahead of the input. In other words, all models are trained under a one-step open-loop setting: the model inputs are ground-truth values, and performance is evaluated against the ground truth. For example,
\[
\widehat{RV_t} = f\!\left(RV_{t-3}, Xs_{t-3},\, RV_{t-2}, Xs_{t-2},\, RV_{t-1}, Xs_{t-1}\right),
\]
where $Xs$ denotes other (exogenous) variables~(see Table~\ref{Table : Macro_features}). If a model name does not include the \texttt{X} suffix, then $Xs$ is empty (i.e., no exogenous variables are used). This is called open-loop strategy in which historic data is used for predicting one step ahead. However, one might be interested to predict the outcomes well  ahead of the current data. For instance, by accessing the historic data one might be interested to know the outcome of $S$ steps ahead. This can be done through a closed-loop procedure which is explained below. To accomplish this, we rely on a conventional trained open-loop model which predict $S=1$ step ahead. For predicting $S>1$ steps in the future, we repeatedly call the model and feed its previous predictions back to the model as part of the next input.  Specifically:

\begin{itemize}
  \item For $S=1$,
  \[
  \widehat{RV_t} = f\!\left(RV_{t-3}, Xs_{t-3},\, RV_{t-2}, Xs_{t-2},\, RV_{t-1}, Xs_{t-1}\right).
  \]

  \item For $S=2$,
  \[
  \widehat{RV_{t+1}} = f\!\left(RV_{t-2}, Xs_{t-2},\, RV_{t-1}, Xs_{t-1},\, \widehat{RV_t}, Xs_t\right).
  \]
  At this step, $\widehat{RV_t}$ is taken from the previous step's output. Since our models predict only $RV$ (and not the other variables), $Xs_t$ is still provided as ground truth. However, if such data is not available then a new predictor has to be trained targeting  $Xs_t$ and use its predicted value (i.e. $\widehat{Xs}_t$) as the input of the closed-loop model. Here, for simplicity we assume that the ground truth of $Xs_t$ is available. 
\end{itemize}

Noted that when $S=1$, the closed-loop setting degenerates to the open-loop setting.
Using the closed-loop strategy allows us to assess a model's ability to forecast into the future, which is crucial in many forecasting scenarios. We consider $S=1$ and $S=5$ to evaluate both short-term and longer-term predictive performance. It is worth emphasizing that by increasing $S$ the accuracy is expected to go down as the inputs come from prediction rather than the actual ground truth data. In other words, at every steps, a small  error can propagate to the next steps affecting the accuracy of the model for predicting distant future outcomes.  

\subsection{Feature Selection for Quantum Reservoir Computing}

For the classical models that we selected, the features listed in Table~\ref{Table : Macro_features} can be easily applied to the models. 
However, considering the performance of currently available quantum computers, we have limited the system size of the quantum reservoir model to $10$ qubits. Among these qubits, we use $n_1$ qubits for encoding the features and reserve the rest (i.e., $n_2=10-n_1$) for hidden qubits. Consequently,  we cannot apply all the features listed in Table~\ref{Table : Macro_features} to the quantum model at the same time. 
Therefore, we need to further select the most significant features for our quantum reservoir computing models \texttt{QR1} and \texttt{QR2}.

To refine the key features that best fit our model, we employ one of the wrapper methods, which are model-related feature selection algorithms. The wrapper method treats the machine learning model as a black box, optimizing it to evaluate different subsets of features. This approach simplifies feature selection by framing it as a search problem to identify the best subset of features. The wrapper method requires a strategy to search through possible feature subsets. Common search strategies include exhaustive search, forward selection, backward elimination, and various heuristic approaches, such as genetic algorithms or simulated annealing.

\begin{figure}
    \centering
    \includegraphics[width=0.35\textwidth]{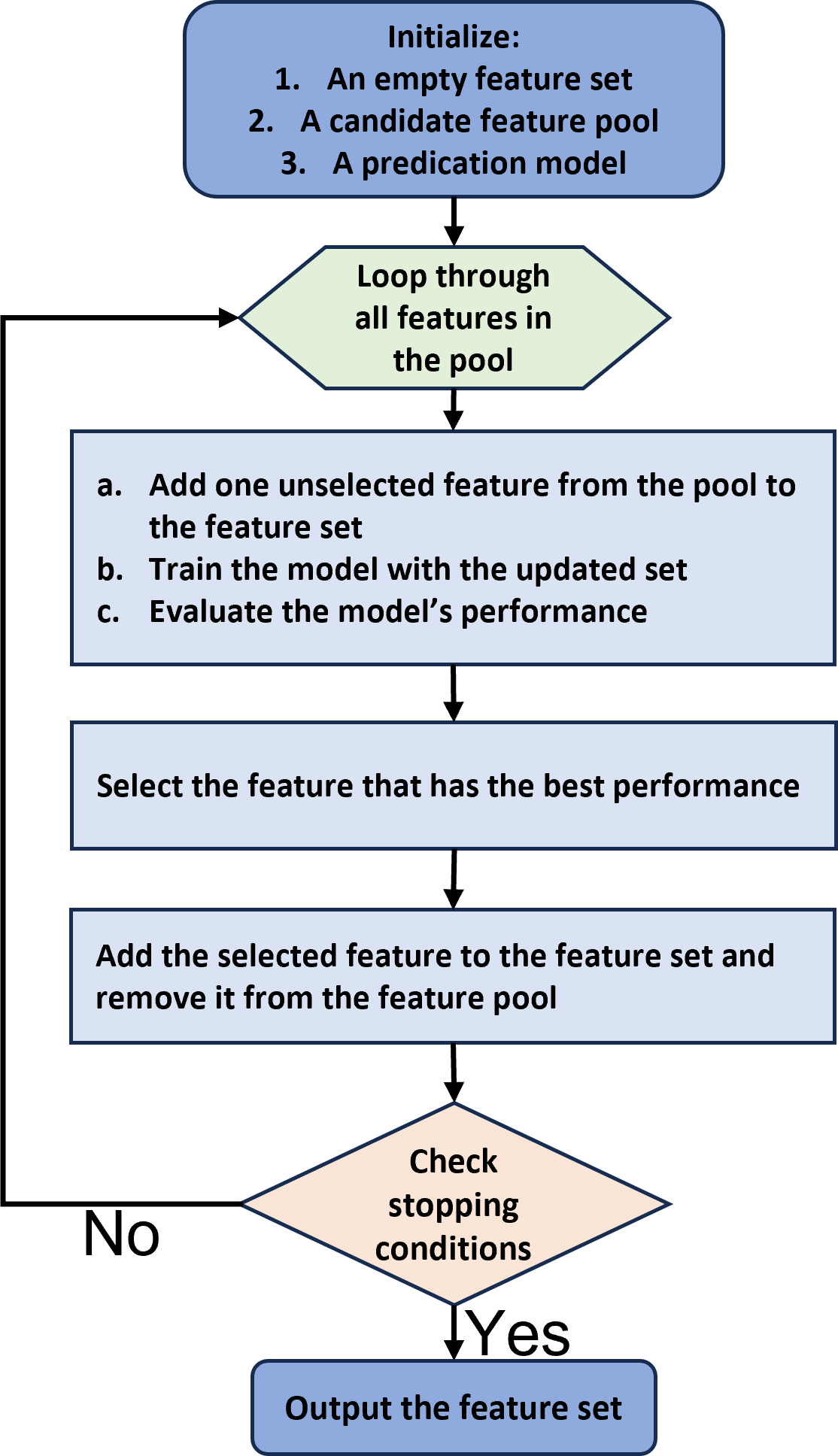}
    \caption{\textbf{Feature selection by wrapper method.} The process begins with an initial pool of features, which contains all candidate features without prior filtering. At each iteration, one feature from the pool is considered as a candidate for inclusion in the optimal feature set (initially empty). Each candidate feature is temporarily added to the current optimal feature set, and the resulting feature set is then used to train a quantum reservoir model. The performance of the trained model is evaluated using the mean squared error (MSE) loss function. Among the candidates, the feature that leads to the lowest MSE is selected and permanently added to the optimal feature set, while the remaining candidates return to the pool. This selection process is repeated iteratively until a predefined stopping criterion is met, such as reaching a target performance threshold or exhausting all candidate features.}\label{Fig: schematic_feature_selection}
\end{figure}

In this study, we use forward selection as the search strategy, which is an iterative method, see Fig.~\ref{Fig: schematic_feature_selection}. 
In forward selection, the process begins with three key components: (i) a prediction model; (ii) an initially empty feature set; and (iii) a feature pool containing all candidate features available for selection. 
At each iteration, the method evaluates the performance of the prediction model by temporarily adding one feature at a time from the candidate pool to the current feature set.
The feature that results in the best performance in the current iteration is permanently added to the optimal feature set and removed from the feature pool. 
The process continues until no feature from the candidate pool can significantly improve the model's performance or until a predefined stopping criterion is met.
The algorithm is schematically shown in Fig.~\ref{Fig: schematic_feature_selection}. 
Through this incremental feature selection process, the dimension of the feature space can be effectively reduced while retaining the most predictive features. This not only improves the explainability and stability of the model, but also avoids overfitting, thus enhancing the accuracy of predicting volatility.

\begin{figure}
    \centering
    \includegraphics[width=0.8\linewidth]{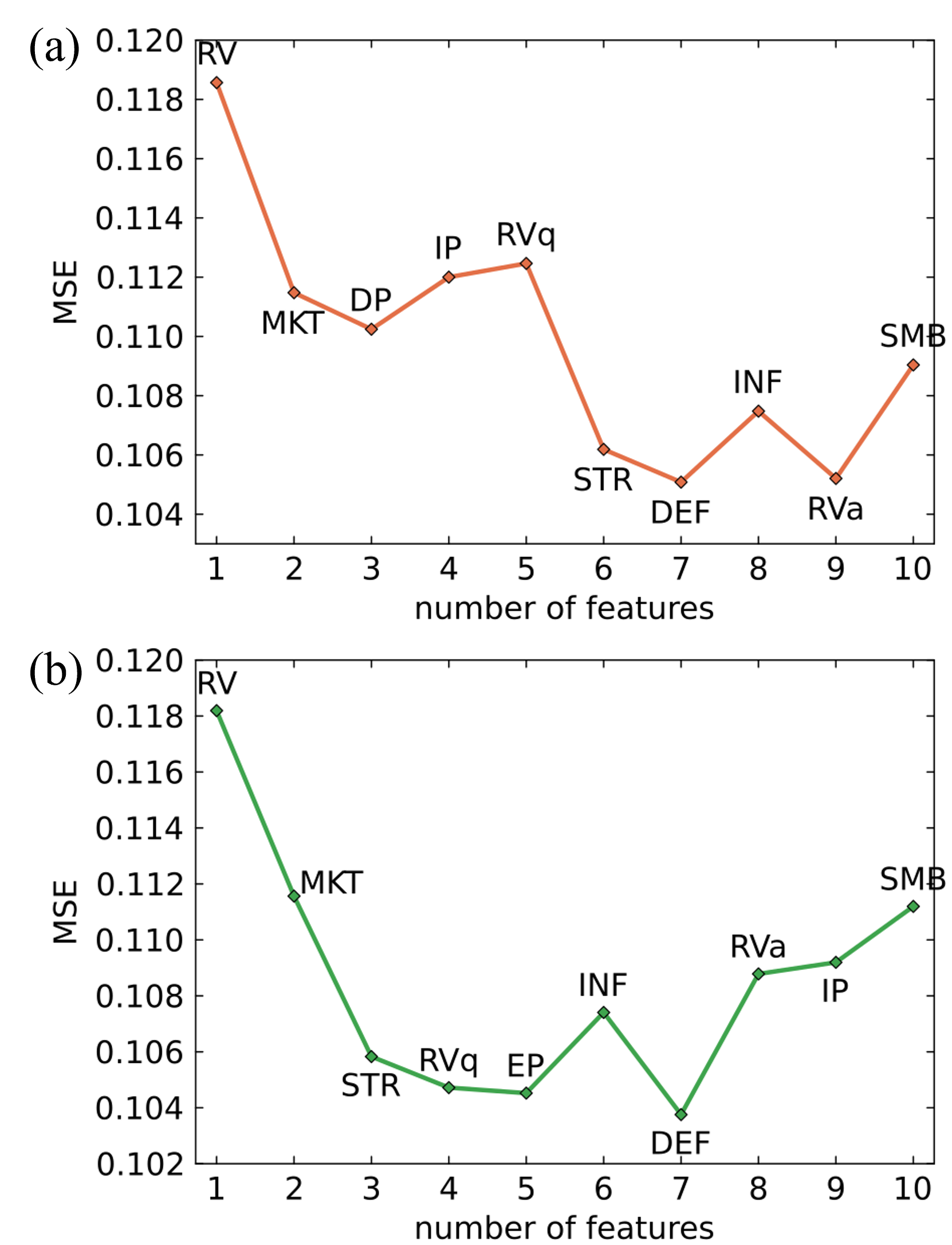}
    \caption{\textbf{Performance (MSE) of \texttt{QR1} and \texttt{QR2} models with their optimal feature sets.} (a) \texttt{QR1} model (orange), the best performance is achieved with the optimal features $\{RV, MKT, DP, IP, RVq, STR, DEF\}$. (b) \texttt{QR2} model (green), the best performance is achieved with the optimal features $\{RV, MKT, STR, RVq, EP, INF, DEF\}$. Order of features in the optimal feature set is left to right along horizontal axis. Note that the features included in these two sets are not the same and have different orderings.}
    \label{Fig: Performance_MSE}
\end{figure}

We use the forward selection method with the stop condiation as the maximum featurs in the optimal features is $10$, see Fig.~\ref{Fig: schematic_feature_selection}, for both the \texttt{QR1} and the \texttt{QR2} models to identify the best subsets of features.
For the \texttt{QR1}, the optimal subset is $F^*_{\texttt{QR1}}=\{RV, MKT, DP, IP, RVq, STR, DEF\}$, and for the \texttt{QR2}, it is $F^*_{\texttt{QR2}}=\{RV, MKT, STR, RVq, EP, INF, DEF\}$. Thus, we have selected $n_1=7$ qubits to input features for each model, allowing $n_2=3$ hidden qubits to carry information from the past data to be used for future prediction. 
We measure the forecast Mean Squared Error (MSE) of \texttt{QR1} and \texttt{QR2} with different subsets of features and present the results in Figs.~\ref{Fig: Performance_MSE}(a) and (b). These figures clearly show that as the number of optimal features increases, the performance of the model initially improves but then deteriorates. One reason for this is that as the number of features increases, more qubits are required to input the values. Since the total size of the quantum system is fixed, the number of qubits for hidden states decreases, preventing the quantum model from capturing the long-memory property of past information. Based on the results in Fig.~\ref{Fig: Performance_MSE}, the optimal number of features for the \texttt{QR1} and \texttt{QR2} models is $n_1=7$. In addition, they share several features, $\{RV, MKT, STR, RVq, DEF\}$. 

\begin{figure*}[t]
     \centering
     \includegraphics[width=1\textwidth]{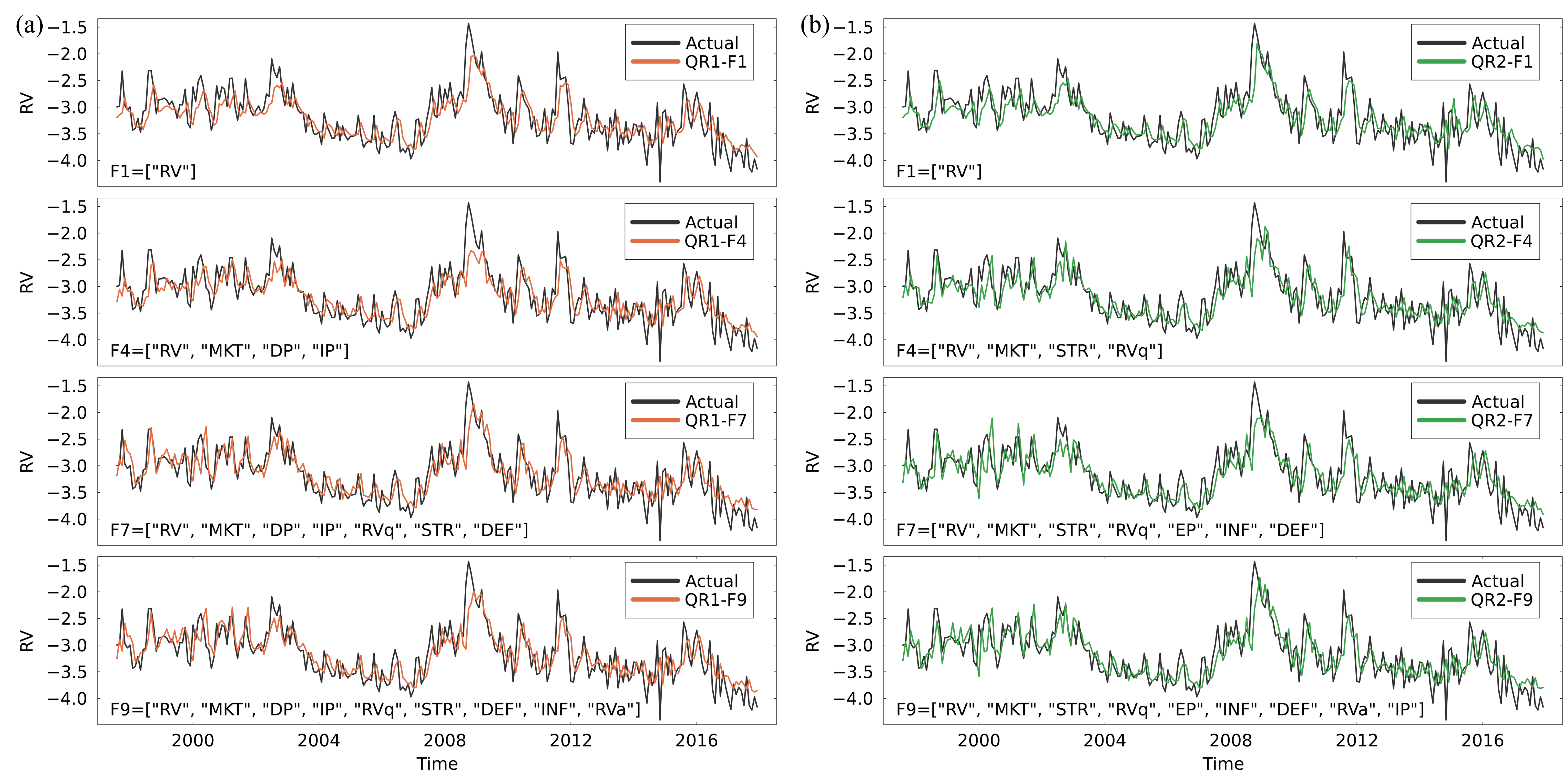}
     \caption{\textbf{Forecasting performance of \texttt{QR1} (a) and \texttt{QR2} (b) models on realized volatility with different feature subsets.} Each subfigure presents the actual realized volatility (black line) of the S\&P 500 index alongside predictions from the quantum reservoir computing models (\texttt{QR1} in orange and \texttt{QR2} in green) using varying feature configurations. \texttt{QR2} demonstrates higher accuracy, especially during volatility peaks, suggesting it has better responsiveness to market fluctuations.}
     \label{Fig: performance with diff F}
\end{figure*}

To see how increasing the features affects the prediction quality, in Fig.~\ref{Fig: performance with diff F}(a), we plot the realized volatility as well as the \texttt{QR1} forecast for the first $1$, $4$, $7$, and $9$ features, which are  shown in Fig.~\ref{Fig: Performance_MSE}. 
As the figure clearly shows by increasing the features, the prediction improves until the number of features reach $n_1=7$. By further increasing the number of features, the prediction quality cannot be further improved because the number of hidden qubits decreases; thus, the reservoir lacks sufficient memory to incorporate past information for future prediction. The corresponding results for the \texttt{QR2} model is also shown in Fig.~\ref{Fig: performance with diff F}(b). The same conclusion about the impact of features on prediction quality can be deduced from the \texttt{QR2} model. 
In addition, by comparing the corresponding panels in Fig.~\ref{Fig: performance with diff F}(a) Fig.~\ref{Fig: performance with diff F}(b) one can clearly see that the \texttt{QR2} outperforms the \texttt{QR1} prediction. This is expected as the  the \texttt{QR2} uses twice more measurement data for predicting the future outcomes. 

\subsection{Model Interpretability  }

Any machine learning based algorithm should not only minimize its loss function, but preferably also explains which features contribute the most towards this goal. In the context of this work, we want to understand which external features contribute the most towards realized volatility. In the previous section devoted to the forward feature selection method, we adopted a constructive approach of adding one feature at a time from scratch. However, this leaves an equally important question untouched - given a set of features already being encoded into our simulation, how do we separate out individual contributions coming from each feature towards the predictive success of our quantum reservoir computing based method? 

\begin{figure*}
    \centering
    \includegraphics[width=1\textwidth]{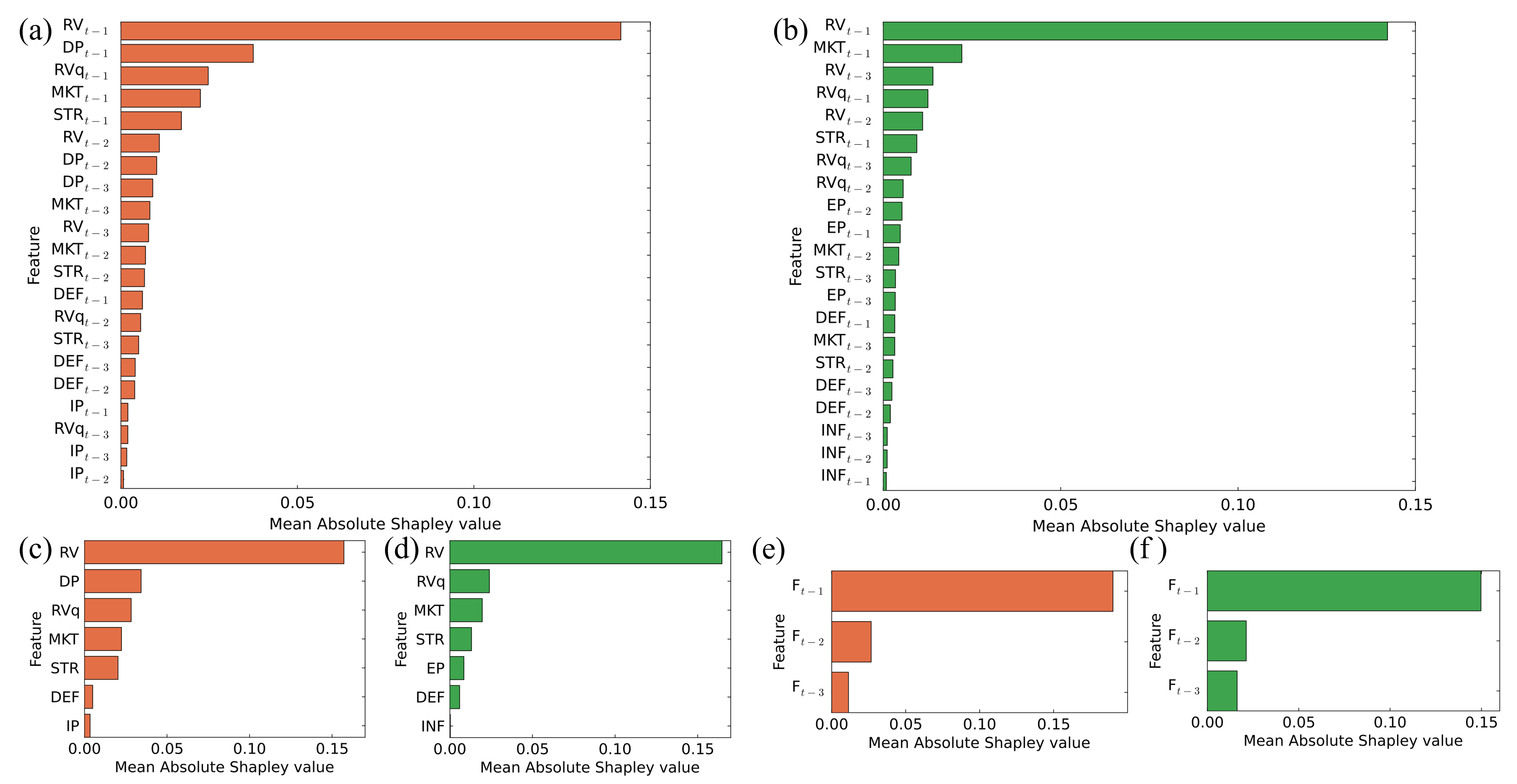}
    \caption{\textbf{Shapley-value based feature importance test.} The Shapley values for the optimal feature set are presented for \texttt{QR1} and \texttt{QR2}. Panels (a), (c), and (e) correspond to \texttt{QR1}, while panels (b), (d), and (f) correspond to \texttt{QR2}.  
    (a) and (b) show the Shapley values for each feature at different time lags.  
    (c) and (d) show the Shapley values for each feature, aggregated across all time lags.  
    (e) and (f) show the Shapley values for each lagged input.}
\label{Fig: Shapley_values}
\end{figure*}

To answer this  we employ the Shapley value method, a model-agnostic technique grounded in cooperative game theory. 
Originally introduced in Ref.~\cite{shapley1953value}, the Shapley value offers a principled framework for fairly distributing the total payoff among participants in a coalition based on their individual contributions. 
This approach considers all possible combinations and permutations of features, helping to quantify the contribution of each feature to the model prediction.  \citet{lundberg2017unified} adapted this concept to machine learning, proposing Shapley values as a unified and interpretable measure of feature importance that satisfies key axioms such as consistency and local accuracy.
In this spirit, our analysis applies Shapley values to the quantum reservoir computing framework to assess how macro-financial variables contribute to realized volatility forecasts, bridging interpretability with quantum-enhanced prediction.
In theory, exact computation of Shapley values is exponential in the number of features; therefore, we typically rely on approximation methods in practice. In this work, we estimate Shapley values for the \texttt{QR1} and \texttt{QR2} models using the Monte Carlo sampling method~\cite{strumbeljExplainingPredictionModels2014}, as implemented in the Julia package \texttt{ShapML}.
The choice of the features is decided according to one of following three strategies: 
\begin{enumerate}
    \item \textbf{Individual feature contribution:} In this strategy, we evaluate the contribution of each feature at each lag separately, e.g. $\{RV_{t-3},MKT_{t-3},\cdots,RV_{t-2},MKT_{t-2},\cdots,\newline RV_{t-1},MKT_{t-1},\cdots\}$. The results are shown in Figs.~\ref{Fig: Shapley_values}(a) and (b), for \texttt{QR1} and \texttt{QR2} models, respectively. As expected, in both models, the $RV_{t-1}$ has the most contribution in predicting the future outcome $RV_t$. \\
    \item  \textbf{Feature-family contribution:} In this strategy, we evaluate the contribution of each feature family irrespective of the time lag. Therefore, during the evaluation of the Shapley value when we evaluate the contribution of one feature type, e.g. for realized volatility $RV$, we consider contributions from all $\lbrace RV_{t-3}, RV_{t-2}, RV_{t-1}\rbrace$ time-lagged steps. The results are shown in Figs.~\ref{Fig: Shapley_values}(c) and (d) for the \texttt{QR1} and \texttt{QR2} models respectively. Interestingly, the Shapley value-based method furnishes a somewhat different ordering of feature importance compared to the forward selection method results depicted in Fig.~\ref{Fig: Performance_MSE}. Note that this is not inconsistent and stems from the fact that the Shapley value is computed when all the features are embedded in the model while the forward selection method works by considering each individual feature in isolation. \\
    \item \textbf{Time lagged feature contribution:} In this strategy, we evaluate the contribution of all the features at a given time-lag, e.g. $F_{t-k}=\{RV_{t-k},MKT_{t-k},\cdots\}$. We expect that the more recent data contributes the most towards future prediction compared to historical data. The results are shown in Figs.~\ref{Fig: Shapley_values}(e) and (f) for \texttt{QR1} and \texttt{QR2} respectively. As expected, in both models our expectation is borne out and the more recent data are indeed more useful. 
\end{enumerate}

\subsection{Performance Metrics and Forecast Evaluation}

The selection of appropriate performance metrics is pivotal in the evaluation of volatility forecasting models~\cite{HansenLunde2005}.  In all the models, either classical or quantum, we have used MSE, defined in Eq.~\eqref{MSE_loss}, as a loss function which is minimized. Although MSE is a popular figure of merit for evaluating the performance of a time series predictor, we stress that it is blind to the direction of error. In other words, overestimation as well as underestimation, both get equally penalized. However, in the concrete case of realized volatility prediction, underestimation is far more dangerous as it can lead to serious undermining of emerging risks in the market. Thus, we specifically need a figure of merit which quantifies the degree of underestimation.  In order to accomplish this task, we employ the widely recognized Quasi-Likelihood (QLIKE) as our figure of merit which captures both symmetric and asymmetric aspects of forecast errors, thereby providing a comprehensive assessment of model performance. The QLIKE is defined as:
\[
\text{QLIKE} = \frac{1}{T} \sum_{t=1}^{T} \left( \log \widehat{RV}_t^2 + \Big(\frac{RV_t}{\widehat{RV}_t}\Big)^2 \right),
\]
As it is evident from the definition, the QLIKE function penalizes under-predictions more heavily, reflecting the higher costs of underestimating volatility in risk management and option pricing. Furthermore, \citet{Patton2011} demonstrates that QLIKE is robust to measurement errors in volatility proxies. 
For statistical evaluation of predictive accuracy, we employ the Model Confidence Set (MCS) procedure proposed by \citet{HansenLundeNason2011} and the Diebold-Mariano (DM) test \citep{DieboldMariano1995}. We utilize MSE as our primary loss function due to its simplicity, while also reporting QLIKE values to assess asymmetry and penalize underestimation of volatility more strongly.\\

\emph{Model Confidence Set.--}  
The MCS procedure is designed to identify a subset of forecasting models whose predictive performance is statistically indistinguishable from that of the best model at a specified confidence level. The procedure begins with an initial set of candidate models, denoted $\mathcal{M}_0$, and iteratively eliminates inferior models based on tests of equal predictive ability. Formally, let $L_{i,t}$ denote the loss (e.g., squared error) incurred by model $i$ at time $t$ and define the loss differential between models $i$ and $j$ as:
\[
d_{ij,t} = L_{i,t} - L_{j,t}, \quad i, j \in \mathcal{M}_0.
1\]
The null hypothesis of equal predictive ability (EPA) is:
\[
H_{0,\mathcal{M}} : \mu_{i,j} = \mathbb{E}[d_{ij,t}] = 0 \quad \forall i, j \in \mathcal{M},
\]
where $\mathcal{M} \subseteq \mathcal{M}_0$ denotes the set of models under consideration at a given iteration. Rejection of $H_{0,\mathcal{M}}$ implies that at least one model in $\mathcal{M}$ exhibits significantly inferior forecasting performance.

To test $H_{0,\mathcal{M}}$, a range-type test statistic is employed:
\[
T_{R,\mathcal{M}} = \max_{i \in \mathcal{M}} t_i,
\]
where $t_i$ denotes a studentized statistic comparing the performance of model $i$ to others (e.g., via average loss differences). If the null is rejected, the model with the most evidence against it is eliminated according to the elimination rule:
\[
e_{\mathcal{M}} = \arg\max_{i \in \mathcal{M}} \left( \sup_{j \in \mathcal{M}} \hat{d}_{i,j} \right),
\]
where $\hat{d}_{i,j}$ is the sample mean of the loss differential. This elimination process is repeated until the null hypothesis is no longer rejected.

The final surviving set of models is denoted by $\widehat{\mathcal{M}}^*_{1-\alpha}$, the $(1-\alpha)$ MCS, which is asymptotically guaranteed to contain the model(s) with the lowest expected loss with probability at least $1 - \alpha$. In this study, we set $\alpha = 0.05$, ensuring that the resulting confidence set contains the best-performing model(s) with at least 95\% probability in large samples. We complement the MCS analysis with pairwise Diebold-Mariano tests to assess the statistical significance of forecast performance differentials between models. This dual approach allows us to validate robustness in model rankings and mitigate the limitations associated with relying on a single evaluation metric.\\

\emph{Diebold-Mariano test (DM test).-- } The DM test is a particular adaptation of the above procedure for time-series data, and evaluates the null hypothesis of equal predictive accuracy between two competing models, formally stated as:
\[
H_0: \mathbb{E}[d_t] = 0,
\]
where \( d_t = L_{1,t} - L_{2,t} \) is the loss differential at time \( t \) between models 1 and 2. Here, \( L_{i,t} \) represents the loss at time \( t \) for model \( i \), and the choice of loss function \( L_{i,t} \) (e.g., MSE in this paper) reflects the predictive objective. The DM test statistic is computed as:
\[
\text{DM statistic} = \frac{\bar{d}}{\sqrt{\widehat{\sigma}^2 / T}},
\]
where \( \bar{d} = \frac{1}{T} \sum_{t=1}^T d_t \) is the sample mean of the loss differential, \( \widehat{\sigma}^2 \) is the Newey-West adjusted variance of \( d_t \), and \( T \) is the sample size. The test accounts for potential autocorrelation and heteroskedasticity in \( d_t \), making it suitable for time-series data with serially correlated forecast errors.

\begin{table*}[hbt]
\centering
    \caption{\textbf{Model Confidence Set.} Performance of classical and quantum Models in predicting realized volatility with predicting $S$-ahead step. The MSE and QLIKE are presented alongside the MCS $p$-values ($P_{\text{MCS}}$). Asterisks ($^*$) indicate inclusion in the model confidence set. The quantum reservoir computing models outperform all classical models, with \texttt{QR2} exhibiting the best performance across all measures.}
\begin{tabularx}{\textwidth}{lcccc} 
    \multicolumn{5}{c}{\textbf{S=1 (entire sample: 1997.08-2017.12) (1950.01- 2017.12)}} \\
    \hline
    Model & \multicolumn{2}{l}{MSE} & \multicolumn{2}{l}{QLike} \\
    & Loss of MSE     & $P_{MCS}$ of MSE    & Loss of QLike     & $P_{MCS}$ of QLike \\
    \hline
    \multicolumn{5}{c}{\textbf{Classical Time Series Models}}  \\
    \hline
    \texttt{HAR}                  & 0.1476    & 0.0004       & 2.0431                  & 0.0008                    \\
    \texttt{HARX}                 & 0.1508    & 0.0004       & 2.2436                  & 0.0008                    \\
    \texttt{AR1}                & 0.1304    & 0.0065       & 1.7279                  & 0.0050                    \\
    \texttt{AR3}                & 0.1178    & 0.0936       & 1.5893                  & 0.0861                    \\
    \texttt{ARMAX}          & 0.1145    & $0.4406^*$   & 1.6196                  & $0.4355^*$                \\
    \texttt{LSTM}                 & 0.1295    & 0.0221       & 1.7909                  & 0.0188                    \\
    \texttt{LSTMX}                & 0.1185    & $0.4406^*$   & 1.7571                  & $0.4355^*$                \\
    \texttt{RC}                   & 0.1441    & 0.0084       & 2.1011                  & 0.0061                    \\
    \texttt{RCX}                  & 0.1089    & $0.6086^*$   & 1.6480                  & $0.6106^*$                \\
    \hline
    \multicolumn{5}{c}{\textbf{Quantum Reservoir Computing}}  \\
    \hline
    \texttt{QR1}                  & 0.105     & $0.7603^*$   & 1.4427                  & $0.7510^*$                \\
    \texttt{QR2}                  & \textbf{0.103} & $1.0000^*$ & \textbf{1.4004}          & $1.0000^*$                \\
    \hline
    \end{tabularx}
    \vspace{1ex}
     \begin{minipage}{\textwidth}
     \footnotesize
    \end{minipage}

    \begin{tabularx}{\textwidth}{lcccc} 
    \multicolumn{5}{c}{\textbf{S=5 (entire sample: 1998.01-2017.12) (1950.01- 2017.12)}} \\
    \hline
    Model & \multicolumn{2}{l}{MSE} & \multicolumn{2}{l}{QLike} \\
    & Loss of MSE     & $P_{MCS}$ of MSE    & Loss of QLike     & $P_{MCS}$ of QLike \\
    \hline
    \multicolumn{5}{c}{\textbf{Classical Time Series Models}}  \\
    \hline
    \texttt{HAR}  & 0.2143  & $0.1291^*$   & 2.9041  & $0.1275^*$  \\
    \texttt{HARX} & 0.2934  & 0.0044 & 4.5800  & 0.0040\\ 
    \texttt{AR1}  & 0.2642  & $0.0938^*$  & 3.4136 & $0.0937^*$ \\
    \texttt{AR3}  & 0.2134  & $0.1302^*$  & 2.8369 & $0.1297^*$ \\
    \texttt{ARMAX}& 0.2134  & $0.1302^*$ & 3.0703 & $0.1297^*$             \\
    \texttt{LSTM} & 0.1831  & $0.1291^*$   & 2.4600 & $0.1275^*$ \\
    \texttt{LSTMX}& 0.2200  & $0.0925^*$  & 3.4512 & $0.0851^*$\\
    \texttt{RC}   & \textbf{0.1528}  & $1.0000^*$     & \textbf{2.0551} & $1.0000^*$  \\
    \texttt{RCX}  & 0.1667  & $0.6333^*$  & 2.4605 & $0.6251^*$ \\
    \hline
    \multicolumn{5}{c}{\textbf{Quantum Reservoir Computing}}  \\
    \hline
    \texttt{QR1} & 0.1556    & $0.7642^*$ & 2.1518 & $0.7703^*$\\
    \texttt{QR2} & 0.1663 & $0.6333^*$ & 2.2332  & $0.6251^*$  \\
    \hline
    \end{tabularx}
    \label{Table: performance_benchmarking}
\end{table*}

\begin{table*}
\centering
\caption{\textbf{Diebold-Mariano Test.}  DM statistic values (lower triangular matrix) and $p$-values (upper triangular matrix) for model comparisons. The statistic values closer to zero indicate that the pair of models have similar predictive accuracy. Lower $p$-values suggest rejection of the null hypothesis of equal predictive ability.}
\begin{tabularx}{\textwidth}{lccccccccccc}

\multicolumn{12}{c}{\textbf{}}                                                             \\
\hline
           & \texttt{HAR}   & \texttt{HARX}  & \texttt{AR1}   & \texttt{AR3}  & \texttt{ARMAX}  & \texttt{LSTM}  & \texttt{LSTMX}   & \texttt{RC}    & \texttt{RCX}  & \texttt{QR1} & \texttt{QR2}  \\
           \hline
\texttt{HAR}        &        & 0.575 & 0.203  & 0.004  & 0.009      & 0.006  & 0.01   & 0.801 & 0.001 & 0     & 0     \\
\texttt{HARX}       & -0.562 &       & 0.116  & 0.001  & 0.003      & 0.004  & 0.002  & 0.597 & 0     & 0     & 0     \\
\texttt{AR1}      & 1.275  & 1.577 &        & 0.002  & 0.077      & 0.925  & 0.301  & 0.154 & 0.009 & 0.001 & 0     \\
\texttt{AR3}      & 2.929  & 3.313 & 3.12   &        & 0.652      & 0.065  & 0.937  & 0.003 & 0.17  & 0.036 & 0.014 \\
$\texttt{ARMAX}$ & 2.636  & 2.999 & 1.775  & 0.451  &            & 0.134  & 0.657  & 0.006 & 0.393 & 0.12  & 0.111 \\
\texttt{LSTM}       & 2.764  & 2.938 & 0.094  & -1.856 & -1.502     &        & 0.268  & 0.168 & 0.017 & 0.006 & 0.002 \\
\texttt{LSTMX}      & 2.583  & 3.168 & 1.037  & -0.079 & -0.445     & 1.109  &        & 0.028 & 0.229 & 0.114 & 0.09  \\
\texttt{RC}         & 0.252  & 0.53  & -1.431 & -3.032 & -2.758     & -1.383 & -2.206 &       & 0     & 0     & 0     \\
\texttt{RCX}        & 3.395  & 4.001 & 2.629  & 1.376  & 0.856      & 2.403  & 1.205  & 3.639 &       & 0.501 & 0.296 \\
\texttt{QR1}      & 3.695  & 4.149 & 3.322  & 2.104  & 1.561      & 2.788  & 1.584  & 3.82  & 0.674 &       & 0.771 \\
\texttt{QR2}      & 3.806  & 4.289 & 3.577  & 2.465  & 1.6        & 3.111  & 1.701  & 4.584 & 1.048 & 0.291 &       \\
\hline
\end{tabularx}
\label{Table: DM_test}
\end{table*}

\emph{MCS results.--} The results of MCS are presented in Table~\ref{Table: performance_benchmarking}. The table reports the performance metrics (MSE and QLIKE) along with the MCS $p$-values for classical time series models and quantum reservoir computing models. 
In particular, when S=1, namely one-step ahead prediction, quantum models, especially \texttt{QR2}, demonstrate superior performance across all measures. From the upper table in Table~\ref{Table: performance_benchmarking}, we further observe that the \texttt{QR2} model achieves the lowest MSE and QLIKE values, indicating its superior predictive accuracy. The MCS $p$-values further reinforce this finding, as \texttt{QR2} attains a $p$-value of 1.0, signifying its inclusion in the superior set of models at the 95\% significance level. In contrast, most classical models exhibit significantly higher loss values and lower MCS $p$-values, suggesting inferior performance. Examining the MCS $p$-values, models with asterisks are included in the Model Confidence Set $\mathcal{M}_{\text{MCS}}$, indicating that their predictive accuracy is statistically indistinguishable from the best-performing model.  Among the classical models, only \texttt{ARMAX}, \texttt{LSTMX}, and \texttt{RCX} are included in the MCS based on their $p$-values, although their loss metrics are still higher than those of the quantum models. Additionally, \texttt{QR1} is also included in the MCS, further highlighting the strong performance of quantum models. Examining the results, we also observe that including additional features sometimes enhances the model performance. For example, \texttt{ARMAX}  and \texttt{LSTMX} exhibit lower loss values and higher MCS $p$-values compared to their counterparts \texttt{AR3} and \texttt{LSTM}, suggesting that the inclusion of characteristics improves the accuracy of forecasting in these models. However, this improvement is not consistent across all models. Notably, \texttt{HARX} performs slightly worse than \texttt{HAR} in both MSE and QLIKE metrics, indicating that additional features do not always contribute to better performance. This suggests that the effectiveness of including features depends on the model structure and the relevance of the features.
When consider $S=5$, namely, the long-term prediction, both Quantum and Classical reservoir demonstrated outstanding performance compared with other models.
The performances of \texttt{QR1} and \texttt{QR2} are very similar to that of the classic reservoir computing. Considering that we only used a simple reservoir model, these indirectly demonstrate the capabilities of Quantum reservoir.
\\

\emph{Diebold-Mariano Test results.-- } Table~\ref{Table: DM_test} presents the Diebold-Mariano test statistic(s) and corresponding $p$-value(s) for pairwise model comparisons, providing further evidence of the quantum models' outperformance. Analyzing the DM test results, we find that the quantum models (\texttt{QR1} and \texttt{QR2}) significantly outperform most classical models. For instance, the $p$-values for comparisons between \texttt{QR2} and classical models like \texttt{HAR} and  \texttt{AR1}  are effectively zero, indicating strong evidence against the null hypothesis of equal predictive accuracy. Additionally, the test statistics are positive and relatively large, further supporting the superiority of the quantum models. Interestingly, the comparison between \texttt{QR1} and \texttt{QR2} yields a high $p$-value (0.771), suggesting no significant difference in predictive accuracy between the two quantum models. This implies that both quantum models perform comparably well, although \texttt{QR2} has a slight edge based on the loss metrics in Table~\ref{Table: performance_benchmarking}. The results also reveal that among classical models, \texttt{AR3},  and \texttt{ARMAX} exhibit relatively better performance, with higher $p$-values when compared to other classical models. However, they still fall short when compared to the quantum models. The results further illustrate the mixed impact of including additional features. For some model pairs, such as \texttt{LSTM} and \texttt{LSTMX}, the inclusion of features leads to significant improvements, as evidenced by lower test statistics and higher $p$-values. In contrast, comparisons between \texttt{AR3} and \texttt{ARMAX} yield $p$-values that indicate no significant difference, implying that the additional features in \texttt{ARMAX} do not provide substantial benefits over \texttt{AR3}. Overall, the results suggest that while incorporating additional features can enhance model performance in certain cases, it does not universally guarantee better forecasts. The effectiveness of additional features appears to be model-specific, highlighting the importance of selecting relevant variables that contribute meaningfully to volatility forecasting. \\

Overall, the combined evidence from the loss functions, MCS procedure, and DM test underscores the superior predictive performance of the quantum reservoir computing models over classical time series and machine learning models in forecasting realized volatility.

\section{Conclusions and Outlook}
\label{sec: conclusions and outlook}
Quantum reservoir computing presents a promising new paradigm for the modeling and forecasting of time series data. In this study, we have demonstrated how a specific instantiation of quantum reservoir computing that utilizes disordered spin systems as reservoirs can be effectively employed for forecasting realized volatility in financial markets. Our empirical evaluation reveals that even with a modest quantum architecture comprising only ten qubits and a moderated number of macroeconomic input features per instance, the proposed framework outperforms classical models in predictive accuracy.  While we do not advance any claim to quantum supremacy in the rigorous sense of the term since this is a phenomenological study, the results obtained herein do indicate a potential future application of quantum reservoir computing towards applying them to other similar financial problems. Our protocol is especially suitable to use in real-life quantum computers based on trapped ion platforms, in particular, \cite{kielpinski2002architecture,erhard2021entangling,akhtar2023high}. These setups offer excellent all-to-all connectivity between spins, as required in our specific choice of the quantum reservoir. We anticipate that alternative quantum reservoir architectures with different connectivity constraints may yield comparable predictive performance, although empirical verification would be necessary to substantiate this speculation. \\

Quantum computing for machine learning still faces significant challenges, both in hardware and algorithmic development. On the hardware side, current quantum devices are still in their infancy, constrained by limited qubit counts, restricted connectivity, finite coherence times, and imperfect readout. Overcoming these limitations will require advances in error correction—a milestone likely still a decade away. Nevertheless, near-term quantum devices provide valuable opportunities to develop and test small-scale algorithms, paving the way for large-scale applications on future error-corrected quantum computers. From an algorithmic perspective, it remains unproven whether quantum computers can outperform classical machine learning methods on classical data. In this work, we demonstrate that quantum computers can indeed solve prediction tasks in stochastic time series. While our quantum reservoir learning approach shows advantages over several classical methods, we do not claim a definitive quantum advantage, as such a conclusion would require rigorous theoretical proof beyond the scope of this paper. Moreover, intrinsic limitations on QRC like the exponential concentration problem \cite{Xiong2025on}, whereby the predictions may become input-independent due to concentration of expectation values for many-qubit reservoirs, exist and have to be addressed in future work.\\

\noindent \textbf{Code Availability.}  The Code used in this paper can be found  \href{https://github.com/LeeQY1996/Quantum-Reservoir-computing-for-Realized-Volatility-Forecasting}{here}  in the Github Repositories. \\

\noindent \textbf{Acknowledgments.}   AB acknowledges support from the National Natural Science Foundation of China (grants No. 12274059,
No. 12574528, No. 1251101297 and No. W2541020).

\bibliography{rv_forecasting_PRR}

\begin{thebibliography}{134}%
\makeatletter
\providecommand \@ifxundefined [1]{%
 \@ifx{#1\undefined}
}%
\providecommand \@ifnum [1]{%
 \ifnum #1\expandafter \@firstoftwo
 \else \expandafter \@secondoftwo
 \fi
}%
\providecommand \@ifx [1]{%
 \ifx #1\expandafter \@firstoftwo
 \else \expandafter \@secondoftwo
 \fi
}%
\providecommand \natexlab [1]{#1}%
\providecommand \enquote  [1]{``#1''}%
\providecommand \bibnamefont  [1]{#1}%
\providecommand \bibfnamefont [1]{#1}%
\providecommand \citenamefont [1]{#1}%
\providecommand \href@noop [0]{\@secondoftwo}%
\providecommand \href [0]{\begingroup \@sanitize@url \@href}%
\providecommand \@href[1]{\@@startlink{#1}\@@href}%
\providecommand \@@href[1]{\endgroup#1\@@endlink}%
\providecommand \@sanitize@url [0]{\catcode `\\12\catcode `\$12\catcode `\&12\catcode `\#12\catcode `\^12\catcode `\_12\catcode `\%12\relax}%
\providecommand \@@startlink[1]{}%
\providecommand \@@endlink[0]{}%
\providecommand \url  [0]{\begingroup\@sanitize@url \@url }%
\providecommand \@url [1]{\endgroup\@href {#1}{\urlprefix }}%
\providecommand \urlprefix  [0]{URL }%
\providecommand \Eprint [0]{\href }%
\providecommand \doibase [0]{https://doi.org/}%
\providecommand \selectlanguage [0]{\@gobble}%
\providecommand \bibinfo  [0]{\@secondoftwo}%
\providecommand \bibfield  [0]{\@secondoftwo}%
\providecommand \translation [1]{[#1]}%
\providecommand \BibitemOpen [0]{}%
\providecommand \bibitemStop [0]{}%
\providecommand \bibitemNoStop [0]{.\EOS\space}%
\providecommand \EOS [0]{\spacefactor3000\relax}%
\providecommand \BibitemShut  [1]{\csname bibitem#1\endcsname}%
\let\auto@bib@innerbib\@empty
\bibitem [{\citenamefont {Shor}(1994)}]{shor1994algorithms}%
  \BibitemOpen
  \bibfield  {author} {\bibinfo {author} {\bibfnamefont {P.}~\bibnamefont {Shor}},\ }\bibfield  {title} {\bibinfo {title} {Algorithms for quantum computation: discrete logarithms and factoring},\ }in\ \href {https://doi.org/10.1109/SFCS.1994.365700} {\emph {\bibinfo {booktitle} {Proceedings 35th Annual Symposium on Foundations of Computer Science}}}\ (\bibinfo {year} {1994})\ pp.\ \bibinfo {pages} {124--134}\BibitemShut {NoStop}%
\bibitem [{\citenamefont {Steane}(1998)}]{steane1998quantum}%
  \BibitemOpen
  \bibfield  {author} {\bibinfo {author} {\bibfnamefont {A.}~\bibnamefont {Steane}},\ }\bibfield  {title} {\bibinfo {title} {Quantum computing},\ }\href {https://doi.org/10.1088/0034-4885/61/2/002} {\bibfield  {journal} {\bibinfo  {journal} {Rep. Prog. Phys.}\ }\textbf {\bibinfo {volume} {61}},\ \bibinfo {pages} {117} (\bibinfo {year} {1998})}\BibitemShut {NoStop}%
\bibitem [{\citenamefont {Kitaev}\ \emph {et~al.}(2002)\citenamefont {Kitaev}, \citenamefont {Shen},\ and\ \citenamefont {Vyalyi}}]{kitaev2002classical}%
  \BibitemOpen
  \bibfield  {author} {\bibinfo {author} {\bibfnamefont {A.~Y.}\ \bibnamefont {Kitaev}}, \bibinfo {author} {\bibfnamefont {A.~H.}\ \bibnamefont {Shen}},\ and\ \bibinfo {author} {\bibfnamefont {M.~N.}\ \bibnamefont {Vyalyi}},\ }\href@noop {} {\emph {\bibinfo {title} {Classical and Quantum Computation}}}\ (\bibinfo  {publisher} {American Mathematical Society},\ \bibinfo {address} {USA},\ \bibinfo {year} {2002})\BibitemShut {NoStop}%
\bibitem [{\citenamefont {Arute}\ \emph {et~al.}(2019)\citenamefont {Arute}, \citenamefont {Arya}, \citenamefont {Babbush}, \citenamefont {Bacon}, \citenamefont {Bardin}, \citenamefont {Barends}, \citenamefont {Biswas}, \citenamefont {Boixo}, \citenamefont {Brandao}, \citenamefont {Buell} \emph {et~al.}}]{arute2019quantum}%
  \BibitemOpen
  \bibfield  {author} {\bibinfo {author} {\bibfnamefont {F.}~\bibnamefont {Arute}}, \bibinfo {author} {\bibfnamefont {K.}~\bibnamefont {Arya}}, \bibinfo {author} {\bibfnamefont {R.}~\bibnamefont {Babbush}}, \bibinfo {author} {\bibfnamefont {D.}~\bibnamefont {Bacon}}, \bibinfo {author} {\bibfnamefont {J.~C.}\ \bibnamefont {Bardin}}, \bibinfo {author} {\bibfnamefont {R.}~\bibnamefont {Barends}}, \bibinfo {author} {\bibfnamefont {R.}~\bibnamefont {Biswas}}, \bibinfo {author} {\bibfnamefont {S.}~\bibnamefont {Boixo}}, \bibinfo {author} {\bibfnamefont {F.~G. S.~L.}\ \bibnamefont {Brandao}}, \bibinfo {author} {\bibfnamefont {D.~A.}\ \bibnamefont {Buell}}, \emph {et~al.},\ }\bibfield  {title} {\bibinfo {title} {Quantum supremacy using a programmable superconducting processor},\ }\href {https://doi.org/10.1038/s41586-019-1666-5} {\bibfield  {journal} {\bibinfo  {journal} {Nature}\ }\textbf {\bibinfo {volume} {574}},\ \bibinfo {pages} {505} (\bibinfo {year} {2019})}\BibitemShut {NoStop}%
\bibitem [{\citenamefont {Wu}\ \emph {et~al.}(2021)\citenamefont {Wu}, \citenamefont {Bao}, \citenamefont {Cao}, \citenamefont {Chen}, \citenamefont {Chen}, \citenamefont {Chen}, \citenamefont {Chung}, \citenamefont {Deng}, \citenamefont {Du}, \citenamefont {Fan} \emph {et~al.}}]{wu2021strong}%
  \BibitemOpen
  \bibfield  {author} {\bibinfo {author} {\bibfnamefont {Y.}~\bibnamefont {Wu}}, \bibinfo {author} {\bibfnamefont {W.-S.}\ \bibnamefont {Bao}}, \bibinfo {author} {\bibfnamefont {S.}~\bibnamefont {Cao}}, \bibinfo {author} {\bibfnamefont {F.}~\bibnamefont {Chen}}, \bibinfo {author} {\bibfnamefont {M.-C.}\ \bibnamefont {Chen}}, \bibinfo {author} {\bibfnamefont {X.}~\bibnamefont {Chen}}, \bibinfo {author} {\bibfnamefont {T.-H.}\ \bibnamefont {Chung}}, \bibinfo {author} {\bibfnamefont {H.}~\bibnamefont {Deng}}, \bibinfo {author} {\bibfnamefont {Y.}~\bibnamefont {Du}}, \bibinfo {author} {\bibfnamefont {D.}~\bibnamefont {Fan}}, \emph {et~al.},\ }\bibfield  {title} {\bibinfo {title} {Strong quantum computational advantage using a superconducting quantum processor},\ }\href {https://doi.org/10.1103/PhysRevLett.127.180501} {\bibfield  {journal} {\bibinfo  {journal} {Phys. Rev. Lett.}\ }\textbf {\bibinfo {volume} {127}},\ \bibinfo {pages} {180501} (\bibinfo {year} {2021})}\BibitemShut {NoStop}%
\bibitem [{\citenamefont {Han}\ \emph {et~al.}(2024)\citenamefont {Han}, \citenamefont {Lyu}, \citenamefont {Zhou}, \citenamefont {Yuan}, \citenamefont {Chu}, \citenamefont {Nuerbolati}, \citenamefont {Jia}, \citenamefont {Nie}, \citenamefont {Wei}, \citenamefont {Yang}, \citenamefont {Zhang}, \citenamefont {Zhang}, \citenamefont {Hu}, \citenamefont {Hu}, \citenamefont {Li}, \citenamefont {Tan}, \citenamefont {Bayat}, \citenamefont {Liu}, \citenamefont {Yan},\ and\ \citenamefont {Yu}}]{zhikun2024multilevel}%
  \BibitemOpen
  \bibfield  {author} {\bibinfo {author} {\bibfnamefont {Z.}~\bibnamefont {Han}}, \bibinfo {author} {\bibfnamefont {C.}~\bibnamefont {Lyu}}, \bibinfo {author} {\bibfnamefont {Y.}~\bibnamefont {Zhou}}, \bibinfo {author} {\bibfnamefont {J.}~\bibnamefont {Yuan}}, \bibinfo {author} {\bibfnamefont {J.}~\bibnamefont {Chu}}, \bibinfo {author} {\bibfnamefont {W.}~\bibnamefont {Nuerbolati}}, \bibinfo {author} {\bibfnamefont {H.}~\bibnamefont {Jia}}, \bibinfo {author} {\bibfnamefont {L.}~\bibnamefont {Nie}}, \bibinfo {author} {\bibfnamefont {W.}~\bibnamefont {Wei}}, \bibinfo {author} {\bibfnamefont {Z.}~\bibnamefont {Yang}}, \bibinfo {author} {\bibfnamefont {L.}~\bibnamefont {Zhang}}, \bibinfo {author} {\bibfnamefont {Z.}~\bibnamefont {Zhang}}, \bibinfo {author} {\bibfnamefont {C.-K.}\ \bibnamefont {Hu}}, \bibinfo {author} {\bibfnamefont {L.}~\bibnamefont {Hu}}, \bibinfo {author} {\bibfnamefont {J.}~\bibnamefont {Li}}, \bibinfo {author} {\bibfnamefont {D.}~\bibnamefont {Tan}}, \bibinfo {author} {\bibfnamefont
  {A.}~\bibnamefont {Bayat}}, \bibinfo {author} {\bibfnamefont {S.}~\bibnamefont {Liu}}, \bibinfo {author} {\bibfnamefont {F.}~\bibnamefont {Yan}},\ and\ \bibinfo {author} {\bibfnamefont {D.}~\bibnamefont {Yu}},\ }\bibfield  {title} {\bibinfo {title} {Multilevel variational spectroscopy using a programmable quantum simulator},\ }\href {https://doi.org/10.1103/PhysRevResearch.6.013015} {\bibfield  {journal} {\bibinfo  {journal} {Phys. Rev. Res.}\ }\textbf {\bibinfo {volume} {6}},\ \bibinfo {pages} {013015} (\bibinfo {year} {2024})}\BibitemShut {NoStop}%
\bibitem [{\citenamefont {Ren}\ \emph {et~al.}(2022)\citenamefont {Ren}, \citenamefont {Li}, \citenamefont {Xu}, \citenamefont {Wang}, \citenamefont {Jiang}, \citenamefont {Jin}, \citenamefont {Zhu}, \citenamefont {Chen}, \citenamefont {Song}, \citenamefont {Zhang} \emph {et~al.}}]{ren2022experimental}%
  \BibitemOpen
  \bibfield  {author} {\bibinfo {author} {\bibfnamefont {W.}~\bibnamefont {Ren}}, \bibinfo {author} {\bibfnamefont {W.}~\bibnamefont {Li}}, \bibinfo {author} {\bibfnamefont {S.}~\bibnamefont {Xu}}, \bibinfo {author} {\bibfnamefont {K.}~\bibnamefont {Wang}}, \bibinfo {author} {\bibfnamefont {W.}~\bibnamefont {Jiang}}, \bibinfo {author} {\bibfnamefont {F.}~\bibnamefont {Jin}}, \bibinfo {author} {\bibfnamefont {X.}~\bibnamefont {Zhu}}, \bibinfo {author} {\bibfnamefont {J.}~\bibnamefont {Chen}}, \bibinfo {author} {\bibfnamefont {Z.}~\bibnamefont {Song}}, \bibinfo {author} {\bibfnamefont {P.}~\bibnamefont {Zhang}}, \emph {et~al.},\ }\bibfield  {title} {\bibinfo {title} {Experimental quantum adversarial learning with programmable superconducting qubits},\ }\href {https://doi.org/https://doi.org/10.1038/s43588-022-00351-9} {\bibfield  {journal} {\bibinfo  {journal} {Nat. Comput. Sci.}\ }\textbf {\bibinfo {volume} {2}},\ \bibinfo {pages} {711} (\bibinfo {year} {2022})}\BibitemShut {NoStop}%
\bibitem [{\citenamefont {Kielpinski}\ \emph {et~al.}(2002)\citenamefont {Kielpinski}, \citenamefont {Monroe},\ and\ \citenamefont {Wineland}}]{kielpinski2002architecture}%
  \BibitemOpen
  \bibfield  {author} {\bibinfo {author} {\bibfnamefont {D.}~\bibnamefont {Kielpinski}}, \bibinfo {author} {\bibfnamefont {C.}~\bibnamefont {Monroe}},\ and\ \bibinfo {author} {\bibfnamefont {D.~J.}\ \bibnamefont {Wineland}},\ }\bibfield  {title} {\bibinfo {title} {Architecture for a large-scale ion-trap quantum computer},\ }\href {https://doi.org/10.1038/nature00784} {\bibfield  {journal} {\bibinfo  {journal} {Nature}\ }\textbf {\bibinfo {volume} {417}},\ \bibinfo {pages} {709} (\bibinfo {year} {2002})}\BibitemShut {NoStop}%
\bibitem [{\citenamefont {Zhang}\ \emph {et~al.}(2017)\citenamefont {Zhang}, \citenamefont {Pagano}, \citenamefont {Hess}, \citenamefont {Kyprianidis}, \citenamefont {Becker}, \citenamefont {Kaplan}, \citenamefont {Gorshkov}, \citenamefont {Gong},\ and\ \citenamefont {Monroe}}]{zhang2017observation}%
  \BibitemOpen
  \bibfield  {author} {\bibinfo {author} {\bibfnamefont {J.}~\bibnamefont {Zhang}}, \bibinfo {author} {\bibfnamefont {G.}~\bibnamefont {Pagano}}, \bibinfo {author} {\bibfnamefont {P.~W.}\ \bibnamefont {Hess}}, \bibinfo {author} {\bibfnamefont {A.}~\bibnamefont {Kyprianidis}}, \bibinfo {author} {\bibfnamefont {P.}~\bibnamefont {Becker}}, \bibinfo {author} {\bibfnamefont {H.}~\bibnamefont {Kaplan}}, \bibinfo {author} {\bibfnamefont {A.~V.}\ \bibnamefont {Gorshkov}}, \bibinfo {author} {\bibfnamefont {Z.-X.}\ \bibnamefont {Gong}},\ and\ \bibinfo {author} {\bibfnamefont {C.}~\bibnamefont {Monroe}},\ }\bibfield  {title} {\bibinfo {title} {Observation of a many-body dynamical phase transition with a 53-qubit quantum simulator},\ }\href {https://doi.org/https://doi.org/10.1038/nature24654} {\bibfield  {journal} {\bibinfo  {journal} {Nature}\ }\textbf {\bibinfo {volume} {551}},\ \bibinfo {pages} {601} (\bibinfo {year} {2017})}\BibitemShut {NoStop}%
\bibitem [{\citenamefont {Ringbauer}\ \emph {et~al.}(2022)\citenamefont {Ringbauer}, \citenamefont {Meth}, \citenamefont {Postler}, \citenamefont {Stricker}, \citenamefont {Blatt}, \citenamefont {Schindler},\ and\ \citenamefont {Monz}}]{ringbauer2022universal}%
  \BibitemOpen
  \bibfield  {author} {\bibinfo {author} {\bibfnamefont {M.}~\bibnamefont {Ringbauer}}, \bibinfo {author} {\bibfnamefont {M.}~\bibnamefont {Meth}}, \bibinfo {author} {\bibfnamefont {L.}~\bibnamefont {Postler}}, \bibinfo {author} {\bibfnamefont {R.}~\bibnamefont {Stricker}}, \bibinfo {author} {\bibfnamefont {R.}~\bibnamefont {Blatt}}, \bibinfo {author} {\bibfnamefont {P.}~\bibnamefont {Schindler}},\ and\ \bibinfo {author} {\bibfnamefont {T.}~\bibnamefont {Monz}},\ }\bibfield  {title} {\bibinfo {title} {A universal qudit quantum processor with trapped ions},\ }\href {https://doi.org/https://doi.org/10.1038/s41567-022-01658-0} {\bibfield  {journal} {\bibinfo  {journal} {Nat. Phys.}\ }\textbf {\bibinfo {volume} {18}},\ \bibinfo {pages} {1053} (\bibinfo {year} {2022})}\BibitemShut {NoStop}%
\bibitem [{\citenamefont {Monroe}\ \emph {et~al.}(2021)\citenamefont {Monroe}, \citenamefont {Campbell}, \citenamefont {Duan}, \citenamefont {Gong}, \citenamefont {Gorshkov}, \citenamefont {Hess}, \citenamefont {Islam}, \citenamefont {Kim}, \citenamefont {Linke}, \citenamefont {Pagano}, \citenamefont {Richerme}, \citenamefont {Senko},\ and\ \citenamefont {Yao}}]{monroe2021Programmable}%
  \BibitemOpen
  \bibfield  {author} {\bibinfo {author} {\bibfnamefont {C.}~\bibnamefont {Monroe}}, \bibinfo {author} {\bibfnamefont {W.~C.}\ \bibnamefont {Campbell}}, \bibinfo {author} {\bibfnamefont {L.-M.}\ \bibnamefont {Duan}}, \bibinfo {author} {\bibfnamefont {Z.-X.}\ \bibnamefont {Gong}}, \bibinfo {author} {\bibfnamefont {A.~V.}\ \bibnamefont {Gorshkov}}, \bibinfo {author} {\bibfnamefont {P.~W.}\ \bibnamefont {Hess}}, \bibinfo {author} {\bibfnamefont {R.}~\bibnamefont {Islam}}, \bibinfo {author} {\bibfnamefont {K.}~\bibnamefont {Kim}}, \bibinfo {author} {\bibfnamefont {N.~M.}\ \bibnamefont {Linke}}, \bibinfo {author} {\bibfnamefont {G.}~\bibnamefont {Pagano}}, \bibinfo {author} {\bibfnamefont {P.}~\bibnamefont {Richerme}}, \bibinfo {author} {\bibfnamefont {C.}~\bibnamefont {Senko}},\ and\ \bibinfo {author} {\bibfnamefont {N.~Y.}\ \bibnamefont {Yao}},\ }\bibfield  {title} {\bibinfo {title} {Programmable quantum simulations of spin systems with trapped ions},\ }\href {https://doi.org/10.1103/RevModPhys.93.025001}
  {\bibfield  {journal} {\bibinfo  {journal} {Rev. Mod. Phys.}\ }\textbf {\bibinfo {volume} {93}},\ \bibinfo {pages} {025001} (\bibinfo {year} {2021})}\BibitemShut {NoStop}%
\bibitem [{\citenamefont {Bernien}\ \emph {et~al.}(2017)\citenamefont {Bernien}, \citenamefont {Schwartz}, \citenamefont {Keesling}, \citenamefont {Levine}, \citenamefont {Omran}, \citenamefont {Pichler}, \citenamefont {Choi}, \citenamefont {Zibrov}, \citenamefont {Endres}, \citenamefont {Greiner} \emph {et~al.}}]{bernien2017probing}%
  \BibitemOpen
  \bibfield  {author} {\bibinfo {author} {\bibfnamefont {H.}~\bibnamefont {Bernien}}, \bibinfo {author} {\bibfnamefont {S.}~\bibnamefont {Schwartz}}, \bibinfo {author} {\bibfnamefont {A.}~\bibnamefont {Keesling}}, \bibinfo {author} {\bibfnamefont {H.}~\bibnamefont {Levine}}, \bibinfo {author} {\bibfnamefont {A.}~\bibnamefont {Omran}}, \bibinfo {author} {\bibfnamefont {H.}~\bibnamefont {Pichler}}, \bibinfo {author} {\bibfnamefont {S.}~\bibnamefont {Choi}}, \bibinfo {author} {\bibfnamefont {A.~S.}\ \bibnamefont {Zibrov}}, \bibinfo {author} {\bibfnamefont {M.}~\bibnamefont {Endres}}, \bibinfo {author} {\bibfnamefont {M.}~\bibnamefont {Greiner}}, \emph {et~al.},\ }\bibfield  {title} {\bibinfo {title} {Probing many-body dynamics on a 51-atom quantum simulator},\ }\href {https://doi.org/https://doi.org/10.1038/nature24622} {\bibfield  {journal} {\bibinfo  {journal} {Nature}\ }\textbf {\bibinfo {volume} {551}},\ \bibinfo {pages} {579} (\bibinfo {year} {2017})}\BibitemShut {NoStop}%
\bibitem [{\citenamefont {Ebadi}\ \emph {et~al.}(2021)\citenamefont {Ebadi}, \citenamefont {Wang}, \citenamefont {Levine}, \citenamefont {Keesling}, \citenamefont {Semeghini}, \citenamefont {Omran}, \citenamefont {Bluvstein}, \citenamefont {Samajdar}, \citenamefont {Pichler}, \citenamefont {Ho} \emph {et~al.}}]{ebadi2021quantum}%
  \BibitemOpen
  \bibfield  {author} {\bibinfo {author} {\bibfnamefont {S.}~\bibnamefont {Ebadi}}, \bibinfo {author} {\bibfnamefont {T.~T.}\ \bibnamefont {Wang}}, \bibinfo {author} {\bibfnamefont {H.}~\bibnamefont {Levine}}, \bibinfo {author} {\bibfnamefont {A.}~\bibnamefont {Keesling}}, \bibinfo {author} {\bibfnamefont {G.}~\bibnamefont {Semeghini}}, \bibinfo {author} {\bibfnamefont {A.}~\bibnamefont {Omran}}, \bibinfo {author} {\bibfnamefont {D.}~\bibnamefont {Bluvstein}}, \bibinfo {author} {\bibfnamefont {R.}~\bibnamefont {Samajdar}}, \bibinfo {author} {\bibfnamefont {H.}~\bibnamefont {Pichler}}, \bibinfo {author} {\bibfnamefont {W.~W.}\ \bibnamefont {Ho}}, \emph {et~al.},\ }\bibfield  {title} {\bibinfo {title} {Quantum phases of matter on a 256-atom programmable quantum simulator},\ }\href {https://doi.org/https://doi.org/10.1038/s41586-021-03582-4} {\bibfield  {journal} {\bibinfo  {journal} {Nature}\ }\textbf {\bibinfo {volume} {595}},\ \bibinfo {pages} {227} (\bibinfo {year} {2021})}\BibitemShut {NoStop}%
\bibitem [{\citenamefont {Zhong}\ \emph {et~al.}(2021)\citenamefont {Zhong}, \citenamefont {Deng}, \citenamefont {Qin}, \citenamefont {Wang}, \citenamefont {Chen}, \citenamefont {Peng}, \citenamefont {Luo}, \citenamefont {Wu}, \citenamefont {Gong}, \citenamefont {Su} \emph {et~al.}}]{zhong2021phase}%
  \BibitemOpen
  \bibfield  {author} {\bibinfo {author} {\bibfnamefont {H.-S.}\ \bibnamefont {Zhong}}, \bibinfo {author} {\bibfnamefont {Y.-H.}\ \bibnamefont {Deng}}, \bibinfo {author} {\bibfnamefont {J.}~\bibnamefont {Qin}}, \bibinfo {author} {\bibfnamefont {H.}~\bibnamefont {Wang}}, \bibinfo {author} {\bibfnamefont {M.-C.}\ \bibnamefont {Chen}}, \bibinfo {author} {\bibfnamefont {L.-C.}\ \bibnamefont {Peng}}, \bibinfo {author} {\bibfnamefont {Y.-H.}\ \bibnamefont {Luo}}, \bibinfo {author} {\bibfnamefont {D.}~\bibnamefont {Wu}}, \bibinfo {author} {\bibfnamefont {S.-Q.}\ \bibnamefont {Gong}}, \bibinfo {author} {\bibfnamefont {H.}~\bibnamefont {Su}}, \emph {et~al.},\ }\bibfield  {title} {\bibinfo {title} {Phase-programmable gaussian boson sampling using stimulated squeezed light},\ }\href {https://doi.org/10.1103/PhysRevLett.127.180502} {\bibfield  {journal} {\bibinfo  {journal} {Phys. Rev. Lett.}\ }\textbf {\bibinfo {volume} {127}},\ \bibinfo {pages} {180502} (\bibinfo {year} {2021})}\BibitemShut {NoStop}%
\bibitem [{\citenamefont {Xiao}\ \emph {et~al.}(2020)\citenamefont {Xiao}, \citenamefont {Deng}, \citenamefont {Wang}, \citenamefont {Zhu}, \citenamefont {Wang}, \citenamefont {Yi},\ and\ \citenamefont {Xue}}]{xiao2020nonHermitian}%
  \BibitemOpen
  \bibfield  {author} {\bibinfo {author} {\bibfnamefont {L.}~\bibnamefont {Xiao}}, \bibinfo {author} {\bibfnamefont {T.}~\bibnamefont {Deng}}, \bibinfo {author} {\bibfnamefont {K.}~\bibnamefont {Wang}}, \bibinfo {author} {\bibfnamefont {G.}~\bibnamefont {Zhu}}, \bibinfo {author} {\bibfnamefont {Z.}~\bibnamefont {Wang}}, \bibinfo {author} {\bibfnamefont {W.}~\bibnamefont {Yi}},\ and\ \bibinfo {author} {\bibfnamefont {P.}~\bibnamefont {Xue}},\ }\bibfield  {title} {\bibinfo {title} {Non-hermitian bulk--boundary correspondence in quantum dynamics},\ }\href {https://doi.org/https://doi.org/10.1038/s41567-020-0836-6} {\bibfield  {journal} {\bibinfo  {journal} {Nat. Phys.}\ }\textbf {\bibinfo {volume} {16}},\ \bibinfo {pages} {761} (\bibinfo {year} {2020})}\BibitemShut {NoStop}%
\bibitem [{\citenamefont {Nemoto}\ \emph {et~al.}(2014)\citenamefont {Nemoto}, \citenamefont {Trupke}, \citenamefont {Devitt}, \citenamefont {Stephens}, \citenamefont {Scharfenberger}, \citenamefont {Buczak}, \citenamefont {N{\"o}bauer}, \citenamefont {Everitt}, \citenamefont {Schmiedmayer},\ and\ \citenamefont {Munro}}]{nemoto2014photonic}%
  \BibitemOpen
  \bibfield  {author} {\bibinfo {author} {\bibfnamefont {K.}~\bibnamefont {Nemoto}}, \bibinfo {author} {\bibfnamefont {M.}~\bibnamefont {Trupke}}, \bibinfo {author} {\bibfnamefont {S.~J.}\ \bibnamefont {Devitt}}, \bibinfo {author} {\bibfnamefont {A.~M.}\ \bibnamefont {Stephens}}, \bibinfo {author} {\bibfnamefont {B.}~\bibnamefont {Scharfenberger}}, \bibinfo {author} {\bibfnamefont {K.}~\bibnamefont {Buczak}}, \bibinfo {author} {\bibfnamefont {T.}~\bibnamefont {N{\"o}bauer}}, \bibinfo {author} {\bibfnamefont {M.~S.}\ \bibnamefont {Everitt}}, \bibinfo {author} {\bibfnamefont {J.}~\bibnamefont {Schmiedmayer}},\ and\ \bibinfo {author} {\bibfnamefont {W.~J.}\ \bibnamefont {Munro}},\ }\bibfield  {title} {\bibinfo {title} {Photonic architecture for scalable quantum information processing in diamond},\ }\href {https://doi.org/10.1103/PhysRevX.4.031022} {\bibfield  {journal} {\bibinfo  {journal} {Phys. Rev. X.}\ }\textbf {\bibinfo {volume} {4}},\ \bibinfo {pages} {031022} (\bibinfo {year} {2014})}\BibitemShut
  {NoStop}%
\bibitem [{\citenamefont {Quantum}\ \emph {et~al.}(2025)\citenamefont {Quantum}, \citenamefont {Aghaee}, \citenamefont {Alcaraz~Ramirez}, \citenamefont {Alam}, \citenamefont {Ali}, \citenamefont {Andrzejczuk}, \citenamefont {Antipov}, \citenamefont {Astafev}, \citenamefont {Barzegar}, \citenamefont {Bauer} \emph {et~al.}}]{microsoft2025interferometric}%
  \BibitemOpen
  \bibfield  {author} {\bibinfo {author} {\bibfnamefont {M.~A.}\ \bibnamefont {Quantum}}, \bibinfo {author} {\bibfnamefont {M.}~\bibnamefont {Aghaee}}, \bibinfo {author} {\bibfnamefont {A.}~\bibnamefont {Alcaraz~Ramirez}}, \bibinfo {author} {\bibfnamefont {Z.}~\bibnamefont {Alam}}, \bibinfo {author} {\bibfnamefont {R.}~\bibnamefont {Ali}}, \bibinfo {author} {\bibfnamefont {M.}~\bibnamefont {Andrzejczuk}}, \bibinfo {author} {\bibfnamefont {A.}~\bibnamefont {Antipov}}, \bibinfo {author} {\bibfnamefont {M.}~\bibnamefont {Astafev}}, \bibinfo {author} {\bibfnamefont {A.}~\bibnamefont {Barzegar}}, \bibinfo {author} {\bibfnamefont {B.}~\bibnamefont {Bauer}}, \emph {et~al.},\ }\bibfield  {title} {\bibinfo {title} {Interferometric single-shot parity measurement in inas--al hybrid devices},\ }\href {https://doi.org/https://doi.org/10.1038/s41586-024-08445-2} {\bibfield  {journal} {\bibinfo  {journal} {Nature}\ }\textbf {\bibinfo {volume} {638}},\ \bibinfo {pages} {651} (\bibinfo {year} {2025})}\BibitemShut {NoStop}%
\bibitem [{\citenamefont {Preskill}(2018)}]{preskill2018quantum}%
  \BibitemOpen
  \bibfield  {author} {\bibinfo {author} {\bibfnamefont {J.}~\bibnamefont {Preskill}},\ }\bibfield  {title} {\bibinfo {title} {Quantum {C}omputing in the {NISQ} era and beyond},\ }\href {https://doi.org/10.22331/q-2018-08-06-79} {\bibfield  {journal} {\bibinfo  {journal} {{Quantum}}\ }\textbf {\bibinfo {volume} {2}},\ \bibinfo {pages} {79} (\bibinfo {year} {2018})}\BibitemShut {NoStop}%
\bibitem [{\citenamefont {Cerezo}\ \emph {et~al.}(2021)\citenamefont {Cerezo}, \citenamefont {Arrasmith}, \citenamefont {Babbush}, \citenamefont {Benjamin}, \citenamefont {Endo}, \citenamefont {Fujii}, \citenamefont {McClean}, \citenamefont {Mitarai}, \citenamefont {Yuan}, \citenamefont {Cincio} \emph {et~al.}}]{cerezo2021variational}%
  \BibitemOpen
  \bibfield  {author} {\bibinfo {author} {\bibfnamefont {M.}~\bibnamefont {Cerezo}}, \bibinfo {author} {\bibfnamefont {A.}~\bibnamefont {Arrasmith}}, \bibinfo {author} {\bibfnamefont {R.}~\bibnamefont {Babbush}}, \bibinfo {author} {\bibfnamefont {S.~C.}\ \bibnamefont {Benjamin}}, \bibinfo {author} {\bibfnamefont {S.}~\bibnamefont {Endo}}, \bibinfo {author} {\bibfnamefont {K.}~\bibnamefont {Fujii}}, \bibinfo {author} {\bibfnamefont {J.~R.}\ \bibnamefont {McClean}}, \bibinfo {author} {\bibfnamefont {K.}~\bibnamefont {Mitarai}}, \bibinfo {author} {\bibfnamefont {X.}~\bibnamefont {Yuan}}, \bibinfo {author} {\bibfnamefont {L.}~\bibnamefont {Cincio}}, \emph {et~al.},\ }\bibfield  {title} {\bibinfo {title} {Variational quantum algorithms},\ }\href {https://doi.org/https://doi.org/10.1038/s42254-021-00348-9} {\bibfield  {journal} {\bibinfo  {journal} {Nat. Rev. Phys.}\ }\textbf {\bibinfo {volume} {3}},\ \bibinfo {pages} {625} (\bibinfo {year} {2021})}\BibitemShut {NoStop}%
\bibitem [{\citenamefont {Farhi}\ \emph {et~al.}(2014)\citenamefont {Farhi}, \citenamefont {Goldstone},\ and\ \citenamefont {Gutmann}}]{farhi2014quantum}%
  \BibitemOpen
  \bibfield  {author} {\bibinfo {author} {\bibfnamefont {E.}~\bibnamefont {Farhi}}, \bibinfo {author} {\bibfnamefont {J.}~\bibnamefont {Goldstone}},\ and\ \bibinfo {author} {\bibfnamefont {S.}~\bibnamefont {Gutmann}},\ }\href {https://arxiv.org/abs/1411.4028} {\bibinfo {title} {A quantum approximate optimization algorithm}} (\bibinfo {year} {2014}),\ \Eprint {https://arxiv.org/abs/1411.4028} {arXiv:1411.4028 [quant-ph]} \BibitemShut {NoStop}%
\bibitem [{\citenamefont {Cao}\ \emph {et~al.}(2019)\citenamefont {Cao}, \citenamefont {Romero}, \citenamefont {Olson}, \citenamefont {Degroote}, \citenamefont {Johnson}, \citenamefont {Kieferov{\'a}}, \citenamefont {Kivlichan}, \citenamefont {Menke}, \citenamefont {Peropadre}, \citenamefont {Sawaya}, \citenamefont {Sim}, \citenamefont {Veis},\ and\ \citenamefont {{Aspuru-Guzik}}}]{cao2019quantum}%
  \BibitemOpen
  \bibfield  {author} {\bibinfo {author} {\bibfnamefont {Y.}~\bibnamefont {Cao}}, \bibinfo {author} {\bibfnamefont {J.}~\bibnamefont {Romero}}, \bibinfo {author} {\bibfnamefont {J.~P.}\ \bibnamefont {Olson}}, \bibinfo {author} {\bibfnamefont {M.}~\bibnamefont {Degroote}}, \bibinfo {author} {\bibfnamefont {P.~D.}\ \bibnamefont {Johnson}}, \bibinfo {author} {\bibfnamefont {M.}~\bibnamefont {Kieferov{\'a}}}, \bibinfo {author} {\bibfnamefont {I.~D.}\ \bibnamefont {Kivlichan}}, \bibinfo {author} {\bibfnamefont {T.}~\bibnamefont {Menke}}, \bibinfo {author} {\bibfnamefont {B.}~\bibnamefont {Peropadre}}, \bibinfo {author} {\bibfnamefont {N.~P.~D.}\ \bibnamefont {Sawaya}}, \bibinfo {author} {\bibfnamefont {S.}~\bibnamefont {Sim}}, \bibinfo {author} {\bibfnamefont {L.}~\bibnamefont {Veis}},\ and\ \bibinfo {author} {\bibfnamefont {A.}~\bibnamefont {{Aspuru-Guzik}}},\ }\bibfield  {title} {\bibinfo {title} {Quantum {{Chemistry}} in the {{Age}} of {{Quantum Computing}}},\ }\href
  {https://doi.org/10.1021/acs.chemrev.8b00803} {\bibfield  {journal} {\bibinfo  {journal} {Chem. Rev.}\ }\textbf {\bibinfo {volume} {119}},\ \bibinfo {pages} {10856} (\bibinfo {year} {2019})}\BibitemShut {NoStop}%
\bibitem [{\citenamefont {Li}\ \emph {et~al.}(2023)\citenamefont {Li}, \citenamefont {Mukhopadhyay},\ and\ \citenamefont {Bayat}}]{li2023fermionic}%
  \BibitemOpen
  \bibfield  {author} {\bibinfo {author} {\bibfnamefont {Q.}~\bibnamefont {Li}}, \bibinfo {author} {\bibfnamefont {C.}~\bibnamefont {Mukhopadhyay}},\ and\ \bibinfo {author} {\bibfnamefont {A.}~\bibnamefont {Bayat}},\ }\bibfield  {title} {\bibinfo {title} {Fermionic simulators for enhanced scalability of variational quantum simulation},\ }\href {https://doi.org/10.1103/PhysRevResearch.5.043175} {\bibfield  {journal} {\bibinfo  {journal} {Phys. Rev. Res.}\ }\textbf {\bibinfo {volume} {5}},\ \bibinfo {pages} {043175} (\bibinfo {year} {2023})}\BibitemShut {NoStop}%
\bibitem [{\citenamefont {Baiardi}\ \emph {et~al.}(2023)\citenamefont {Baiardi}, \citenamefont {Christandl},\ and\ \citenamefont {Reiher}}]{baiardi2023quantum}%
  \BibitemOpen
  \bibfield  {author} {\bibinfo {author} {\bibfnamefont {A.}~\bibnamefont {Baiardi}}, \bibinfo {author} {\bibfnamefont {M.}~\bibnamefont {Christandl}},\ and\ \bibinfo {author} {\bibfnamefont {M.}~\bibnamefont {Reiher}},\ }\bibfield  {title} {\bibinfo {title} {Quantum computing for molecular biology},\ }\href {https://doi.org/https://doi.org/10.1002/cbic.202300120} {\bibfield  {journal} {\bibinfo  {journal} {ChemBioChem}\ }\textbf {\bibinfo {volume} {24}},\ \bibinfo {pages} {e202300120} (\bibinfo {year} {2023})}\BibitemShut {NoStop}%
\bibitem [{\citenamefont {Paulson}\ \emph {et~al.}(2021)\citenamefont {Paulson}, \citenamefont {Dellantonio}, \citenamefont {Haase}, \citenamefont {Celi}, \citenamefont {Kan}, \citenamefont {Jena}, \citenamefont {Kokail}, \citenamefont {van Bijnen}, \citenamefont {Jansen}, \citenamefont {Zoller},\ and\ \citenamefont {Muschik}}]{paulson2021simulating}%
  \BibitemOpen
  \bibfield  {author} {\bibinfo {author} {\bibfnamefont {D.}~\bibnamefont {Paulson}}, \bibinfo {author} {\bibfnamefont {L.}~\bibnamefont {Dellantonio}}, \bibinfo {author} {\bibfnamefont {J.~F.}\ \bibnamefont {Haase}}, \bibinfo {author} {\bibfnamefont {A.}~\bibnamefont {Celi}}, \bibinfo {author} {\bibfnamefont {A.}~\bibnamefont {Kan}}, \bibinfo {author} {\bibfnamefont {A.}~\bibnamefont {Jena}}, \bibinfo {author} {\bibfnamefont {C.}~\bibnamefont {Kokail}}, \bibinfo {author} {\bibfnamefont {R.}~\bibnamefont {van Bijnen}}, \bibinfo {author} {\bibfnamefont {K.}~\bibnamefont {Jansen}}, \bibinfo {author} {\bibfnamefont {P.}~\bibnamefont {Zoller}},\ and\ \bibinfo {author} {\bibfnamefont {C.~A.}\ \bibnamefont {Muschik}},\ }\bibfield  {title} {\bibinfo {title} {Simulating 2d effects in lattice gauge theories on a quantum computer},\ }\href {https://doi.org/10.1103/PRXQuantum.2.030334} {\bibfield  {journal} {\bibinfo  {journal} {PRX Quantum}\ }\textbf {\bibinfo {volume} {2}},\ \bibinfo {pages} {030334} (\bibinfo {year}
  {2021})}\BibitemShut {NoStop}%
\bibitem [{\citenamefont {Crutchfield}\ and\ \citenamefont {Young}(1989)}]{crutchfield1989inferring}%
  \BibitemOpen
  \bibfield  {author} {\bibinfo {author} {\bibfnamefont {J.~P.}\ \bibnamefont {Crutchfield}}\ and\ \bibinfo {author} {\bibfnamefont {K.}~\bibnamefont {Young}},\ }\bibfield  {title} {\bibinfo {title} {Inferring statistical complexity},\ }\href {https://doi.org/10.1103/PhysRevLett.63.105} {\bibfield  {journal} {\bibinfo  {journal} {Phys. Rev. Lett.}\ }\textbf {\bibinfo {volume} {63}},\ \bibinfo {pages} {105} (\bibinfo {year} {1989})}\BibitemShut {NoStop}%
\bibitem [{\citenamefont {Shalizi}\ and\ \citenamefont {Crutchfield}(2001)}]{shalizi2001computational}%
  \BibitemOpen
  \bibfield  {author} {\bibinfo {author} {\bibfnamefont {C.~R.}\ \bibnamefont {Shalizi}}\ and\ \bibinfo {author} {\bibfnamefont {J.~P.}\ \bibnamefont {Crutchfield}},\ }\bibfield  {title} {\bibinfo {title} {Computational mechanics: Pattern and prediction, structure and simplicity},\ }\href {https://doi.org/https://doi.org/10.1023/A:1010388907793} {\bibfield  {journal} {\bibinfo  {journal} {J. Stat. Phys.}\ }\textbf {\bibinfo {volume} {104}},\ \bibinfo {pages} {817} (\bibinfo {year} {2001})}\BibitemShut {NoStop}%
\bibitem [{\citenamefont {Gu}\ \emph {et~al.}(2012)\citenamefont {Gu}, \citenamefont {Wiesner}, \citenamefont {Rieper},\ and\ \citenamefont {Vedral}}]{gu2012quantum}%
  \BibitemOpen
  \bibfield  {author} {\bibinfo {author} {\bibfnamefont {M.}~\bibnamefont {Gu}}, \bibinfo {author} {\bibfnamefont {K.}~\bibnamefont {Wiesner}}, \bibinfo {author} {\bibfnamefont {E.}~\bibnamefont {Rieper}},\ and\ \bibinfo {author} {\bibfnamefont {V.}~\bibnamefont {Vedral}},\ }\bibfield  {title} {\bibinfo {title} {Quantum mechanics can reduce the complexity of classical models},\ }\href {https://doi.org/https://doi.org/10.1038/ncomms1761} {\bibfield  {journal} {\bibinfo  {journal} {Nat. Commun.}\ }\textbf {\bibinfo {volume} {3}},\ \bibinfo {pages} {762} (\bibinfo {year} {2012})}\BibitemShut {NoStop}%
\bibitem [{\citenamefont {da~Silva~Coelho}\ \emph {et~al.}(2023)\citenamefont {da~Silva~Coelho}, \citenamefont {Henriet},\ and\ \citenamefont {Henry}}]{da2023quantum}%
  \BibitemOpen
  \bibfield  {author} {\bibinfo {author} {\bibfnamefont {W.}~\bibnamefont {da~Silva~Coelho}}, \bibinfo {author} {\bibfnamefont {L.}~\bibnamefont {Henriet}},\ and\ \bibinfo {author} {\bibfnamefont {L.-P.}\ \bibnamefont {Henry}},\ }\bibfield  {title} {\bibinfo {title} {Quantum pricing-based column-generation framework for hard combinatorial problems},\ }\href {https://doi.org/10.1103/PhysRevA.107.032426} {\bibfield  {journal} {\bibinfo  {journal} {Phys. Rev. A}\ }\textbf {\bibinfo {volume} {107}},\ \bibinfo {pages} {032426} (\bibinfo {year} {2023})}\BibitemShut {NoStop}%
\bibitem [{\citenamefont {Mugel}\ \emph {et~al.}(2021)\citenamefont {Mugel}, \citenamefont {Abad}, \citenamefont {Bermejo}, \citenamefont {S{\'a}nchez}, \citenamefont {Lizaso},\ and\ \citenamefont {Or{\'u}s}}]{mugel2021hybrid}%
  \BibitemOpen
  \bibfield  {author} {\bibinfo {author} {\bibfnamefont {S.}~\bibnamefont {Mugel}}, \bibinfo {author} {\bibfnamefont {M.}~\bibnamefont {Abad}}, \bibinfo {author} {\bibfnamefont {M.}~\bibnamefont {Bermejo}}, \bibinfo {author} {\bibfnamefont {J.}~\bibnamefont {S{\'a}nchez}}, \bibinfo {author} {\bibfnamefont {E.}~\bibnamefont {Lizaso}},\ and\ \bibinfo {author} {\bibfnamefont {R.}~\bibnamefont {Or{\'u}s}},\ }\bibfield  {title} {\bibinfo {title} {Hybrid quantum investment optimization with minimal holding period},\ }\href {https://doi.org/https://doi.org/10.1038/s41598-021-98297-x} {\bibfield  {journal} {\bibinfo  {journal} {Sci. Rep.}\ }\textbf {\bibinfo {volume} {11}},\ \bibinfo {pages} {19587} (\bibinfo {year} {2021})}\BibitemShut {NoStop}%
\bibitem [{\citenamefont {Mugel}\ \emph {et~al.}(2022)\citenamefont {Mugel}, \citenamefont {Kuchkovsky}, \citenamefont {S{\'a}nchez}, \citenamefont {Fern{\'a}ndez-Lorenzo}, \citenamefont {Luis-Hita}, \citenamefont {Lizaso},\ and\ \citenamefont {Or{\'u}s}}]{mugel2022dynamic}%
  \BibitemOpen
  \bibfield  {author} {\bibinfo {author} {\bibfnamefont {S.}~\bibnamefont {Mugel}}, \bibinfo {author} {\bibfnamefont {C.}~\bibnamefont {Kuchkovsky}}, \bibinfo {author} {\bibfnamefont {E.}~\bibnamefont {S{\'a}nchez}}, \bibinfo {author} {\bibfnamefont {S.}~\bibnamefont {Fern{\'a}ndez-Lorenzo}}, \bibinfo {author} {\bibfnamefont {J.}~\bibnamefont {Luis-Hita}}, \bibinfo {author} {\bibfnamefont {E.}~\bibnamefont {Lizaso}},\ and\ \bibinfo {author} {\bibfnamefont {R.}~\bibnamefont {Or{\'u}s}},\ }\bibfield  {title} {\bibinfo {title} {Dynamic portfolio optimization with real datasets using quantum processors and quantum-inspired tensor networks},\ }\href {https://doi.org/10.1103/PhysRevResearch.4.013006} {\bibfield  {journal} {\bibinfo  {journal} {Phys. Rev. Res.}\ }\textbf {\bibinfo {volume} {4}},\ \bibinfo {pages} {013006} (\bibinfo {year} {2022})}\BibitemShut {NoStop}%
\bibitem [{\citenamefont {Wi{\'s}niewska}\ and\ \citenamefont {Sawerwain}(2023)}]{wisniewska2023variational}%
  \BibitemOpen
  \bibfield  {author} {\bibinfo {author} {\bibfnamefont {J.}~\bibnamefont {Wi{\'s}niewska}}\ and\ \bibinfo {author} {\bibfnamefont {M.}~\bibnamefont {Sawerwain}},\ }\bibfield  {title} {\bibinfo {title} {Variational quantum eigensolver for classification in credit sales risk},\ }\bibfield  {journal} {\bibinfo  {journal} {arXiv preprint arXiv:2303.02797}\ }\href {https://doi.org/https://doi.org/10.48550/arXiv.2303.02797} {https://doi.org/10.48550/arXiv.2303.02797} (\bibinfo {year} {2023})\BibitemShut {NoStop}%
\bibitem [{\citenamefont {Leclerc}\ \emph {et~al.}(2023)\citenamefont {Leclerc}, \citenamefont {Ortiz-Guti{\'e}rrez}, \citenamefont {Grijalva}, \citenamefont {Albrecht}, \citenamefont {Cline}, \citenamefont {Elfving}, \citenamefont {Signoles}, \citenamefont {Henriet}, \citenamefont {Del~Bimbo}, \citenamefont {Sheikh} \emph {et~al.}}]{leclerc2023financial}%
  \BibitemOpen
  \bibfield  {author} {\bibinfo {author} {\bibfnamefont {L.}~\bibnamefont {Leclerc}}, \bibinfo {author} {\bibfnamefont {L.}~\bibnamefont {Ortiz-Guti{\'e}rrez}}, \bibinfo {author} {\bibfnamefont {S.}~\bibnamefont {Grijalva}}, \bibinfo {author} {\bibfnamefont {B.}~\bibnamefont {Albrecht}}, \bibinfo {author} {\bibfnamefont {J.~R.}\ \bibnamefont {Cline}}, \bibinfo {author} {\bibfnamefont {V.~E.}\ \bibnamefont {Elfving}}, \bibinfo {author} {\bibfnamefont {A.}~\bibnamefont {Signoles}}, \bibinfo {author} {\bibfnamefont {L.}~\bibnamefont {Henriet}}, \bibinfo {author} {\bibfnamefont {G.}~\bibnamefont {Del~Bimbo}}, \bibinfo {author} {\bibfnamefont {U.~A.}\ \bibnamefont {Sheikh}}, \emph {et~al.},\ }\bibfield  {title} {\bibinfo {title} {Financial risk management on a neutral atom quantum processor},\ }\href {https://doi.org/10.1103/PhysRevResearch.5.043117} {\bibfield  {journal} {\bibinfo  {journal} {Phys. Rev. Res.}\ }\textbf {\bibinfo {volume} {5}},\ \bibinfo {pages} {043117} (\bibinfo {year} {2023})}\BibitemShut
  {NoStop}%
\bibitem [{\citenamefont {Innan}\ \emph {et~al.}(2024)\citenamefont {Innan}, \citenamefont {Marchisio}, \citenamefont {Bennai},\ and\ \citenamefont {Shafique}}]{innan2024lep}%
  \BibitemOpen
  \bibfield  {author} {\bibinfo {author} {\bibfnamefont {N.}~\bibnamefont {Innan}}, \bibinfo {author} {\bibfnamefont {A.}~\bibnamefont {Marchisio}}, \bibinfo {author} {\bibfnamefont {M.}~\bibnamefont {Bennai}},\ and\ \bibinfo {author} {\bibfnamefont {M.}~\bibnamefont {Shafique}},\ }\bibfield  {title} {\bibinfo {title} {Lep-qnn: Loan eligibility prediction using quantum neural networks},\ }\bibfield  {journal} {\bibinfo  {journal} {arXiv preprint arXiv:2412.03158}\ }\href {https://doi.org/https://doi.org/10.48550/arXiv.2412.03158} {https://doi.org/10.48550/arXiv.2412.03158} (\bibinfo {year} {2024})\BibitemShut {NoStop}%
\bibitem [{\citenamefont {Or{\'u}s}\ \emph {et~al.}(2019)\citenamefont {Or{\'u}s}, \citenamefont {Mugel},\ and\ \citenamefont {Lizaso}}]{orus2019quantum}%
  \BibitemOpen
  \bibfield  {author} {\bibinfo {author} {\bibfnamefont {R.}~\bibnamefont {Or{\'u}s}}, \bibinfo {author} {\bibfnamefont {S.}~\bibnamefont {Mugel}},\ and\ \bibinfo {author} {\bibfnamefont {E.}~\bibnamefont {Lizaso}},\ }\bibfield  {title} {\bibinfo {title} {Quantum computing for finance: Overview and prospects},\ }\href {https://doi.org/https://doi.org/10.1016/j.revip.2019.100028} {\bibfield  {journal} {\bibinfo  {journal} {Phys. Rev.}\ }\textbf {\bibinfo {volume} {4}},\ \bibinfo {pages} {100028} (\bibinfo {year} {2019})}\BibitemShut {NoStop}%
\bibitem [{\citenamefont {Herman}\ \emph {et~al.}(2022)\citenamefont {Herman}, \citenamefont {Googin}, \citenamefont {Liu}, \citenamefont {Galda}, \citenamefont {Safro}, \citenamefont {Sun}, \citenamefont {Pistoia},\ and\ \citenamefont {Alexeev}}]{herman2022survey}%
  \BibitemOpen
  \bibfield  {author} {\bibinfo {author} {\bibfnamefont {D.}~\bibnamefont {Herman}}, \bibinfo {author} {\bibfnamefont {C.}~\bibnamefont {Googin}}, \bibinfo {author} {\bibfnamefont {X.}~\bibnamefont {Liu}}, \bibinfo {author} {\bibfnamefont {A.}~\bibnamefont {Galda}}, \bibinfo {author} {\bibfnamefont {I.}~\bibnamefont {Safro}}, \bibinfo {author} {\bibfnamefont {Y.}~\bibnamefont {Sun}}, \bibinfo {author} {\bibfnamefont {M.}~\bibnamefont {Pistoia}},\ and\ \bibinfo {author} {\bibfnamefont {Y.}~\bibnamefont {Alexeev}},\ }\bibfield  {title} {\bibinfo {title} {A survey of quantum computing for finance},\ }\bibfield  {journal} {\bibinfo  {journal} {arXiv preprint arXiv:2201.02773}\ }\href {https://doi.org/https://doi.org/10.48550/arXiv.2201.02773} {https://doi.org/10.48550/arXiv.2201.02773} (\bibinfo {year} {2022})\BibitemShut {NoStop}%
\bibitem [{\citenamefont {Naik}\ \emph {et~al.}(2025)\citenamefont {Naik}, \citenamefont {Yeniaras}, \citenamefont {Hellstern}, \citenamefont {Prasad},\ and\ \citenamefont {Vishwakarma}}]{naik2025portfolio}%
  \BibitemOpen
  \bibfield  {author} {\bibinfo {author} {\bibfnamefont {A.~S.}\ \bibnamefont {Naik}}, \bibinfo {author} {\bibfnamefont {E.}~\bibnamefont {Yeniaras}}, \bibinfo {author} {\bibfnamefont {G.}~\bibnamefont {Hellstern}}, \bibinfo {author} {\bibfnamefont {G.}~\bibnamefont {Prasad}},\ and\ \bibinfo {author} {\bibfnamefont {S.~K. L.~P.}\ \bibnamefont {Vishwakarma}},\ }\bibfield  {title} {\bibinfo {title} {From portfolio optimization to quantum blockchain and security: A systematic review of quantum computing in finance},\ }\href {https://doi.org/https://doi.org/10.1186/s40854-025-00751-6} {\bibfield  {journal} {\bibinfo  {journal} {Financial Innovation}\ }\textbf {\bibinfo {volume} {11}},\ \bibinfo {pages} {1} (\bibinfo {year} {2025})}\BibitemShut {NoStop}%
\bibitem [{\citenamefont {Jaeger}\ and\ \citenamefont {Haas}(2004)}]{jaeger2004harnessing}%
  \BibitemOpen
  \bibfield  {author} {\bibinfo {author} {\bibfnamefont {H.}~\bibnamefont {Jaeger}}\ and\ \bibinfo {author} {\bibfnamefont {H.}~\bibnamefont {Haas}},\ }\bibfield  {title} {\bibinfo {title} {Harnessing nonlinearity: Predicting chaotic systems and saving energy in wireless communication},\ }\href {https://doi.org/10.1126/science.1091277} {\bibfield  {journal} {\bibinfo  {journal} {Science}\ }\textbf {\bibinfo {volume} {304}},\ \bibinfo {pages} {78} (\bibinfo {year} {2004})}\BibitemShut {NoStop}%
\bibitem [{\citenamefont {Li}\ and\ \citenamefont {Law}(2024)}]{LiDeep2024}%
  \BibitemOpen
  \bibfield  {author} {\bibinfo {author} {\bibfnamefont {W.}~\bibnamefont {Li}}\ and\ \bibinfo {author} {\bibfnamefont {K.~L.~E.}\ \bibnamefont {Law}},\ }\bibfield  {title} {\bibinfo {title} {Deep learning models for time series forecasting: A review},\ }\href {https://doi.org/10.1109/ACCESS.2024.3422528} {\bibfield  {journal} {\bibinfo  {journal} {IEEE Access}\ }\textbf {\bibinfo {volume} {12}},\ \bibinfo {pages} {92306} (\bibinfo {year} {2024})}\BibitemShut {NoStop}%
\bibitem [{\citenamefont {Kim}\ \emph {et~al.}(2025)\citenamefont {Kim}, \citenamefont {Kim}, \citenamefont {Kim}, \citenamefont {Lee},\ and\ \citenamefont {Yoon}}]{kimComprehensiveSurveyDeep2025}%
  \BibitemOpen
  \bibfield  {author} {\bibinfo {author} {\bibfnamefont {J.}~\bibnamefont {Kim}}, \bibinfo {author} {\bibfnamefont {H.}~\bibnamefont {Kim}}, \bibinfo {author} {\bibfnamefont {H.}~\bibnamefont {Kim}}, \bibinfo {author} {\bibfnamefont {D.}~\bibnamefont {Lee}},\ and\ \bibinfo {author} {\bibfnamefont {S.}~\bibnamefont {Yoon}},\ }\bibfield  {title} {\bibinfo {title} {A comprehensive survey of deep learning for time series forecasting: Architectural diversity and open challenges},\ }\href {https://doi.org/10.1007/s10462-025-11223-9} {\bibfield  {journal} {\bibinfo  {journal} {Artif. Intell. Rev.}\ }\textbf {\bibinfo {volume} {58}},\ \bibinfo {pages} {216} (\bibinfo {year} {2025})}\BibitemShut {NoStop}%
\bibitem [{\citenamefont {Fujii}\ and\ \citenamefont {Nakajima}(2017)}]{fujii2017harnessing}%
  \BibitemOpen
  \bibfield  {author} {\bibinfo {author} {\bibfnamefont {K.}~\bibnamefont {Fujii}}\ and\ \bibinfo {author} {\bibfnamefont {K.}~\bibnamefont {Nakajima}},\ }\bibfield  {title} {\bibinfo {title} {Harnessing disordered-ensemble quantum dynamics for machine learning},\ }\href {https://doi.org/10.1103/PhysRevApplied.8.024030} {\bibfield  {journal} {\bibinfo  {journal} {Phys. Rev. Appl.}\ }\textbf {\bibinfo {volume} {8}},\ \bibinfo {pages} {024030} (\bibinfo {year} {2017})}\BibitemShut {NoStop}%
\bibitem [{\citenamefont {Govia}\ \emph {et~al.}(2021)\citenamefont {Govia}, \citenamefont {Ribeill}, \citenamefont {Rowlands}, \citenamefont {Krovi},\ and\ \citenamefont {Ohki}}]{govia2021quantum}%
  \BibitemOpen
  \bibfield  {author} {\bibinfo {author} {\bibfnamefont {L.~C.~G.}\ \bibnamefont {Govia}}, \bibinfo {author} {\bibfnamefont {G.~J.}\ \bibnamefont {Ribeill}}, \bibinfo {author} {\bibfnamefont {G.~E.}\ \bibnamefont {Rowlands}}, \bibinfo {author} {\bibfnamefont {H.~K.}\ \bibnamefont {Krovi}},\ and\ \bibinfo {author} {\bibfnamefont {T.~A.}\ \bibnamefont {Ohki}},\ }\bibfield  {title} {\bibinfo {title} {Quantum reservoir computing with a single nonlinear oscillator},\ }\href {https://doi.org/10.1103/PhysRevResearch.3.013077} {\bibfield  {journal} {\bibinfo  {journal} {Phys. Rev. Res.}\ }\textbf {\bibinfo {volume} {3}},\ \bibinfo {pages} {013077} (\bibinfo {year} {2021})}\BibitemShut {NoStop}%
\bibitem [{\citenamefont {Das}\ \emph {et~al.}(2025)\citenamefont {Das}, \citenamefont {Giorgi},\ and\ \citenamefont {Zambrini}}]{das2025quantum}%
  \BibitemOpen
  \bibfield  {author} {\bibinfo {author} {\bibfnamefont {S.}~\bibnamefont {Das}}, \bibinfo {author} {\bibfnamefont {G.~L.}\ \bibnamefont {Giorgi}},\ and\ \bibinfo {author} {\bibfnamefont {R.}~\bibnamefont {Zambrini}},\ }\href@noop {} {\bibinfo {title} {Quantum reservoir computing using jaynes-cummings model}} (\bibinfo {year} {2025}),\ \Eprint {https://arxiv.org/abs/2510.00171} {arXiv:2510.00171 [quant-ph]} \BibitemShut {NoStop}%
\bibitem [{\citenamefont {Yasuda}\ \emph {et~al.}(2023)\citenamefont {Yasuda}, \citenamefont {Suzuki}, \citenamefont {Kubota}, \citenamefont {Nakajima}, \citenamefont {Gao}, \citenamefont {Zhang}, \citenamefont {Shimono}, \citenamefont {Nurdin},\ and\ \citenamefont {Yamamoto}}]{yasuda2023quantum}%
  \BibitemOpen
  \bibfield  {author} {\bibinfo {author} {\bibfnamefont {T.}~\bibnamefont {Yasuda}}, \bibinfo {author} {\bibfnamefont {Y.}~\bibnamefont {Suzuki}}, \bibinfo {author} {\bibfnamefont {T.}~\bibnamefont {Kubota}}, \bibinfo {author} {\bibfnamefont {K.}~\bibnamefont {Nakajima}}, \bibinfo {author} {\bibfnamefont {Q.}~\bibnamefont {Gao}}, \bibinfo {author} {\bibfnamefont {W.}~\bibnamefont {Zhang}}, \bibinfo {author} {\bibfnamefont {S.}~\bibnamefont {Shimono}}, \bibinfo {author} {\bibfnamefont {H.~I.}\ \bibnamefont {Nurdin}},\ and\ \bibinfo {author} {\bibfnamefont {N.}~\bibnamefont {Yamamoto}},\ }\bibfield  {title} {\bibinfo {title} {Quantum reservoir computing with repeated measurements on superconducting devices},\ }\bibfield  {journal} {\bibinfo  {journal} {arXiv preprint arXiv:2310.06706}\ }\href {https://doi.org/https://doi.org/10.48550/arXiv.2310.06706} {https://doi.org/10.48550/arXiv.2310.06706} (\bibinfo {year} {2023})\BibitemShut {NoStop}%
\bibitem [{\citenamefont {Mackey}\ and\ \citenamefont {Glass}(1977)}]{mackey1977oscillation}%
  \BibitemOpen
  \bibfield  {author} {\bibinfo {author} {\bibfnamefont {M.~C.}\ \bibnamefont {Mackey}}\ and\ \bibinfo {author} {\bibfnamefont {L.}~\bibnamefont {Glass}},\ }\bibfield  {title} {\bibinfo {title} {Oscillation and chaos in physiological control systems},\ }\href {https://doi.org/10.1126/science.267326} {\bibfield  {journal} {\bibinfo  {journal} {Science}\ }\textbf {\bibinfo {volume} {197}},\ \bibinfo {pages} {287} (\bibinfo {year} {1977})}\BibitemShut {NoStop}%
\bibitem [{\citenamefont {Moran}\ and\ \citenamefont {Whittle}(1951)}]{whittle1951hypothesis}%
  \BibitemOpen
  \bibfield  {author} {\bibinfo {author} {\bibfnamefont {P.~A.}\ \bibnamefont {Moran}}\ and\ \bibinfo {author} {\bibfnamefont {P.}~\bibnamefont {Whittle}},\ }\bibfield  {title} {\bibinfo {title} {Hypothesis {{Testing}} in {{Time Series Analysis}}.},\ }\href {https://doi.org/10.2307/2981095} {\bibfield  {journal} {\bibinfo  {journal} {J. R. Stat. Soc}\ }\textbf {\bibinfo {volume} {114}},\ \bibinfo {pages} {579} (\bibinfo {year} {1951})},\ \Eprint {https://arxiv.org/abs/10.2307/2981095} {10.2307/2981095} \BibitemShut {NoStop}%
\bibitem [{\citenamefont {Black}\ and\ \citenamefont {Scholes}(1973)}]{black1973pricing}%
  \BibitemOpen
  \bibfield  {author} {\bibinfo {author} {\bibfnamefont {F.}~\bibnamefont {Black}}\ and\ \bibinfo {author} {\bibfnamefont {M.}~\bibnamefont {Scholes}},\ }\bibfield  {title} {\bibinfo {title} {The pricing of options and corporate liabilities},\ }\href {http://www.jstor.org/stable/1831029} {\bibfield  {journal} {\bibinfo  {journal} {J. Political Econ.}\ }\textbf {\bibinfo {volume} {81}},\ \bibinfo {pages} {637} (\bibinfo {year} {1973})}\BibitemShut {NoStop}%
\bibitem [{\citenamefont {Poon}\ and\ \citenamefont {Granger}(2003)}]{poon2003forecasting}%
  \BibitemOpen
  \bibfield  {author} {\bibinfo {author} {\bibfnamefont {S.-H.}\ \bibnamefont {Poon}}\ and\ \bibinfo {author} {\bibfnamefont {C.~W.~J.}\ \bibnamefont {Granger}},\ }\bibfield  {title} {\bibinfo {title} {Forecasting volatility in financial markets: A review},\ }\href {http://www.jstor.org/stable/3216966} {\bibfield  {journal} {\bibinfo  {journal} {Econ. Lit.}\ }\textbf {\bibinfo {volume} {41}},\ \bibinfo {pages} {478} (\bibinfo {year} {2003})}\BibitemShut {NoStop}%
\bibitem [{\citenamefont {Bollerslev}(1986)}]{bollerslev1986generalized}%
  \BibitemOpen
  \bibfield  {author} {\bibinfo {author} {\bibfnamefont {T.}~\bibnamefont {Bollerslev}},\ }\bibfield  {title} {\bibinfo {title} {Generalized autoregressive conditional heteroskedasticity},\ }\href {https://doi.org/10.1016/0304-4076(86)90063-1} {\bibfield  {journal} {\bibinfo  {journal} {J. Econom.}\ }\textbf {\bibinfo {volume} {31}},\ \bibinfo {pages} {307} (\bibinfo {year} {1986})}\BibitemShut {NoStop}%
\bibitem [{\citenamefont {Andersen}\ \emph {et~al.}(2001)\citenamefont {Andersen}, \citenamefont {Bollerslev}, \citenamefont {Diebold},\ and\ \citenamefont {Ebens}}]{andersen2001distribution}%
  \BibitemOpen
  \bibfield  {author} {\bibinfo {author} {\bibfnamefont {T.~G.}\ \bibnamefont {Andersen}}, \bibinfo {author} {\bibfnamefont {T.}~\bibnamefont {Bollerslev}}, \bibinfo {author} {\bibfnamefont {F.~X.}\ \bibnamefont {Diebold}},\ and\ \bibinfo {author} {\bibfnamefont {H.}~\bibnamefont {Ebens}},\ }\bibfield  {title} {\bibinfo {title} {The distribution of realized stock return volatility},\ }\href {https://doi.org/10.1016/S0304-405X(01)00055-1} {\bibfield  {journal} {\bibinfo  {journal} {J. financ. econ.}\ }\textbf {\bibinfo {volume} {61}},\ \bibinfo {pages} {43} (\bibinfo {year} {2001})}\BibitemShut {NoStop}%
\bibitem [{\citenamefont {Barndorff-Nielsen}\ and\ \citenamefont {Shephard}(2002)}]{barndorff2002econometric}%
  \BibitemOpen
  \bibfield  {author} {\bibinfo {author} {\bibfnamefont {O.~E.}\ \bibnamefont {Barndorff-Nielsen}}\ and\ \bibinfo {author} {\bibfnamefont {N.}~\bibnamefont {Shephard}},\ }\bibfield  {title} {\bibinfo {title} {Econometric analysis of realized volatility and its use in estimating stochastic volatility models},\ }\href {http://www.jstor.org/stable/3088799} {\bibfield  {journal} {\bibinfo  {journal} {J. R. Stat. Soc}\ }\textbf {\bibinfo {volume} {64}},\ \bibinfo {pages} {253} (\bibinfo {year} {2002})}\BibitemShut {NoStop}%
\bibitem [{\citenamefont {Corsi}(2009)}]{corsi2009simple}%
  \BibitemOpen
  \bibfield  {author} {\bibinfo {author} {\bibfnamefont {F.}~\bibnamefont {Corsi}},\ }\bibfield  {title} {\bibinfo {title} {A {{Simple Approximate Long-Memory Model}} of {{Realized Volatility}}},\ }\href {https://doi.org/10.1093/jjfinec/nbp001} {\bibfield  {journal} {\bibinfo  {journal} {J. financ. econ.}\ }\textbf {\bibinfo {volume} {7}},\ \bibinfo {pages} {174} (\bibinfo {year} {2009})}\BibitemShut {NoStop}%
\bibitem [{\citenamefont {Andersen}\ \emph {et~al.}(2007)\citenamefont {Andersen}, \citenamefont {Bollerslev},\ and\ \citenamefont {Diebold}}]{andersen2007roughing}%
  \BibitemOpen
  \bibfield  {author} {\bibinfo {author} {\bibfnamefont {T.~G.}\ \bibnamefont {Andersen}}, \bibinfo {author} {\bibfnamefont {T.}~\bibnamefont {Bollerslev}},\ and\ \bibinfo {author} {\bibfnamefont {F.~X.}\ \bibnamefont {Diebold}},\ }\bibfield  {title} {\bibinfo {title} {Roughing it up: {{Including}} jump components in the measurement, modeling, and forecasting of return volatility},\ }\href {https://doi.org/10.1162/rest.89.4.701} {\bibfield  {journal} {\bibinfo  {journal} {Rev. Econ. Stat.}\ }\textbf {\bibinfo {volume} {89}},\ \bibinfo {pages} {701} (\bibinfo {year} {2007})}\BibitemShut {NoStop}%
\bibitem [{\citenamefont {Patton}\ and\ \citenamefont {Sheppard}(2015{\natexlab{a}})}]{patton2015good}%
  \BibitemOpen
  \bibfield  {author} {\bibinfo {author} {\bibfnamefont {A.~J.}\ \bibnamefont {Patton}}\ and\ \bibinfo {author} {\bibfnamefont {K.}~\bibnamefont {Sheppard}},\ }\bibfield  {title} {\bibinfo {title} {Good volatility, bad volatility: Signed jumps and the persistence of volatility},\ }\href {http://www.jstor.org/stable/43555003} {\bibfield  {journal} {\bibinfo  {journal} {Rev. Econ. Stat.}\ }\textbf {\bibinfo {volume} {97}},\ \bibinfo {pages} {683} (\bibinfo {year} {2015}{\natexlab{a}})}\BibitemShut {NoStop}%
\bibitem [{\citenamefont {Bollerslev}\ \emph {et~al.}(2016)\citenamefont {Bollerslev}, \citenamefont {Patton},\ and\ \citenamefont {Quaedvlieg}}]{bollerslev2016exploiting}%
  \BibitemOpen
  \bibfield  {author} {\bibinfo {author} {\bibfnamefont {T.}~\bibnamefont {Bollerslev}}, \bibinfo {author} {\bibfnamefont {A.~J.}\ \bibnamefont {Patton}},\ and\ \bibinfo {author} {\bibfnamefont {R.}~\bibnamefont {Quaedvlieg}},\ }\bibfield  {title} {\bibinfo {title} {Exploiting the errors: {{A}} simple approach for improved volatility forecasting},\ }\href {https://doi.org/10.1016/j.jeconom.2015.10.007} {\bibfield  {journal} {\bibinfo  {journal} {J. Econom.}\ }\textbf {\bibinfo {volume} {192}},\ \bibinfo {pages} {1} (\bibinfo {year} {2016})}\BibitemShut {NoStop}%
\bibitem [{\citenamefont {Kuan}\ and\ \citenamefont {White}(1994)}]{kuan1994artificial}%
  \BibitemOpen
  \bibfield  {author} {\bibinfo {author} {\bibfnamefont {C.-M.}\ \bibnamefont {Kuan}}\ and\ \bibinfo {author} {\bibfnamefont {H.}~\bibnamefont {White}},\ }\bibfield  {title} {\bibinfo {title} {Artificial neural networks: An econometric perspective},\ }\href {https://doi.org/https://doi.org/10.1080/07474939408800273} {\bibfield  {journal} {\bibinfo  {journal} {Econom. Rev.}\ }\textbf {\bibinfo {volume} {13}},\ \bibinfo {pages} {1} (\bibinfo {year} {1994})}\BibitemShut {NoStop}%
\bibitem [{\citenamefont {Habibnia}(2016)}]{habibnia2016essays}%
  \BibitemOpen
  \bibfield  {author} {\bibinfo {author} {\bibfnamefont {A.}~\bibnamefont {Habibnia}},\ }\emph {\bibinfo {title} {Essays in high-dimensional nonlinear time series analysis}},\ \href {http://etheses.lse.ac.uk/id/eprint/3485} {Ph.D. thesis},\ \bibinfo  {school} {London School of Economics and Political Science} (\bibinfo {year} {2016})\BibitemShut {NoStop}%
\bibitem [{\citenamefont {Gu}\ \emph {et~al.}(2020)\citenamefont {Gu}, \citenamefont {Kelly},\ and\ \citenamefont {Xiu}}]{gu2020empirical}%
  \BibitemOpen
  \bibfield  {author} {\bibinfo {author} {\bibfnamefont {S.}~\bibnamefont {Gu}}, \bibinfo {author} {\bibfnamefont {B.}~\bibnamefont {Kelly}},\ and\ \bibinfo {author} {\bibfnamefont {D.}~\bibnamefont {Xiu}},\ }\bibfield  {title} {\bibinfo {title} {Empirical asset pricing via machine learning},\ }\href {https://doi.org/10.1093/rfs/hhaa009} {\bibfield  {journal} {\bibinfo  {journal} {Rev. Financ. Stud.}\ }\textbf {\bibinfo {volume} {33}},\ \bibinfo {pages} {2223} (\bibinfo {year} {2020})}\BibitemShut {NoStop}%
\bibitem [{\citenamefont {Bucci}(2020)}]{bucci2020realized}%
  \BibitemOpen
  \bibfield  {author} {\bibinfo {author} {\bibfnamefont {A.}~\bibnamefont {Bucci}},\ }\bibfield  {title} {\bibinfo {title} {{Realized Volatility Forecasting with Neural Networks}},\ }\href {https://doi.org/10.1093/jjfinec/nbaa008} {\bibfield  {journal} {\bibinfo  {journal} {J. financ. econ.}\ }\textbf {\bibinfo {volume} {18}},\ \bibinfo {pages} {502} (\bibinfo {year} {2020})}\BibitemShut {NoStop}%
\bibitem [{\citenamefont {Gu}\ \emph {et~al.}(2021)\citenamefont {Gu}, \citenamefont {Kelly},\ and\ \citenamefont {Xiu}}]{gu2021autoencoder}%
  \BibitemOpen
  \bibfield  {author} {\bibinfo {author} {\bibfnamefont {S.}~\bibnamefont {Gu}}, \bibinfo {author} {\bibfnamefont {B.}~\bibnamefont {Kelly}},\ and\ \bibinfo {author} {\bibfnamefont {D.}~\bibnamefont {Xiu}},\ }\bibfield  {title} {\bibinfo {title} {Autoencoder asset pricing models},\ }\href {https://doi.org/10.1016/j.jeconom.2020.07.009} {\bibfield  {journal} {\bibinfo  {journal} {J. Econom.}\ }\bibinfo {series} {Annals {{Issue}}: {{Financial Econometrics}} in the {{Age}} of the {{Digital Economy}}},\ \textbf {\bibinfo {volume} {222}},\ \bibinfo {pages} {429} (\bibinfo {year} {2021})}\BibitemShut {NoStop}%
\bibitem [{\citenamefont {Habibnia}\ and\ \citenamefont {Maasoumi}(2021)}]{habibnia2021forecasting}%
  \BibitemOpen
  \bibfield  {author} {\bibinfo {author} {\bibfnamefont {A.}~\bibnamefont {Habibnia}}\ and\ \bibinfo {author} {\bibfnamefont {E.}~\bibnamefont {Maasoumi}},\ }\bibfield  {title} {\bibinfo {title} {Forecasting in {{Big Data Environments}}: {{An Adaptable}} and {{Automated Shrinkage Estimation}} of {{Neural Networks}} ({{AAShNet}})},\ }\href {https://doi.org/10.1007/s40953-021-00275-7} {\bibfield  {journal} {\bibinfo  {journal} {Quant. Econ. J.}\ }\textbf {\bibinfo {volume} {19}},\ \bibinfo {pages} {363} (\bibinfo {year} {2021})}\BibitemShut {NoStop}%
\bibitem [{\citenamefont {Zhu}\ \emph {et~al.}(2023)\citenamefont {Zhu}, \citenamefont {Bai}, \citenamefont {He},\ and\ \citenamefont {Liu}}]{zhu2023forecasting}%
  \BibitemOpen
  \bibfield  {author} {\bibinfo {author} {\bibfnamefont {H.}~\bibnamefont {Zhu}}, \bibinfo {author} {\bibfnamefont {L.}~\bibnamefont {Bai}}, \bibinfo {author} {\bibfnamefont {L.}~\bibnamefont {He}},\ and\ \bibinfo {author} {\bibfnamefont {Z.}~\bibnamefont {Liu}},\ }\bibfield  {title} {\bibinfo {title} {Forecasting realized volatility with machine learning: {{Panel}} data perspective},\ }\href {https://doi.org/10.1016/j.jempfin.2023.07.003} {\bibfield  {journal} {\bibinfo  {journal} {J. Empir. Finance}\ }\textbf {\bibinfo {volume} {73}},\ \bibinfo {pages} {251} (\bibinfo {year} {2023})}\BibitemShut {NoStop}%
\bibitem [{\citenamefont {Jiang}\ \emph {et~al.}(2023)\citenamefont {Jiang}, \citenamefont {Kelly},\ and\ \citenamefont {Xiu}}]{jiang2023reimaging}%
  \BibitemOpen
  \bibfield  {author} {\bibinfo {author} {\bibfnamefont {J.}~\bibnamefont {Jiang}}, \bibinfo {author} {\bibfnamefont {B.}~\bibnamefont {Kelly}},\ and\ \bibinfo {author} {\bibfnamefont {D.}~\bibnamefont {Xiu}},\ }\bibfield  {title} {\bibinfo {title} {({{Re-}}){{Imag}}(in)ing {{Price Trends}}},\ }\href {https://doi.org/10.1111/jofi.13268} {\bibfield  {journal} {\bibinfo  {journal} {J. Finance.}\ }\textbf {\bibinfo {volume} {78}},\ \bibinfo {pages} {3193} (\bibinfo {year} {2023})}\BibitemShut {NoStop}%
\bibitem [{\citenamefont {Chen}\ \emph {et~al.}(2024)\citenamefont {Chen}, \citenamefont {Pelger},\ and\ \citenamefont {Zhu}}]{chen2024deep}%
  \BibitemOpen
  \bibfield  {author} {\bibinfo {author} {\bibfnamefont {L.}~\bibnamefont {Chen}}, \bibinfo {author} {\bibfnamefont {M.}~\bibnamefont {Pelger}},\ and\ \bibinfo {author} {\bibfnamefont {J.}~\bibnamefont {Zhu}},\ }\bibfield  {title} {\bibinfo {title} {Deep {{Learning}} in {{Asset Pricing}}},\ }\href {https://doi.org/10.1287/mnsc.2023.4695} {\bibfield  {journal} {\bibinfo  {journal} {Manag. Sci.}\ }\textbf {\bibinfo {volume} {70}},\ \bibinfo {pages} {714} (\bibinfo {year} {2024})}\BibitemShut {NoStop}%
\bibitem [{\citenamefont {Hillebrand}\ and\ \citenamefont {{and Medeiros}}(2010)}]{hillebrand2010bagging}%
  \BibitemOpen
  \bibfield  {author} {\bibinfo {author} {\bibfnamefont {E.}~\bibnamefont {Hillebrand}}\ and\ \bibinfo {author} {\bibfnamefont {M.~C.}\ \bibnamefont {{and Medeiros}}},\ }\bibfield  {title} {\bibinfo {title} {The {{Benefits}} of {{Bagging}} for {{Forecast Models}} of {{Realized Volatility}}},\ }\href {https://doi.org/10.1080/07474938.2010.481554} {\bibfield  {journal} {\bibinfo  {journal} {Econom. Rev.}\ }\textbf {\bibinfo {volume} {29}},\ \bibinfo {pages} {571} (\bibinfo {year} {2010})}\BibitemShut {NoStop}%
\bibitem [{\citenamefont {Fernandes}\ \emph {et~al.}(2014)\citenamefont {Fernandes}, \citenamefont {Medeiros},\ and\ \citenamefont {Scharth}}]{fernandes2014modeling}%
  \BibitemOpen
  \bibfield  {author} {\bibinfo {author} {\bibfnamefont {M.}~\bibnamefont {Fernandes}}, \bibinfo {author} {\bibfnamefont {M.~C.}\ \bibnamefont {Medeiros}},\ and\ \bibinfo {author} {\bibfnamefont {M.}~\bibnamefont {Scharth}},\ }\bibfield  {title} {\bibinfo {title} {Modeling and predicting the {{CBOE}} market volatility index},\ }\href {https://doi.org/10.1016/j.jbankfin.2013.11.004} {\bibfield  {journal} {\bibinfo  {journal} {Journal of Banking \& Finance}\ }\textbf {\bibinfo {volume} {40}},\ \bibinfo {pages} {1} (\bibinfo {year} {2014})}\BibitemShut {NoStop}%
\bibitem [{\citenamefont {Audrino}\ and\ \citenamefont {{and Knaus}}(2016)}]{audrino2016lassoing}%
  \BibitemOpen
  \bibfield  {author} {\bibinfo {author} {\bibfnamefont {F.}~\bibnamefont {Audrino}}\ and\ \bibinfo {author} {\bibfnamefont {S.~D.}\ \bibnamefont {{and Knaus}}},\ }\bibfield  {title} {\bibinfo {title} {Lassoing the {{HAR Model}}: {{A Model Selection Perspective}} on {{Realized Volatility Dynamics}}},\ }\href {https://doi.org/10.1080/07474938.2015.1092801} {\bibfield  {journal} {\bibinfo  {journal} {Econom. Rev.}\ }\textbf {\bibinfo {volume} {35}},\ \bibinfo {pages} {1485} (\bibinfo {year} {2016})}\BibitemShut {NoStop}%
\bibitem [{\citenamefont {Branco}\ \emph {et~al.}(2024)\citenamefont {Branco}, \citenamefont {Rubesam},\ and\ \citenamefont {Zevallos}}]{branco2022forecasting}%
  \BibitemOpen
  \bibfield  {author} {\bibinfo {author} {\bibfnamefont {R.~R.}\ \bibnamefont {Branco}}, \bibinfo {author} {\bibfnamefont {A.}~\bibnamefont {Rubesam}},\ and\ \bibinfo {author} {\bibfnamefont {M.}~\bibnamefont {Zevallos}},\ }\bibfield  {title} {\bibinfo {title} {Forecasting realized volatility: {{Does}} anything beat linear models?},\ }\href {https://doi.org/10.1016/j.jempfin.2024.101524} {\bibfield  {journal} {\bibinfo  {journal} {J. Empir. Finance}\ }\textbf {\bibinfo {volume} {78}},\ \bibinfo {pages} {101524} (\bibinfo {year} {2024})}\BibitemShut {NoStop}%
\bibitem [{\citenamefont {Christensen}\ \emph {et~al.}(2023)\citenamefont {Christensen}, \citenamefont {Siggaard},\ and\ \citenamefont {Veliyev}}]{christensen2023machine}%
  \BibitemOpen
  \bibfield  {author} {\bibinfo {author} {\bibfnamefont {K.}~\bibnamefont {Christensen}}, \bibinfo {author} {\bibfnamefont {M.}~\bibnamefont {Siggaard}},\ and\ \bibinfo {author} {\bibfnamefont {B.}~\bibnamefont {Veliyev}},\ }\bibfield  {title} {\bibinfo {title} {A machine learning approach to volatility forecasting},\ }\href {https://doi.org/https://doi.org/10.1093/jjfinec/nbac020} {\bibfield  {journal} {\bibinfo  {journal} {J. financ. econ.}\ }\textbf {\bibinfo {volume} {21}},\ \bibinfo {pages} {1680} (\bibinfo {year} {2023})}\BibitemShut {NoStop}%
\bibitem [{\citenamefont {Zhang}\ \emph {et~al.}(2024)\citenamefont {Zhang}, \citenamefont {Zhang}, \citenamefont {Cucuringu},\ and\ \citenamefont {Qian}}]{zhang2024volatility}%
  \BibitemOpen
  \bibfield  {author} {\bibinfo {author} {\bibfnamefont {C.}~\bibnamefont {Zhang}}, \bibinfo {author} {\bibfnamefont {Y.}~\bibnamefont {Zhang}}, \bibinfo {author} {\bibfnamefont {M.}~\bibnamefont {Cucuringu}},\ and\ \bibinfo {author} {\bibfnamefont {Z.}~\bibnamefont {Qian}},\ }\bibfield  {title} {\bibinfo {title} {{Volatility Forecasting with Machine Learning and Intraday Commonality}},\ }\href {https://doi.org/10.1093/jjfinec/nbad005} {\bibfield  {journal} {\bibinfo  {journal} {J. financ. econ.}\ }\textbf {\bibinfo {volume} {22}},\ \bibinfo {pages} {492} (\bibinfo {year} {2024})}\BibitemShut {NoStop}%
\bibitem [{\citenamefont {Ghysels}\ \emph {et~al.}(2006)\citenamefont {Ghysels}, \citenamefont {{Santa-Clara}},\ and\ \citenamefont {Valkanov}}]{ghysels2006predicting}%
  \BibitemOpen
  \bibfield  {author} {\bibinfo {author} {\bibfnamefont {E.}~\bibnamefont {Ghysels}}, \bibinfo {author} {\bibfnamefont {P.}~\bibnamefont {{Santa-Clara}}},\ and\ \bibinfo {author} {\bibfnamefont {R.}~\bibnamefont {Valkanov}},\ }\bibfield  {title} {\bibinfo {title} {Predicting volatility: Getting the most out of return data sampled at different frequencies},\ }\href {https://doi.org/10.1016/j.jeconom.2005.01.004} {\bibfield  {journal} {\bibinfo  {journal} {J. Econom.}\ }\textbf {\bibinfo {volume} {131}},\ \bibinfo {pages} {59} (\bibinfo {year} {2006})}\BibitemShut {NoStop}%
\bibitem [{\citenamefont {Gunnarsson}\ \emph {et~al.}(2024)\citenamefont {Gunnarsson}, \citenamefont {Isern}, \citenamefont {Kaloudis}, \citenamefont {Risstad}, \citenamefont {Vigdel},\ and\ \citenamefont {Westgaard}}]{gunnarsson2024prediction}%
  \BibitemOpen
  \bibfield  {author} {\bibinfo {author} {\bibfnamefont {E.~S.}\ \bibnamefont {Gunnarsson}}, \bibinfo {author} {\bibfnamefont {H.~R.}\ \bibnamefont {Isern}}, \bibinfo {author} {\bibfnamefont {A.}~\bibnamefont {Kaloudis}}, \bibinfo {author} {\bibfnamefont {M.}~\bibnamefont {Risstad}}, \bibinfo {author} {\bibfnamefont {B.}~\bibnamefont {Vigdel}},\ and\ \bibinfo {author} {\bibfnamefont {S.}~\bibnamefont {Westgaard}},\ }\bibfield  {title} {\bibinfo {title} {Prediction of realized volatility and implied volatility indices using {{AI}} and machine learning: {{A}} review},\ }\href {https://doi.org/10.1016/j.irfa.2024.103221} {\bibfield  {journal} {\bibinfo  {journal} {International Review of Financial Analysis}\ }\textbf {\bibinfo {volume} {93}},\ \bibinfo {pages} {103221} (\bibinfo {year} {2024})}\BibitemShut {NoStop}%
\bibitem [{\citenamefont {Hochreiter}\ and\ \citenamefont {Schmidhuber}(1997)}]{hochreiter1997long}%
  \BibitemOpen
  \bibfield  {author} {\bibinfo {author} {\bibfnamefont {S.}~\bibnamefont {Hochreiter}}\ and\ \bibinfo {author} {\bibfnamefont {J.}~\bibnamefont {Schmidhuber}},\ }\bibfield  {title} {\bibinfo {title} {Long {{Short-Term Memory}}},\ }\href {https://doi.org/10.1162/neco.1997.9.8.1735} {\bibfield  {journal} {\bibinfo  {journal} {Neural Comput.}\ }\textbf {\bibinfo {volume} {9}},\ \bibinfo {pages} {1735} (\bibinfo {year} {1997})}\BibitemShut {NoStop}%
\bibitem [{\citenamefont {Goodfellow}\ \emph {et~al.}(2016)\citenamefont {Goodfellow}, \citenamefont {Bengio},\ and\ \citenamefont {Courville}}]{goodfellow2016deep}%
  \BibitemOpen
  \bibfield  {author} {\bibinfo {author} {\bibfnamefont {I.}~\bibnamefont {Goodfellow}}, \bibinfo {author} {\bibfnamefont {Y.}~\bibnamefont {Bengio}},\ and\ \bibinfo {author} {\bibfnamefont {A.}~\bibnamefont {Courville}},\ }\href@noop {} {\emph {\bibinfo {title} {Deep Learning}}}\ (\bibinfo  {publisher} {MIT Press},\ \bibinfo {year} {2016})\ \bibinfo {note} {\url{http://www.deeplearningbook.org}}\BibitemShut {NoStop}%
\bibitem [{\citenamefont {Luko{\v{s}}evi{\v{c}}ius}\ and\ \citenamefont {Jaeger}(2009)}]{lukovsevivcius2009reservoir}%
  \BibitemOpen
  \bibfield  {author} {\bibinfo {author} {\bibfnamefont {M.}~\bibnamefont {Luko{\v{s}}evi{\v{c}}ius}}\ and\ \bibinfo {author} {\bibfnamefont {H.}~\bibnamefont {Jaeger}},\ }\bibfield  {title} {\bibinfo {title} {Reservoir computing approaches to recurrent neural network training},\ }\href {https://doi.org/https://doi.org/10.1016/j.cosrev.2009.03.005} {\bibfield  {journal} {\bibinfo  {journal} {Comput. Sci. Rev.}\ }\textbf {\bibinfo {volume} {3}},\ \bibinfo {pages} {127} (\bibinfo {year} {2009})}\BibitemShut {NoStop}%
\bibitem [{\citenamefont {Mujal}\ \emph {et~al.}(2023)\citenamefont {Mujal}, \citenamefont {Mart{\'\i}nez-Pe{\~n}a}, \citenamefont {Giorgi}, \citenamefont {Soriano},\ and\ \citenamefont {Zambrini}}]{mujal2023time}%
  \BibitemOpen
  \bibfield  {author} {\bibinfo {author} {\bibfnamefont {P.}~\bibnamefont {Mujal}}, \bibinfo {author} {\bibfnamefont {R.}~\bibnamefont {Mart{\'\i}nez-Pe{\~n}a}}, \bibinfo {author} {\bibfnamefont {G.~L.}\ \bibnamefont {Giorgi}}, \bibinfo {author} {\bibfnamefont {M.~C.}\ \bibnamefont {Soriano}},\ and\ \bibinfo {author} {\bibfnamefont {R.}~\bibnamefont {Zambrini}},\ }\bibfield  {title} {\bibinfo {title} {Time-series quantum reservoir computing with weak and projective measurements},\ }\href {https://doi.org/https://doi.org/10.1038/s41534-023-00682-z} {\bibfield  {journal} {\bibinfo  {journal} {Npj Quantum Inf.}\ }\textbf {\bibinfo {volume} {9}},\ \bibinfo {pages} {16} (\bibinfo {year} {2023})}\BibitemShut {NoStop}%
\bibitem [{\citenamefont {Garc\'{\i}a-Beni}\ \emph {et~al.}(2023)\citenamefont {Garc\'{\i}a-Beni}, \citenamefont {Giorgi}, \citenamefont {Soriano},\ and\ \citenamefont {Zambrini}}]{garcia2023scalable}%
  \BibitemOpen
  \bibfield  {author} {\bibinfo {author} {\bibfnamefont {J.}~\bibnamefont {Garc\'{\i}a-Beni}}, \bibinfo {author} {\bibfnamefont {G.~L.}\ \bibnamefont {Giorgi}}, \bibinfo {author} {\bibfnamefont {M.~C.}\ \bibnamefont {Soriano}},\ and\ \bibinfo {author} {\bibfnamefont {R.}~\bibnamefont {Zambrini}},\ }\bibfield  {title} {\bibinfo {title} {Scalable photonic platform for real-time quantum reservoir computing},\ }\href {https://doi.org/10.1103/PhysRevApplied.20.014051} {\bibfield  {journal} {\bibinfo  {journal} {Phys. Rev. Appl.}\ }\textbf {\bibinfo {volume} {20}},\ \bibinfo {pages} {014051} (\bibinfo {year} {2023})}\BibitemShut {NoStop}%
\bibitem [{\citenamefont {Llodr{\`a}}\ \emph {et~al.}(2025)\citenamefont {Llodr{\`a}}, \citenamefont {Mujal}, \citenamefont {Zambrini},\ and\ \citenamefont {Giorgi}}]{llodra2025quantum}%
  \BibitemOpen
  \bibfield  {author} {\bibinfo {author} {\bibfnamefont {G.}~\bibnamefont {Llodr{\`a}}}, \bibinfo {author} {\bibfnamefont {P.}~\bibnamefont {Mujal}}, \bibinfo {author} {\bibfnamefont {R.}~\bibnamefont {Zambrini}},\ and\ \bibinfo {author} {\bibfnamefont {G.~L.}\ \bibnamefont {Giorgi}},\ }\bibfield  {title} {\bibinfo {title} {Quantum reservoir computing in atomic lattices},\ }\href {https://doi.org/https://doi.org/10.1016/j.chaos.2025.116289} {\bibfield  {journal} {\bibinfo  {journal} {Chaos, Solitons \& Fractals}\ }\textbf {\bibinfo {volume} {195}},\ \bibinfo {pages} {116289} (\bibinfo {year} {2025})}\BibitemShut {NoStop}%
\bibitem [{\citenamefont {Garcia-Beni}\ \emph {et~al.}(2024)\citenamefont {Garcia-Beni}, \citenamefont {Giorgi}, \citenamefont {Soriano},\ and\ \citenamefont {Zambrini}}]{garcia2024quantum}%
  \BibitemOpen
  \bibfield  {author} {\bibinfo {author} {\bibfnamefont {J.}~\bibnamefont {Garcia-Beni}}, \bibinfo {author} {\bibfnamefont {G.~L.}\ \bibnamefont {Giorgi}}, \bibinfo {author} {\bibfnamefont {M.~C.}\ \bibnamefont {Soriano}},\ and\ \bibinfo {author} {\bibfnamefont {R.}~\bibnamefont {Zambrini}},\ }\bibfield  {title} {\bibinfo {title} {{Quantum reservoir computing for time series processing}},\ }in\ \href {https://doi.org/10.1117/12.3027999} {\emph {\bibinfo {booktitle} {Quantum Communications and Quantum Imaging XXII}}},\ Vol.\ \bibinfo {volume} {PC13148},\ \bibinfo {editor} {edited by\ \bibinfo {editor} {\bibfnamefont {K.~S.}\ \bibnamefont {Deacon}}\ and\ \bibinfo {editor} {\bibfnamefont {R.~E.}\ \bibnamefont {Meyers}}},\ \bibinfo {organization} {International Society for Optics and Photonics}\ (\bibinfo  {publisher} {SPIE},\ \bibinfo {year} {2024})\ p.\ \bibinfo {pages} {PC131480E}\BibitemShut {NoStop}%
\bibitem [{\citenamefont {Thakkar}\ \emph {et~al.}(2024)\citenamefont {Thakkar}, \citenamefont {Kazdaghli}, \citenamefont {Mathur}, \citenamefont {Kerenidis}, \citenamefont {{Ferreira--Martins}},\ and\ \citenamefont {Brito}}]{thakkar2023improved}%
  \BibitemOpen
  \bibfield  {author} {\bibinfo {author} {\bibfnamefont {S.}~\bibnamefont {Thakkar}}, \bibinfo {author} {\bibfnamefont {S.}~\bibnamefont {Kazdaghli}}, \bibinfo {author} {\bibfnamefont {N.}~\bibnamefont {Mathur}}, \bibinfo {author} {\bibfnamefont {I.}~\bibnamefont {Kerenidis}}, \bibinfo {author} {\bibfnamefont {A.~J.}\ \bibnamefont {{Ferreira--Martins}}},\ and\ \bibinfo {author} {\bibfnamefont {S.}~\bibnamefont {Brito}},\ }\bibfield  {title} {\bibinfo {title} {Improved financial forecasting via quantum machine learning},\ }\href {https://doi.org/10.1007/s42484-024-00157-0} {\bibfield  {journal} {\bibinfo  {journal} {Quantum Mach. Intell.}\ }\textbf {\bibinfo {volume} {6}},\ \bibinfo {pages} {27} (\bibinfo {year} {2024})}\BibitemShut {NoStop}%
\bibitem [{\citenamefont {{Rivera-Ruiz}}\ \emph {et~al.}(2022)\citenamefont {{Rivera-Ruiz}}, \citenamefont {{Mendez-Vazquez}},\ and\ \citenamefont {{L{\'o}pez-Romero}}}]{rivera2022time}%
  \BibitemOpen
  \bibfield  {author} {\bibinfo {author} {\bibfnamefont {M.~A.}\ \bibnamefont {{Rivera-Ruiz}}}, \bibinfo {author} {\bibfnamefont {A.}~\bibnamefont {{Mendez-Vazquez}}},\ and\ \bibinfo {author} {\bibfnamefont {J.~M.}\ \bibnamefont {{L{\'o}pez-Romero}}},\ }\bibfield  {title} {\bibinfo {title} {Time {{Series Forecasting}} with~{{Quantum Machine Learning Architectures}}},\ }in\ \href {https://doi.org/10.1007/978-3-031-19493-1_6} {\emph {\bibinfo {booktitle} {Advances in {{Computational Intelligence}}}}},\ \bibinfo {editor} {edited by\ \bibinfo {editor} {\bibfnamefont {O.}~\bibnamefont {Pichardo~Lagunas}}, \bibinfo {editor} {\bibfnamefont {J.}~\bibnamefont {{Mart{\'i}nez-Miranda}}},\ and\ \bibinfo {editor} {\bibfnamefont {B.}~\bibnamefont {Mart{\'i}nez~Seis}}}\ (\bibinfo  {publisher} {Springer Nature Switzerland},\ \bibinfo {address} {Cham},\ \bibinfo {year} {2022})\ pp.\ \bibinfo {pages} {66--82}\BibitemShut {NoStop}%
\bibitem [{\citenamefont {Settino}\ \emph {et~al.}(2024)\citenamefont {Settino}, \citenamefont {Salatino}, \citenamefont {Mariani}, \citenamefont {Channab}, \citenamefont {Bozzolo}, \citenamefont {Vallisa}, \citenamefont {Barill{\`a}}, \citenamefont {Policicchio}, \citenamefont {Gullo}, \citenamefont {Giordano} \emph {et~al.}}]{settino2024memory}%
  \BibitemOpen
  \bibfield  {author} {\bibinfo {author} {\bibfnamefont {J.}~\bibnamefont {Settino}}, \bibinfo {author} {\bibfnamefont {L.}~\bibnamefont {Salatino}}, \bibinfo {author} {\bibfnamefont {L.}~\bibnamefont {Mariani}}, \bibinfo {author} {\bibfnamefont {M.}~\bibnamefont {Channab}}, \bibinfo {author} {\bibfnamefont {L.}~\bibnamefont {Bozzolo}}, \bibinfo {author} {\bibfnamefont {S.}~\bibnamefont {Vallisa}}, \bibinfo {author} {\bibfnamefont {P.}~\bibnamefont {Barill{\`a}}}, \bibinfo {author} {\bibfnamefont {A.}~\bibnamefont {Policicchio}}, \bibinfo {author} {\bibfnamefont {N.~L.}\ \bibnamefont {Gullo}}, \bibinfo {author} {\bibfnamefont {A.}~\bibnamefont {Giordano}}, \emph {et~al.},\ }\bibfield  {title} {\bibinfo {title} {Memory-augmented hybrid quantum reservoir computing},\ }\bibfield  {journal} {\bibinfo  {journal} {arXiv preprint arXiv:2409.09886}\ }\href {https://doi.org/https://doi.org/10.48550/arXiv.2409.09886} {https://doi.org/10.48550/arXiv.2409.09886} (\bibinfo {year} {2024})\BibitemShut {NoStop}%
\bibitem [{\citenamefont {Nerenberg}\ \emph {et~al.}(2025)\citenamefont {Nerenberg}, \citenamefont {Neill}, \citenamefont {Marcucci},\ and\ \citenamefont {Faccio}}]{nerenberg2024photon}%
  \BibitemOpen
  \bibfield  {author} {\bibinfo {author} {\bibfnamefont {S.}~\bibnamefont {Nerenberg}}, \bibinfo {author} {\bibfnamefont {O.~D.}\ \bibnamefont {Neill}}, \bibinfo {author} {\bibfnamefont {G.}~\bibnamefont {Marcucci}},\ and\ \bibinfo {author} {\bibfnamefont {D.}~\bibnamefont {Faccio}},\ }\bibfield  {title} {\bibinfo {title} {Photon number-resolving quantum reservoir computing},\ }\href {https://doi.org/10.1364/OPTICAQ.553294} {\bibfield  {journal} {\bibinfo  {journal} {Optica Quantum}\ }\textbf {\bibinfo {volume} {3}},\ \bibinfo {pages} {201} (\bibinfo {year} {2025})}\BibitemShut {NoStop}%
\bibitem [{\citenamefont {Suprano}\ \emph {et~al.}(2024)\citenamefont {Suprano}, \citenamefont {Zia}, \citenamefont {Innocenti}, \citenamefont {Lorenzo}, \citenamefont {Cimini}, \citenamefont {Giordani}, \citenamefont {Palmisano}, \citenamefont {Polino}, \citenamefont {Spagnolo}, \citenamefont {Sciarrino}, \citenamefont {Palma}, \citenamefont {Ferraro},\ and\ \citenamefont {Paternostro}}]{suprano2024experimental}%
  \BibitemOpen
  \bibfield  {author} {\bibinfo {author} {\bibfnamefont {A.}~\bibnamefont {Suprano}}, \bibinfo {author} {\bibfnamefont {D.}~\bibnamefont {Zia}}, \bibinfo {author} {\bibfnamefont {L.}~\bibnamefont {Innocenti}}, \bibinfo {author} {\bibfnamefont {S.}~\bibnamefont {Lorenzo}}, \bibinfo {author} {\bibfnamefont {V.}~\bibnamefont {Cimini}}, \bibinfo {author} {\bibfnamefont {T.}~\bibnamefont {Giordani}}, \bibinfo {author} {\bibfnamefont {I.}~\bibnamefont {Palmisano}}, \bibinfo {author} {\bibfnamefont {E.}~\bibnamefont {Polino}}, \bibinfo {author} {\bibfnamefont {N.}~\bibnamefont {Spagnolo}}, \bibinfo {author} {\bibfnamefont {F.}~\bibnamefont {Sciarrino}}, \bibinfo {author} {\bibfnamefont {G.~M.}\ \bibnamefont {Palma}}, \bibinfo {author} {\bibfnamefont {A.}~\bibnamefont {Ferraro}},\ and\ \bibinfo {author} {\bibfnamefont {M.}~\bibnamefont {Paternostro}},\ }\bibfield  {title} {\bibinfo {title} {Experimental property reconstruction in a photonic quantum extreme learning machine},\ }\href
  {https://doi.org/10.1103/PhysRevLett.132.160802} {\bibfield  {journal} {\bibinfo  {journal} {Phys. Rev. Lett.}\ }\textbf {\bibinfo {volume} {132}},\ \bibinfo {pages} {160802} (\bibinfo {year} {2024})}\BibitemShut {NoStop}%
\bibitem [{\citenamefont {Abbas}\ and\ \citenamefont {Maksymov}(2024)}]{abbas2024reservoir}%
  \BibitemOpen
  \bibfield  {author} {\bibinfo {author} {\bibfnamefont {A.~H.}\ \bibnamefont {Abbas}}\ and\ \bibinfo {author} {\bibfnamefont {I.~S.}\ \bibnamefont {Maksymov}},\ }\bibfield  {title} {\bibinfo {title} {Reservoir {{Computing Using Measurement-Controlled Quantum Dynamics}}},\ }\href {https://doi.org/10.3390/electronics13061164} {\bibfield  {journal} {\bibinfo  {journal} {Electronics}\ }\textbf {\bibinfo {volume} {13}},\ \bibinfo {pages} {1164} (\bibinfo {year} {2024})}\BibitemShut {NoStop}%
\bibitem [{\citenamefont {Alaminos}\ \emph {et~al.}(2022)\citenamefont {Alaminos}, \citenamefont {Belen~Salas},\ and\ \citenamefont {Fernandez-Gamez}}]{alaminos2022forecasting}%
  \BibitemOpen
  \bibfield  {author} {\bibinfo {author} {\bibfnamefont {D.}~\bibnamefont {Alaminos}}, \bibinfo {author} {\bibfnamefont {M.}~\bibnamefont {Belen~Salas}},\ and\ \bibinfo {author} {\bibfnamefont {M.~A.}\ \bibnamefont {Fernandez-Gamez}},\ }\bibfield  {title} {\bibinfo {title} {Forecasting stock market crashes via real-time recession probabilities: a quantum computing approach},\ }\href {https://doi.org/https://doi.org/10.1142/S0218348X22401624} {\bibfield  {journal} {\bibinfo  {journal} {Fractals}\ }\textbf {\bibinfo {volume} {30}},\ \bibinfo {pages} {2240162} (\bibinfo {year} {2022})}\BibitemShut {NoStop}%
\bibitem [{\citenamefont {Andersen}\ and\ \citenamefont {Bollerslev}(1998)}]{andersen1998answering}%
  \BibitemOpen
  \bibfield  {author} {\bibinfo {author} {\bibfnamefont {T.~G.}\ \bibnamefont {Andersen}}\ and\ \bibinfo {author} {\bibfnamefont {T.}~\bibnamefont {Bollerslev}},\ }\bibfield  {title} {\bibinfo {title} {Answering the {{Skeptics}}: {{Yes}}, {{Standard Volatility Models}} do {{Provide Accurate Forecasts}}},\ }\href {https://doi.org/10.2307/2527343} {\bibfield  {journal} {\bibinfo  {journal} {Int. Rev. Econ.}\ }\textbf {\bibinfo {volume} {39}},\ \bibinfo {pages} {885} (\bibinfo {year} {1998})},\ \Eprint {https://arxiv.org/abs/2527343} {2527343} \BibitemShut {NoStop}%
\bibitem [{\citenamefont {Newey}\ and\ \citenamefont {West}(1986)}]{newey1986simple}%
  \BibitemOpen
  \bibfield  {author} {\bibinfo {author} {\bibfnamefont {W.~K.}\ \bibnamefont {Newey}}\ and\ \bibinfo {author} {\bibfnamefont {K.~D.}\ \bibnamefont {West}},\ }\bibfield  {title} {\bibinfo {title} {A simple, positive semi-definite, heteroskedasticity and autocorrelationconsistent covariance matrix},\ }\href@noop {} {\  (\bibinfo {year} {1986})}\BibitemShut {NoStop}%
\bibitem [{\citenamefont {Hansen}\ \emph {et~al.}(2012)\citenamefont {Hansen}, \citenamefont {Huang},\ and\ \citenamefont {Shek}}]{hansen2012realized}%
  \BibitemOpen
  \bibfield  {author} {\bibinfo {author} {\bibfnamefont {P.~R.}\ \bibnamefont {Hansen}}, \bibinfo {author} {\bibfnamefont {Z.}~\bibnamefont {Huang}},\ and\ \bibinfo {author} {\bibfnamefont {H.~H.}\ \bibnamefont {Shek}},\ }\bibfield  {title} {\bibinfo {title} {Realized garch: a joint model for returns and realized measures of volatility},\ }\href {https://doi.org/https://doi.org/10.1002/jae.1234} {\bibfield  {journal} {\bibinfo  {journal} {Journal of Applied Econometrics}\ }\textbf {\bibinfo {volume} {27}},\ \bibinfo {pages} {877} (\bibinfo {year} {2012})}\BibitemShut {NoStop}%
\bibitem [{\citenamefont {McAleer}\ and\ \citenamefont {Medeiros}(2008)}]{mcaleer2008realized}%
  \BibitemOpen
  \bibfield  {author} {\bibinfo {author} {\bibfnamefont {M.}~\bibnamefont {McAleer}}\ and\ \bibinfo {author} {\bibfnamefont {M.~C.}\ \bibnamefont {Medeiros}},\ }\bibfield  {title} {\bibinfo {title} {Realized volatility: A review},\ }\href {https://doi.org/https://doi.org/10.1080/07474930701853509} {\bibfield  {journal} {\bibinfo  {journal} {Econom. Rev.}\ }\textbf {\bibinfo {volume} {27}},\ \bibinfo {pages} {10} (\bibinfo {year} {2008})}\BibitemShut {NoStop}%
\bibitem [{\citenamefont {Fischer}\ and\ \citenamefont {Krauss}(2018)}]{fischer2018deep}%
  \BibitemOpen
  \bibfield  {author} {\bibinfo {author} {\bibfnamefont {T.}~\bibnamefont {Fischer}}\ and\ \bibinfo {author} {\bibfnamefont {C.}~\bibnamefont {Krauss}},\ }\bibfield  {title} {\bibinfo {title} {Deep learning with long short-term memory networks for financial market predictions},\ }\href {https://doi.org/10.1016/j.ejor.2017.11.054} {\bibfield  {journal} {\bibinfo  {journal} {Eur. J. Oper.}\ }\textbf {\bibinfo {volume} {270}},\ \bibinfo {pages} {654} (\bibinfo {year} {2018})}\BibitemShut {NoStop}%
\bibitem [{\citenamefont {Butcher}\ \emph {et~al.}(2013)\citenamefont {Butcher}, \citenamefont {Verstraeten}, \citenamefont {Schrauwen}, \citenamefont {Day},\ and\ \citenamefont {Haycock}}]{butcher2013reservoir}%
  \BibitemOpen
  \bibfield  {author} {\bibinfo {author} {\bibfnamefont {J.~B.}\ \bibnamefont {Butcher}}, \bibinfo {author} {\bibfnamefont {D.}~\bibnamefont {Verstraeten}}, \bibinfo {author} {\bibfnamefont {B.}~\bibnamefont {Schrauwen}}, \bibinfo {author} {\bibfnamefont {C.~R.}\ \bibnamefont {Day}},\ and\ \bibinfo {author} {\bibfnamefont {P.~W.}\ \bibnamefont {Haycock}},\ }\bibfield  {title} {\bibinfo {title} {Reservoir computing and extreme learning machines for non-linear time-series data analysis},\ }\href {https://doi.org/10.1016/j.neunet.2012.11.011} {\bibfield  {journal} {\bibinfo  {journal} {Neural Networks}\ }\textbf {\bibinfo {volume} {38}},\ \bibinfo {pages} {76} (\bibinfo {year} {2013})}\BibitemShut {NoStop}%
\bibitem [{\citenamefont {Zou}\ \emph {et~al.}(2009)\citenamefont {Zou}, \citenamefont {Han},\ and\ \citenamefont {So}}]{zou2009overview}%
  \BibitemOpen
  \bibfield  {author} {\bibinfo {author} {\bibfnamefont {J.}~\bibnamefont {Zou}}, \bibinfo {author} {\bibfnamefont {Y.}~\bibnamefont {Han}},\ and\ \bibinfo {author} {\bibfnamefont {S.-S.}\ \bibnamefont {So}},\ }\bibfield  {title} {\bibinfo {title} {Overview of {{Artificial Neural Networks}}},\ }in\ \href {https://doi.org/10.1007/978-1-60327-101-1_2} {\emph {\bibinfo {booktitle} {Artificial {{Neural Networks}}: {{Methods}} and {{Applications}}}}},\ \bibinfo {editor} {edited by\ \bibinfo {editor} {\bibfnamefont {D.~J.}\ \bibnamefont {Livingstone}}}\ (\bibinfo  {publisher} {Humana Press},\ \bibinfo {address} {Totowa, NJ},\ \bibinfo {year} {2009})\ pp.\ \bibinfo {pages} {14--22}\BibitemShut {NoStop}%
\bibitem [{\citenamefont {Maass}\ \emph {et~al.}(2002)\citenamefont {Maass}, \citenamefont {Natschl{\"a}ger},\ and\ \citenamefont {Markram}}]{maass2002real}%
  \BibitemOpen
  \bibfield  {author} {\bibinfo {author} {\bibfnamefont {W.}~\bibnamefont {Maass}}, \bibinfo {author} {\bibfnamefont {T.}~\bibnamefont {Natschl{\"a}ger}},\ and\ \bibinfo {author} {\bibfnamefont {H.}~\bibnamefont {Markram}},\ }\bibfield  {title} {\bibinfo {title} {Real-time computing without stable states: A new framework for neural computation based on perturbations},\ }\href {https://doi.org/10.1162/089976602760407955} {\bibfield  {journal} {\bibinfo  {journal} {Neural Comput.}\ }\textbf {\bibinfo {volume} {14}},\ \bibinfo {pages} {2531} (\bibinfo {year} {2002})}\BibitemShut {NoStop}%
\bibitem [{\citenamefont {Yilmaz}(2015)}]{yilmazSymbolicComputationUsing2015}%
  \BibitemOpen
  \bibfield  {author} {\bibinfo {author} {\bibfnamefont {O.}~\bibnamefont {Yilmaz}},\ }\bibfield  {title} {\bibinfo {title} {Symbolic {{Computation Using Cellular Automata-Based Hyperdimensional Computing}}},\ }\href {https://doi.org/10.1162/NECO_a_00787} {\bibfield  {journal} {\bibinfo  {journal} {Neural Comput.}\ }\textbf {\bibinfo {volume} {27}},\ \bibinfo {pages} {2661} (\bibinfo {year} {2015})}\BibitemShut {NoStop}%
\bibitem [{\citenamefont {McDonald}(2017)}]{mcdonaldReservoirComputingExtreme2017}%
  \BibitemOpen
  \bibfield  {author} {\bibinfo {author} {\bibfnamefont {N.}~\bibnamefont {McDonald}},\ }\bibfield  {title} {\bibinfo {title} {Reservoir computing \& extreme learning machines using pairs of cellular automata rules},\ }in\ \href {https://doi.org/10.1109/IJCNN.2017.7966151} {\emph {\bibinfo {booktitle} {2017 {{International Joint Conference}} on {{Neural Networks}} ({{IJCNN}})}}}\ (\bibinfo {year} {2017})\ pp.\ \bibinfo {pages} {2429--2436}\BibitemShut {NoStop}%
\bibitem [{\citenamefont {Yamane}\ \emph {et~al.}(2015)\citenamefont {Yamane}, \citenamefont {Katayama}, \citenamefont {Nakane}, \citenamefont {Tanaka},\ and\ \citenamefont {Nakano}}]{yamaneWaveBasedReservoirComputing2015}%
  \BibitemOpen
  \bibfield  {author} {\bibinfo {author} {\bibfnamefont {T.}~\bibnamefont {Yamane}}, \bibinfo {author} {\bibfnamefont {Y.}~\bibnamefont {Katayama}}, \bibinfo {author} {\bibfnamefont {R.}~\bibnamefont {Nakane}}, \bibinfo {author} {\bibfnamefont {G.}~\bibnamefont {Tanaka}},\ and\ \bibinfo {author} {\bibfnamefont {D.}~\bibnamefont {Nakano}},\ }\bibfield  {title} {\bibinfo {title} {Wave-{{Based Reservoir Computing}} by {{Synchronization}} of {{Coupled Oscillators}}},\ }in\ \href {https://doi.org/10.1007/978-3-319-26555-1_23} {\emph {\bibinfo {booktitle} {Neural {{Information Processing}}}}},\ \bibinfo {editor} {edited by\ \bibinfo {editor} {\bibfnamefont {S.}~\bibnamefont {Arik}}, \bibinfo {editor} {\bibfnamefont {T.}~\bibnamefont {Huang}}, \bibinfo {editor} {\bibfnamefont {W.~K.}\ \bibnamefont {Lai}},\ and\ \bibinfo {editor} {\bibfnamefont {Q.}~\bibnamefont {Liu}}}\ (\bibinfo  {publisher} {Springer International Publishing},\ \bibinfo {address} {Cham},\ \bibinfo {year} {2015})\ pp.\ \bibinfo {pages}
  {198--205}\BibitemShut {NoStop}%
\bibitem [{\citenamefont {Roy}\ \emph {et~al.}(2014)\citenamefont {Roy}, \citenamefont {Banerjee},\ and\ \citenamefont {Basu}}]{royLiquidStateMachine2014}%
  \BibitemOpen
  \bibfield  {author} {\bibinfo {author} {\bibfnamefont {S.}~\bibnamefont {Roy}}, \bibinfo {author} {\bibfnamefont {A.}~\bibnamefont {Banerjee}},\ and\ \bibinfo {author} {\bibfnamefont {A.}~\bibnamefont {Basu}},\ }\bibfield  {title} {\bibinfo {title} {Liquid {{State Machine With Dendritically Enhanced Readout}} for {{Low-Power}}, {{Neuromorphic VLSI Implementations}}},\ }\href {https://doi.org/10.1109/TBCAS.2014.2362969} {\bibfield  {journal} {\bibinfo  {journal} {IEEE Transactions on Biomedical Circuits and Systems}\ }\textbf {\bibinfo {volume} {8}},\ \bibinfo {pages} {681} (\bibinfo {year} {2014})}\BibitemShut {NoStop}%
\bibitem [{\citenamefont {Katumba}\ \emph {et~al.}(2018)\citenamefont {Katumba}, \citenamefont {Heyvaert}, \citenamefont {Schneider}, \citenamefont {Uvin}, \citenamefont {Dambre},\ and\ \citenamefont {Bienstman}}]{katumbaLowLossPhotonicReservoir2018}%
  \BibitemOpen
  \bibfield  {author} {\bibinfo {author} {\bibfnamefont {A.}~\bibnamefont {Katumba}}, \bibinfo {author} {\bibfnamefont {J.}~\bibnamefont {Heyvaert}}, \bibinfo {author} {\bibfnamefont {B.}~\bibnamefont {Schneider}}, \bibinfo {author} {\bibfnamefont {S.}~\bibnamefont {Uvin}}, \bibinfo {author} {\bibfnamefont {J.}~\bibnamefont {Dambre}},\ and\ \bibinfo {author} {\bibfnamefont {P.}~\bibnamefont {Bienstman}},\ }\bibfield  {title} {\bibinfo {title} {Low-{{Loss Photonic Reservoir Computing}} with {{Multimode Photonic Integrated Circuits}}},\ }\href {https://doi.org/10.1038/s41598-018-21011-x} {\bibfield  {journal} {\bibinfo  {journal} {Sci. Rep.}\ }\textbf {\bibinfo {volume} {8}},\ \bibinfo {pages} {2653} (\bibinfo {year} {2018})}\BibitemShut {NoStop}%
\bibitem [{\citenamefont {Duport}\ \emph {et~al.}(2012)\citenamefont {Duport}, \citenamefont {Schneider}, \citenamefont {Smerieri}, \citenamefont {Haelterman},\ and\ \citenamefont {Massar}}]{duportAllopticalReservoirComputing2012}%
  \BibitemOpen
  \bibfield  {author} {\bibinfo {author} {\bibfnamefont {F.}~\bibnamefont {Duport}}, \bibinfo {author} {\bibfnamefont {B.}~\bibnamefont {Schneider}}, \bibinfo {author} {\bibfnamefont {A.}~\bibnamefont {Smerieri}}, \bibinfo {author} {\bibfnamefont {M.}~\bibnamefont {Haelterman}},\ and\ \bibinfo {author} {\bibfnamefont {S.}~\bibnamefont {Massar}},\ }\bibfield  {title} {\bibinfo {title} {All-optical reservoir computing},\ }\href {https://doi.org/10.1364/OE.20.022783} {\bibfield  {journal} {\bibinfo  {journal} {Optics Express}\ }\textbf {\bibinfo {volume} {20}},\ \bibinfo {pages} {22783} (\bibinfo {year} {2012})}\BibitemShut {NoStop}%
\bibitem [{\citenamefont {Mesaritakis}\ \emph {et~al.}(2013)\citenamefont {Mesaritakis}, \citenamefont {Papataxiarhis},\ and\ \citenamefont {Syvridis}}]{mesaritakisMicroRingResonators2013}%
  \BibitemOpen
  \bibfield  {author} {\bibinfo {author} {\bibfnamefont {C.}~\bibnamefont {Mesaritakis}}, \bibinfo {author} {\bibfnamefont {V.}~\bibnamefont {Papataxiarhis}},\ and\ \bibinfo {author} {\bibfnamefont {D.}~\bibnamefont {Syvridis}},\ }\bibfield  {title} {\bibinfo {title} {Micro ring resonators as building blocks for an all-optical high-speed reservoir-computing bit-pattern-recognition system},\ }\href {https://doi.org/10.1364/JOSAB.30.003048} {\bibfield  {journal} {\bibinfo  {journal} {JOSA B}\ }\textbf {\bibinfo {volume} {30}},\ \bibinfo {pages} {3048} (\bibinfo {year} {2013})}\BibitemShut {NoStop}%
\bibitem [{\citenamefont {Dejonckheere}\ \emph {et~al.}(2014)\citenamefont {Dejonckheere}, \citenamefont {Duport}, \citenamefont {Smerieri}, \citenamefont {Fang}, \citenamefont {Oudar}, \citenamefont {Haelterman},\ and\ \citenamefont {Massar}}]{dejonckheereAllopticalReservoirComputer2014}%
  \BibitemOpen
  \bibfield  {author} {\bibinfo {author} {\bibfnamefont {A.}~\bibnamefont {Dejonckheere}}, \bibinfo {author} {\bibfnamefont {F.}~\bibnamefont {Duport}}, \bibinfo {author} {\bibfnamefont {A.}~\bibnamefont {Smerieri}}, \bibinfo {author} {\bibfnamefont {L.}~\bibnamefont {Fang}}, \bibinfo {author} {\bibfnamefont {J.-L.}\ \bibnamefont {Oudar}}, \bibinfo {author} {\bibfnamefont {M.}~\bibnamefont {Haelterman}},\ and\ \bibinfo {author} {\bibfnamefont {S.}~\bibnamefont {Massar}},\ }\bibfield  {title} {\bibinfo {title} {All-optical reservoir computer based on saturation of absorption},\ }\href {https://doi.org/10.1364/OE.22.010868} {\bibfield  {journal} {\bibinfo  {journal} {Optics Express}\ }\textbf {\bibinfo {volume} {22}},\ \bibinfo {pages} {10868} (\bibinfo {year} {2014})}\BibitemShut {NoStop}%
\bibitem [{\citenamefont {Goudarzi}\ \emph {et~al.}(2013)\citenamefont {Goudarzi}, \citenamefont {Lakin},\ and\ \citenamefont {Stefanovic}}]{goudarziDNAReservoirComputing2013}%
  \BibitemOpen
  \bibfield  {author} {\bibinfo {author} {\bibfnamefont {A.}~\bibnamefont {Goudarzi}}, \bibinfo {author} {\bibfnamefont {M.~R.}\ \bibnamefont {Lakin}},\ and\ \bibinfo {author} {\bibfnamefont {D.}~\bibnamefont {Stefanovic}},\ }\bibfield  {title} {\bibinfo {title} {{{DNA Reservoir Computing}}: {{A Novel Molecular Computing Approach}}},\ }in\ \href {https://doi.org/10.1007/978-3-319-01928-4_6} {\emph {\bibinfo {booktitle} {{{DNA Computing}} and {{Molecular Programming}}}}},\ \bibinfo {editor} {edited by\ \bibinfo {editor} {\bibfnamefont {D.}~\bibnamefont {Soloveichik}}\ and\ \bibinfo {editor} {\bibfnamefont {B.}~\bibnamefont {Yurke}}}\ (\bibinfo  {publisher} {Springer International Publishing},\ \bibinfo {address} {Cham},\ \bibinfo {year} {2013})\ pp.\ \bibinfo {pages} {76--89}\BibitemShut {NoStop}%
\bibitem [{\citenamefont {Shor}(1997)}]{shor1999polynomial}%
  \BibitemOpen
  \bibfield  {author} {\bibinfo {author} {\bibfnamefont {P.~W.}\ \bibnamefont {Shor}},\ }\bibfield  {title} {\bibinfo {title} {Polynomial-time algorithms for prime factorization and discrete logarithms on a quantum computer},\ }\href {https://doi.org/10.1137/S0097539795293172} {\bibfield  {journal} {\bibinfo  {journal} {SIAM J. Comput.}\ }\textbf {\bibinfo {volume} {26}},\ \bibinfo {pages} {1484} (\bibinfo {year} {1997})}\BibitemShut {NoStop}%
\bibitem [{\citenamefont {Lloyd}(1996)}]{sethUniversal1996}%
  \BibitemOpen
  \bibfield  {author} {\bibinfo {author} {\bibfnamefont {S.}~\bibnamefont {Lloyd}},\ }\bibfield  {title} {\bibinfo {title} {Universal quantum simulators},\ }\href {https://doi.org/10.1126/science.273.5278.1073} {\bibfield  {journal} {\bibinfo  {journal} {Science}\ }\textbf {\bibinfo {volume} {273}},\ \bibinfo {pages} {1073} (\bibinfo {year} {1996})}\BibitemShut {NoStop}%
\bibitem [{\citenamefont {Ghosh}\ \emph {et~al.}(2019)\citenamefont {Ghosh}, \citenamefont {Opala}, \citenamefont {Matuszewski}, \citenamefont {Paterek},\ and\ \citenamefont {Liew}}]{ghosh2019quantum}%
  \BibitemOpen
  \bibfield  {author} {\bibinfo {author} {\bibfnamefont {S.}~\bibnamefont {Ghosh}}, \bibinfo {author} {\bibfnamefont {A.}~\bibnamefont {Opala}}, \bibinfo {author} {\bibfnamefont {M.}~\bibnamefont {Matuszewski}}, \bibinfo {author} {\bibfnamefont {T.}~\bibnamefont {Paterek}},\ and\ \bibinfo {author} {\bibfnamefont {T.~C.}\ \bibnamefont {Liew}},\ }\bibfield  {title} {\bibinfo {title} {Quantum reservoir processing},\ }\href {https://doi.org/https://doi.org/10.1038/s41534-019-0149-8} {\bibfield  {journal} {\bibinfo  {journal} {Npj Quantum Inf.}\ }\textbf {\bibinfo {volume} {5}},\ \bibinfo {pages} {35} (\bibinfo {year} {2019})}\BibitemShut {NoStop}%
\bibitem [{\citenamefont {Bravo}\ \emph {et~al.}(2022)\citenamefont {Bravo}, \citenamefont {Najafi}, \citenamefont {Gao},\ and\ \citenamefont {Yelin}}]{bravo2022quantum}%
  \BibitemOpen
  \bibfield  {author} {\bibinfo {author} {\bibfnamefont {R.~A.}\ \bibnamefont {Bravo}}, \bibinfo {author} {\bibfnamefont {K.}~\bibnamefont {Najafi}}, \bibinfo {author} {\bibfnamefont {X.}~\bibnamefont {Gao}},\ and\ \bibinfo {author} {\bibfnamefont {S.~F.}\ \bibnamefont {Yelin}},\ }\bibfield  {title} {\bibinfo {title} {Quantum reservoir computing using arrays of rydberg atoms},\ }\href {https://doi.org/10.1103/PRXQuantum.3.030325} {\bibfield  {journal} {\bibinfo  {journal} {PRX Quantum}\ }\textbf {\bibinfo {volume} {3}},\ \bibinfo {pages} {030325} (\bibinfo {year} {2022})}\BibitemShut {NoStop}%
\bibitem [{\citenamefont {Negoro}\ \emph {et~al.}(2018)\citenamefont {Negoro}, \citenamefont {Mitarai}, \citenamefont {Fujii}, \citenamefont {Nakajima},\ and\ \citenamefont {Kitagawa}}]{negoro2018machine}%
  \BibitemOpen
  \bibfield  {author} {\bibinfo {author} {\bibfnamefont {M.}~\bibnamefont {Negoro}}, \bibinfo {author} {\bibfnamefont {K.}~\bibnamefont {Mitarai}}, \bibinfo {author} {\bibfnamefont {K.}~\bibnamefont {Fujii}}, \bibinfo {author} {\bibfnamefont {K.}~\bibnamefont {Nakajima}},\ and\ \bibinfo {author} {\bibfnamefont {M.}~\bibnamefont {Kitagawa}},\ }\bibfield  {title} {\bibinfo {title} {Machine learning with controllable quantum dynamics of a nuclear spin ensemble in a solid},\ }\href {https://api.semanticscholar.org/CorpusID:51807422} {\bibfield  {journal} {\bibinfo  {journal} {arXiv: Quantum Physics}\ } (\bibinfo {year} {2018})}\BibitemShut {NoStop}%
\bibitem [{\citenamefont {Kubota}\ \emph {et~al.}(2023)\citenamefont {Kubota}, \citenamefont {Suzuki}, \citenamefont {Kobayashi}, \citenamefont {Tran}, \citenamefont {Yamamoto},\ and\ \citenamefont {Nakajima}}]{chen2020temporal}%
  \BibitemOpen
  \bibfield  {author} {\bibinfo {author} {\bibfnamefont {T.}~\bibnamefont {Kubota}}, \bibinfo {author} {\bibfnamefont {Y.}~\bibnamefont {Suzuki}}, \bibinfo {author} {\bibfnamefont {S.}~\bibnamefont {Kobayashi}}, \bibinfo {author} {\bibfnamefont {Q.~H.}\ \bibnamefont {Tran}}, \bibinfo {author} {\bibfnamefont {N.}~\bibnamefont {Yamamoto}},\ and\ \bibinfo {author} {\bibfnamefont {K.}~\bibnamefont {Nakajima}},\ }\bibfield  {title} {\bibinfo {title} {Temporal information processing induced by quantum noise},\ }\href {https://doi.org/10.1103/PhysRevResearch.5.023057} {\bibfield  {journal} {\bibinfo  {journal} {Phys. Rev. Res.}\ }\textbf {\bibinfo {volume} {5}},\ \bibinfo {pages} {023057} (\bibinfo {year} {2023})}\BibitemShut {NoStop}%
\bibitem [{\citenamefont {Mujal}\ \emph {et~al.}(2021)\citenamefont {Mujal}, \citenamefont {Mart{\'\i}nez-Pe{\~n}a}, \citenamefont {Nokkala}, \citenamefont {Garc{\'\i}a-Beni}, \citenamefont {Giorgi}, \citenamefont {Soriano},\ and\ \citenamefont {Zambrini}}]{mujal2021opportunities}%
  \BibitemOpen
  \bibfield  {author} {\bibinfo {author} {\bibfnamefont {P.}~\bibnamefont {Mujal}}, \bibinfo {author} {\bibfnamefont {R.}~\bibnamefont {Mart{\'\i}nez-Pe{\~n}a}}, \bibinfo {author} {\bibfnamefont {J.}~\bibnamefont {Nokkala}}, \bibinfo {author} {\bibfnamefont {J.}~\bibnamefont {Garc{\'\i}a-Beni}}, \bibinfo {author} {\bibfnamefont {G.~L.}\ \bibnamefont {Giorgi}}, \bibinfo {author} {\bibfnamefont {M.~C.}\ \bibnamefont {Soriano}},\ and\ \bibinfo {author} {\bibfnamefont {R.}~\bibnamefont {Zambrini}},\ }\bibfield  {title} {\bibinfo {title} {Opportunities in quantum reservoir computing and extreme learning machines},\ }\href {https://doi.org/https://doi.org/10.1002/qute.202100027} {\bibfield  {journal} {\bibinfo  {journal} {Advanced Quantum Technologies}\ }\textbf {\bibinfo {volume} {4}},\ \bibinfo {pages} {2100027} (\bibinfo {year} {2021})}\BibitemShut {NoStop}%
\bibitem [{\citenamefont {Xiong}\ \emph {et~al.}(2025{\natexlab{a}})\citenamefont {Xiong}, \citenamefont {Facelli}, \citenamefont {Sahebi}, \citenamefont {Agnel}, \citenamefont {Chotibut}, \citenamefont {Thanasilp},\ and\ \citenamefont {Holmes}}]{xiongFundamentalAspectsQuantum2025}%
  \BibitemOpen
  \bibfield  {author} {\bibinfo {author} {\bibfnamefont {W.}~\bibnamefont {Xiong}}, \bibinfo {author} {\bibfnamefont {G.}~\bibnamefont {Facelli}}, \bibinfo {author} {\bibfnamefont {M.}~\bibnamefont {Sahebi}}, \bibinfo {author} {\bibfnamefont {O.}~\bibnamefont {Agnel}}, \bibinfo {author} {\bibfnamefont {T.}~\bibnamefont {Chotibut}}, \bibinfo {author} {\bibfnamefont {S.}~\bibnamefont {Thanasilp}},\ and\ \bibinfo {author} {\bibfnamefont {Z.}~\bibnamefont {Holmes}},\ }\bibfield  {title} {\bibinfo {title} {On fundamental aspects of quantum extreme learning machines},\ }\href {https://doi.org/10.1007/s42484-025-00239-7} {\bibfield  {journal} {\bibinfo  {journal} {Quantum Mach. Intell.}\ }\textbf {\bibinfo {volume} {7}},\ \bibinfo {pages} {20} (\bibinfo {year} {2025}{\natexlab{a}})}\BibitemShut {NoStop}%
\bibitem [{Yfi()}]{Yfinance}%
  \BibitemOpen
  \href@noop {} {\bibinfo {title} {Ranaroussi/yfinance: Download market data from yahoo! {{Finance}}'s {{API}}}},\ \bibinfo {howpublished} {https://github.com/ranaroussi/yfinance}\BibitemShut {NoStop}%
\bibitem [{\citenamefont {Schwert}(1989)}]{schwert1989}%
  \BibitemOpen
  \bibfield  {author} {\bibinfo {author} {\bibfnamefont {G.~W.}\ \bibnamefont {Schwert}},\ }\bibfield  {title} {\bibinfo {title} {Why does stock market volatility change over time?},\ }\href {https://doi.org/https://doi.org/10.1111/j.1540-6261.1989.tb02647.x} {\bibfield  {journal} {\bibinfo  {journal} {J. Finance.}\ }\textbf {\bibinfo {volume} {44}},\ \bibinfo {pages} {1115} (\bibinfo {year} {1989})}\BibitemShut {NoStop}%
\bibitem [{\citenamefont {Engle}\ \emph {et~al.}(2008)\citenamefont {Engle}, \citenamefont {Ghysels},\ and\ \citenamefont {Sohn}}]{engle2009economic}%
  \BibitemOpen
  \bibfield  {author} {\bibinfo {author} {\bibfnamefont {R.~F.}\ \bibnamefont {Engle}}, \bibinfo {author} {\bibfnamefont {E.}~\bibnamefont {Ghysels}},\ and\ \bibinfo {author} {\bibfnamefont {B.}~\bibnamefont {Sohn}},\ }\bibfield  {title} {\bibinfo {title} {On the {{Economic Sources}} of {{Stock Market Volatility}}}\ }\href {https://doi.org/10.2139/ssrn.971310} {10.2139/ssrn.971310} (\bibinfo {year} {2008})\BibitemShut {NoStop}%
\bibitem [{\citenamefont {Welch}\ and\ \citenamefont {Goyal}(2008)}]{welch2008comprehensive}%
  \BibitemOpen
  \bibfield  {author} {\bibinfo {author} {\bibfnamefont {I.}~\bibnamefont {Welch}}\ and\ \bibinfo {author} {\bibfnamefont {A.}~\bibnamefont {Goyal}},\ }\bibfield  {title} {\bibinfo {title} {A comprehensive look at the empirical performance of equity premium prediction},\ }\href {https://doi.org/https://doi.org/10.1093/rfs/hhm014} {\bibfield  {journal} {\bibinfo  {journal} {Rev. Financ. Stud.}\ }\textbf {\bibinfo {volume} {21}},\ \bibinfo {pages} {1455} (\bibinfo {year} {2008})}\BibitemShut {NoStop}%
\bibitem [{\citenamefont {Fama}\ and\ \citenamefont {French}(1993)}]{fama1993common}%
  \BibitemOpen
  \bibfield  {author} {\bibinfo {author} {\bibfnamefont {E.~F.}\ \bibnamefont {Fama}}\ and\ \bibinfo {author} {\bibfnamefont {K.~R.}\ \bibnamefont {French}},\ }\bibfield  {title} {\bibinfo {title} {Common risk factors in the returns on stocks and bonds},\ }\href {https://doi.org/https://doi.org/10.1016/0304-405X(93)90023-5} {\bibfield  {journal} {\bibinfo  {journal} {J. financ. econ.}\ }\textbf {\bibinfo {volume} {33}},\ \bibinfo {pages} {3} (\bibinfo {year} {1993})}\BibitemShut {NoStop}%
\bibitem [{\citenamefont {Fama}\ and\ \citenamefont {French}(1996)}]{fama1996multifactor}%
  \BibitemOpen
  \bibfield  {author} {\bibinfo {author} {\bibfnamefont {E.~F.}\ \bibnamefont {Fama}}\ and\ \bibinfo {author} {\bibfnamefont {K.~R.}\ \bibnamefont {French}},\ }\bibfield  {title} {\bibinfo {title} {Multifactor explanations of asset pricing anomalies},\ }\href {https://doi.org/https://doi.org/10.1111/j.1540-6261.1996.tb05202.x} {\bibfield  {journal} {\bibinfo  {journal} {J. Finance.}\ }\textbf {\bibinfo {volume} {51}},\ \bibinfo {pages} {55} (\bibinfo {year} {1996})}\BibitemShut {NoStop}%
\bibitem [{\citenamefont {Feng}\ \emph {et~al.}(2024)\citenamefont {Feng}, \citenamefont {Zhang},\ and\ \citenamefont {Wang}}]{Feng2024}%
  \BibitemOpen
  \bibfield  {author} {\bibinfo {author} {\bibfnamefont {X.}~\bibnamefont {Feng}}, \bibinfo {author} {\bibfnamefont {H.}~\bibnamefont {Zhang}},\ and\ \bibinfo {author} {\bibfnamefont {C.}~\bibnamefont {Wang}},\ }\bibfield  {title} {\bibinfo {title} {Out-of-sample volatility prediction: Rolling window, expanding window, or both?},\ }\href@noop {} {\bibfield  {journal} {\bibinfo  {journal} {Journal of Forecasting}\ } (\bibinfo {year} {2024})}\BibitemShut {NoStop}%
\bibitem [{\citenamefont {Patton}\ and\ \citenamefont {Sheppard}(2015{\natexlab{b}})}]{PattonSheppard2015}%
  \BibitemOpen
  \bibfield  {author} {\bibinfo {author} {\bibfnamefont {A.~J.}\ \bibnamefont {Patton}}\ and\ \bibinfo {author} {\bibfnamefont {K.}~\bibnamefont {Sheppard}},\ }\bibfield  {title} {\bibinfo {title} {Good volatility, bad volatility: Signed jumps and the persistence of volatility},\ }\href@noop {} {\bibfield  {journal} {\bibinfo  {journal} {J. Bus. Econ. Stat.}\ }\textbf {\bibinfo {volume} {33}},\ \bibinfo {pages} {631} (\bibinfo {year} {2015}{\natexlab{b}})}\BibitemShut {NoStop}%
\bibitem [{\citenamefont {Pesaran}\ and\ \citenamefont {Timmermann}(2007)}]{PesaranTimmermann2007}%
  \BibitemOpen
  \bibfield  {author} {\bibinfo {author} {\bibfnamefont {M.~H.}\ \bibnamefont {Pesaran}}\ and\ \bibinfo {author} {\bibfnamefont {A.}~\bibnamefont {Timmermann}},\ }\bibfield  {title} {\bibinfo {title} {Selection of estimation window in the presence of breaks},\ }\href@noop {} {\bibfield  {journal} {\bibinfo  {journal} {J. Econom.}\ }\textbf {\bibinfo {volume} {137}},\ \bibinfo {pages} {134} (\bibinfo {year} {2007})}\BibitemShut {NoStop}%
\bibitem [{\citenamefont {Shapley}(1953)}]{shapley1953value}%
  \BibitemOpen
  \bibfield  {author} {\bibinfo {author} {\bibfnamefont {L.~S.}\ \bibnamefont {Shapley}},\ }\bibfield  {title} {\bibinfo {title} {A value for n-person games},\ }\href@noop {} {\bibfield  {journal} {\bibinfo  {journal} {Contribution to the Theory of Games}\ }\textbf {\bibinfo {volume} {2}} (\bibinfo {year} {1953})}\BibitemShut {NoStop}%
\bibitem [{\citenamefont {Lundberg}\ and\ \citenamefont {Lee}(2017)}]{lundberg2017unified}%
  \BibitemOpen
  \bibfield  {author} {\bibinfo {author} {\bibfnamefont {S.~M.}\ \bibnamefont {Lundberg}}\ and\ \bibinfo {author} {\bibfnamefont {S.-I.}\ \bibnamefont {Lee}},\ }\bibfield  {title} {\bibinfo {title} {A unified approach to interpreting model predictions},\ }in\ \href {https://doi.org/dl.acm.org/doi/10.5555/3295222.3295230} {\emph {\bibinfo {booktitle} {Proceedings of the 31st International Conference on Neural Information Processing Systems}}},\ \bibinfo {series and number} {NIPS'17}\ (\bibinfo  {publisher} {Curran Associates Inc.},\ \bibinfo {address} {Red Hook, NY, USA},\ \bibinfo {year} {2017})\ p.\ \bibinfo {pages} {4768–4777}\BibitemShut {NoStop}%
\bibitem [{\citenamefont {{\v S}trumbelj}\ and\ \citenamefont {Kononenko}(2014)}]{strumbeljExplainingPredictionModels2014}%
  \BibitemOpen
  \bibfield  {author} {\bibinfo {author} {\bibfnamefont {E.}~\bibnamefont {{\v S}trumbelj}}\ and\ \bibinfo {author} {\bibfnamefont {I.}~\bibnamefont {Kononenko}},\ }\bibfield  {title} {\bibinfo {title} {Explaining prediction models and individual predictions with feature contributions},\ }\href {https://doi.org/10.1007/s10115-013-0679-x} {\bibfield  {journal} {\bibinfo  {journal} {Knowledge and Information Systems}\ }\textbf {\bibinfo {volume} {41}},\ \bibinfo {pages} {647} (\bibinfo {year} {2014})}\BibitemShut {NoStop}%
\bibitem [{\citenamefont {Hansen}\ and\ \citenamefont {Lunde}(2005)}]{HansenLunde2005}%
  \BibitemOpen
  \bibfield  {author} {\bibinfo {author} {\bibfnamefont {P.~R.}\ \bibnamefont {Hansen}}\ and\ \bibinfo {author} {\bibfnamefont {A.}~\bibnamefont {Lunde}},\ }\bibfield  {title} {\bibinfo {title} {A forecast comparison of volatility models: Does anything beat a {GARCH}(1,1)?},\ }\href {https://doi.org/10.1002/jae.800} {\bibfield  {journal} {\bibinfo  {journal} {Journal of Applied Econometrics}\ }\textbf {\bibinfo {volume} {20}},\ \bibinfo {pages} {873} (\bibinfo {year} {2005})}\BibitemShut {NoStop}%
\bibitem [{\citenamefont {Patton}(2011)}]{Patton2011}%
  \BibitemOpen
  \bibfield  {author} {\bibinfo {author} {\bibfnamefont {A.~J.}\ \bibnamefont {Patton}},\ }\bibfield  {title} {\bibinfo {title} {Volatility forecast comparison using imperfect volatility proxies},\ }\href {https://doi.org/10.1016/j.jeconom.2010.03.034} {\bibfield  {journal} {\bibinfo  {journal} {J. Econom.}\ }\textbf {\bibinfo {volume} {160}},\ \bibinfo {pages} {246} (\bibinfo {year} {2011})}\BibitemShut {NoStop}%
\bibitem [{\citenamefont {Hansen}\ \emph {et~al.}(2011)\citenamefont {Hansen}, \citenamefont {Lunde},\ and\ \citenamefont {Nason}}]{HansenLundeNason2011}%
  \BibitemOpen
  \bibfield  {author} {\bibinfo {author} {\bibfnamefont {P.~R.}\ \bibnamefont {Hansen}}, \bibinfo {author} {\bibfnamefont {A.}~\bibnamefont {Lunde}},\ and\ \bibinfo {author} {\bibfnamefont {J.~M.}\ \bibnamefont {Nason}},\ }\bibfield  {title} {\bibinfo {title} {The model confidence set},\ }\href {https://doi.org/10.3982/ECTA5771} {\bibfield  {journal} {\bibinfo  {journal} {Econometrica}\ }\textbf {\bibinfo {volume} {79}},\ \bibinfo {pages} {453} (\bibinfo {year} {2011})}\BibitemShut {NoStop}%
\bibitem [{\citenamefont {Diebold}\ and\ \citenamefont {Mariano}(1995)}]{DieboldMariano1995}%
  \BibitemOpen
  \bibfield  {author} {\bibinfo {author} {\bibfnamefont {F.~X.}\ \bibnamefont {Diebold}}\ and\ \bibinfo {author} {\bibfnamefont {R.~S.}\ \bibnamefont {Mariano}},\ }\bibfield  {title} {\bibinfo {title} {Comparing predictive accuracy},\ }\href {https://doi.org/10.1080/07350015.1995.10524599} {\bibfield  {journal} {\bibinfo  {journal} {J. Bus. Econ. Stat.}\ }\textbf {\bibinfo {volume} {13}},\ \bibinfo {pages} {253} (\bibinfo {year} {1995})}\BibitemShut {NoStop}%
\bibitem [{\citenamefont {Erhard}\ \emph {et~al.}(2021)\citenamefont {Erhard}, \citenamefont {Poulsen~Nautrup}, \citenamefont {Meth}, \citenamefont {Postler}, \citenamefont {Stricker}, \citenamefont {Stadler}, \citenamefont {Negnevitsky}, \citenamefont {Ringbauer}, \citenamefont {Schindler}, \citenamefont {Briegel} \emph {et~al.}}]{erhard2021entangling}%
  \BibitemOpen
  \bibfield  {author} {\bibinfo {author} {\bibfnamefont {A.}~\bibnamefont {Erhard}}, \bibinfo {author} {\bibfnamefont {H.}~\bibnamefont {Poulsen~Nautrup}}, \bibinfo {author} {\bibfnamefont {M.}~\bibnamefont {Meth}}, \bibinfo {author} {\bibfnamefont {L.}~\bibnamefont {Postler}}, \bibinfo {author} {\bibfnamefont {R.}~\bibnamefont {Stricker}}, \bibinfo {author} {\bibfnamefont {M.}~\bibnamefont {Stadler}}, \bibinfo {author} {\bibfnamefont {V.}~\bibnamefont {Negnevitsky}}, \bibinfo {author} {\bibfnamefont {M.}~\bibnamefont {Ringbauer}}, \bibinfo {author} {\bibfnamefont {P.}~\bibnamefont {Schindler}}, \bibinfo {author} {\bibfnamefont {H.~J.}\ \bibnamefont {Briegel}}, \emph {et~al.},\ }\bibfield  {title} {\bibinfo {title} {Entangling logical qubits with lattice surgery},\ }\href {https://doi.org/https://doi.org/10.1038/s41586-020-03079-6} {\bibfield  {journal} {\bibinfo  {journal} {Nature}\ }\textbf {\bibinfo {volume} {589}},\ \bibinfo {pages} {220} (\bibinfo {year} {2021})}\BibitemShut {NoStop}%
\bibitem [{\citenamefont {Akhtar}\ \emph {et~al.}(2023)\citenamefont {Akhtar}, \citenamefont {Bonus}, \citenamefont {Lebrun-Gallagher}, \citenamefont {Johnson}, \citenamefont {Siegele-Brown}, \citenamefont {Hong}, \citenamefont {Hile}, \citenamefont {Kulmiya},\ and\ \citenamefont {Hensinger}}]{akhtar2023high}%
  \BibitemOpen
  \bibfield  {author} {\bibinfo {author} {\bibfnamefont {M.}~\bibnamefont {Akhtar}}, \bibinfo {author} {\bibfnamefont {F.}~\bibnamefont {Bonus}}, \bibinfo {author} {\bibfnamefont {F.}~\bibnamefont {Lebrun-Gallagher}}, \bibinfo {author} {\bibfnamefont {N.}~\bibnamefont {Johnson}}, \bibinfo {author} {\bibfnamefont {M.}~\bibnamefont {Siegele-Brown}}, \bibinfo {author} {\bibfnamefont {S.}~\bibnamefont {Hong}}, \bibinfo {author} {\bibfnamefont {S.}~\bibnamefont {Hile}}, \bibinfo {author} {\bibfnamefont {S.}~\bibnamefont {Kulmiya}},\ and\ \bibinfo {author} {\bibfnamefont {W.}~\bibnamefont {Hensinger}},\ }\bibfield  {title} {\bibinfo {title} {A high-fidelity quantum matter-link between ion-trap microchip modules},\ }\href {https://doi.org/https://doi.org/10.1038/s41467-022-35285-3} {\bibfield  {journal} {\bibinfo  {journal} {Nat. Commun.}\ }\textbf {\bibinfo {volume} {14}},\ \bibinfo {pages} {531} (\bibinfo {year} {2023})}\BibitemShut {NoStop}%
\bibitem [{\citenamefont {Xiong}\ \emph {et~al.}(2025{\natexlab{b}})\citenamefont {Xiong}, \citenamefont {Facelli}, \citenamefont {Sahebi}, \citenamefont {Agnel}, \citenamefont {Chotibut}, \citenamefont {Thanasilp},\ and\ \citenamefont {Holmes}}]{Xiong2025on}%
  \BibitemOpen
  \bibfield  {author} {\bibinfo {author} {\bibfnamefont {W.}~\bibnamefont {Xiong}}, \bibinfo {author} {\bibfnamefont {G.}~\bibnamefont {Facelli}}, \bibinfo {author} {\bibfnamefont {M.}~\bibnamefont {Sahebi}}, \bibinfo {author} {\bibfnamefont {O.}~\bibnamefont {Agnel}}, \bibinfo {author} {\bibfnamefont {T.}~\bibnamefont {Chotibut}}, \bibinfo {author} {\bibfnamefont {S.}~\bibnamefont {Thanasilp}},\ and\ \bibinfo {author} {\bibfnamefont {Z.}~\bibnamefont {Holmes}},\ }\bibfield  {title} {\bibinfo {title} {On fundamental aspects of quantum extreme learning machines},\ }\bibfield  {journal} {\bibinfo  {journal} {Quantum Mach. Intell.}\ }\textbf {\bibinfo {volume} {7}},\ \href {https://doi.org/10.1007/s42484-025-00239-7} {10.1007/s42484-025-00239-7} (\bibinfo {year} {2025}{\natexlab{b}})\BibitemShut {NoStop}%
\bibitem [{\citenamefont {Dirac}(2010)}]{dirac1981principles}%
  \BibitemOpen
  \bibfield  {author} {\bibinfo {author} {\bibfnamefont {P.~A.~M.}\ \bibnamefont {Dirac}},\ }\href@noop {} {\emph {\bibinfo {title} {The Principles of Quantum Mechanics}}},\ \bibinfo {edition} {4th}\ ed.,\ \bibinfo {series} {International Series of Monographs on Physics}\ No.~\bibinfo {number} {27}\ (\bibinfo  {publisher} {Clarendon Press, Oxford University Press},\ \bibinfo {address} {Oxford},\ \bibinfo {year} {2010})\BibitemShut {NoStop}%
\bibitem [{\citenamefont {Nielsen}\ and\ \citenamefont {Chuang}(2012)}]{nielsen2010quantum}%
  \BibitemOpen
  \bibfield  {author} {\bibinfo {author} {\bibfnamefont {M.~A.}\ \bibnamefont {Nielsen}}\ and\ \bibinfo {author} {\bibfnamefont {I.~L.}\ \bibnamefont {Chuang}},\ }\href@noop {} {\emph {\bibinfo {title} {Quantum Computation and Quantum Information}}},\ \bibinfo {edition} {10th}\ ed.\ (\bibinfo  {publisher} {Cambridge University Press},\ \bibinfo {address} {Cambridge},\ \bibinfo {year} {2012})\BibitemShut {NoStop}%
\bibitem [{\citenamefont {Abadir}\ and\ \citenamefont {Magnus}(2002)}]{abadir2002notation}%
  \BibitemOpen
  \bibfield  {author} {\bibinfo {author} {\bibfnamefont {K.}~\bibnamefont {Abadir}}\ and\ \bibinfo {author} {\bibfnamefont {J.}~\bibnamefont {Magnus}},\ }\bibfield  {title} {\bibinfo {title} {Notation in econometrics: a proposal for a standard},\ }\href {https://doi.org/https://doi.org/10.1111/1368-423X.t01-1-00074} {\bibfield  {journal} {\bibinfo  {journal} {The Econometrics Journal}\ }\textbf {\bibinfo {volume} {5}},\ \bibinfo {pages} {76} (\bibinfo {year} {2002})}\BibitemShut {NoStop}%
\bibitem [{\citenamefont {Grover}(1996)}]{grover1996fast}%
  \BibitemOpen
  \bibfield  {author} {\bibinfo {author} {\bibfnamefont {L.~K.}\ \bibnamefont {Grover}},\ }\bibfield  {title} {\bibinfo {title} {A fast quantum mechanical algorithm for database search},\ }in\ \href {https://doi.org/10.1145/237814.237866} {\emph {\bibinfo {booktitle} {Proceedings of the Twenty-Eighth Annual ACM Symposium on Theory of Computing}}},\ \bibinfo {series and number} {STOC '96}\ (\bibinfo  {publisher} {Association for Computing Machinery},\ \bibinfo {address} {New York, NY, USA},\ \bibinfo {year} {1996})\ p.\ \bibinfo {pages} {212–219}\BibitemShut {NoStop}%
\bibitem [{\citenamefont {Acharya}\ \emph {et~al.}(2025)\citenamefont {Acharya}, \citenamefont {Abanin}, \citenamefont {{Aghababaie-Beni}}, \citenamefont {Aleiner}, \citenamefont {Andersen}, \citenamefont {Ansmann}, \citenamefont {Arute}, \citenamefont {Arya}, \citenamefont {Asfaw}, \citenamefont {Astrakhantsev}, \citenamefont {Atalaya}, \citenamefont {Babbush}, \citenamefont {Bacon}, \citenamefont {Ballard}, \citenamefont {Bardin}, \citenamefont {Bausch}, \citenamefont {Bengtsson},\ and\ \citenamefont {other}}]{acharya2024quantum}%
  \BibitemOpen
  \bibfield  {author} {\bibinfo {author} {\bibfnamefont {R.}~\bibnamefont {Acharya}}, \bibinfo {author} {\bibfnamefont {D.~A.}\ \bibnamefont {Abanin}}, \bibinfo {author} {\bibfnamefont {L.}~\bibnamefont {{Aghababaie-Beni}}}, \bibinfo {author} {\bibfnamefont {I.}~\bibnamefont {Aleiner}}, \bibinfo {author} {\bibfnamefont {T.~I.}\ \bibnamefont {Andersen}}, \bibinfo {author} {\bibfnamefont {M.}~\bibnamefont {Ansmann}}, \bibinfo {author} {\bibfnamefont {F.}~\bibnamefont {Arute}}, \bibinfo {author} {\bibfnamefont {K.}~\bibnamefont {Arya}}, \bibinfo {author} {\bibfnamefont {A.}~\bibnamefont {Asfaw}}, \bibinfo {author} {\bibfnamefont {N.}~\bibnamefont {Astrakhantsev}}, \bibinfo {author} {\bibfnamefont {J.}~\bibnamefont {Atalaya}}, \bibinfo {author} {\bibfnamefont {R.}~\bibnamefont {Babbush}}, \bibinfo {author} {\bibfnamefont {D.}~\bibnamefont {Bacon}}, \bibinfo {author} {\bibfnamefont {B.}~\bibnamefont {Ballard}}, \bibinfo {author} {\bibfnamefont {J.~C.}\ \bibnamefont {Bardin}}, \bibinfo {author} {\bibfnamefont
  {J.}~\bibnamefont {Bausch}}, \bibinfo {author} {\bibfnamefont {A.}~\bibnamefont {Bengtsson}},\ and\ \bibinfo {author} {\bibnamefont {other}},\ }\bibfield  {title} {\bibinfo {title} {Quantum error correction below the surface code threshold},\ }\href {https://doi.org/10.1038/s41586-024-08449-y} {\bibfield  {journal} {\bibinfo  {journal} {Nature}\ }\textbf {\bibinfo {volume} {638}},\ \bibinfo {pages} {920} (\bibinfo {year} {2025})}\BibitemShut {NoStop}%
\end{thebibliography}%

\appendix

\section{Quantum Computation Preliminaries}
\label{sec: quantum computing}

Let us briefly provide an overview of the quantum theory necessary for understanding the subsequent sections of this paper. Quantum mechanics offers a mathematical framework originally developed to explain several experimental phenomena that challenged classical Newtonian physics in the late 19th and early 20th centuries. Readers interested in a more comprehensive exploration are encouraged to consult standard references such as \citet{dirac1981principles} and \citet{nielsen2010quantum}. Throughout this paper, we employ the Dirac notation commonly used in quantum physics. To assist readers from financial and econometric backgrounds, we present a minidictionary in Table~\ref{table:notation_table_adjusted}, where the departures of the Dirac notation from the standard econometric notation as laid out in \cite{abadir2002notation} are codified. In modern axiomatic form, quantum mechanics is completely described through the following four postulates.

\begin{table*}[h!]
\caption{\textbf{A mini-dictionary of Dirac notation for econometricians.}}
\centering
\resizebox{1\textwidth}{!}{
\begin{tabular}{l c c l}
\hline
\textbf{Object} & \textbf{Econometric Notation} & \textbf{Dirac Notation} & \textbf{Comments} \\
\hline
\textbf{Scalar Variable} & $a$ & $a$ &  \\
\textbf{Complex Conjugate of Scalar} & $\bar{a}$ or $a^*$ & $a^*$ &  \\
\textbf{Vector (Column)} & $\textbf{\textit{a}} = \begin{bmatrix} a_1 \\ a_2 \\ \vdots \\ a_n \end{bmatrix}$ & $|a\rangle$ &  \\
\textbf{Dual Vector (Row)} & $\textbf{\textit{a}}^\top = \begin{bmatrix} a_1^* & a_2^* & \dots & a_n^* \end{bmatrix}$ & $\langle a|$ & $\langle a| = (|a\rangle)^\dagger$ \\
\textbf{Complex Conjugate of Vector} & $\bar{\textbf{\textit{a}}}$ & $|a\rangle^*$ & Element-wise complex conjugate \\
\textbf{Adjoint (Conjugate Transpose) of Vector} & $\textbf{\textit{a}}^\dagger = (\textbf{\textit{a}}^*)^\top$ & $\langle a|$ &  \\
\textbf{Norm of a Vector} & $\|\textbf{\textit{a}}\| = \sqrt{\textbf{\textit{a}}^\dagger \textbf{\textit{a}}}$ & $\|a\| = \sqrt{\langle a | a \rangle}$ &  \\
\textbf{Inner Product} & $\textbf{\textit{a}}' \textbf{\textit{b}}$ & $\langle a|b \rangle$ & Results in a scalar \\
\textbf{Outer Product} & $\textbf{\textit{a}} \textbf{\textit{b}}'$ & $|a\rangle \langle b|$ & Results in a matrix \\
\textbf{Composite Vector} & $\textbf{\textit{a}} \otimes \textbf{\textit{b}}$ & $|a\rangle \otimes |b\rangle \equiv |ab\rangle$ & Produces a higher-dimensional vector \\
\textbf{Composite Vector*} & $\textbf{\textit{a}}^{\otimes n}=\textbf{\textit{a}}\otimes \textbf{\textit{a}}\otimes\cdots\otimes \textbf{\textit{a}}$ & $|a\rangle ^{\otimes n} \equiv \underbrace{|a\rangle\otimes |a\rangle\otimes\cdots\otimes|a\rangle}_n$ & Tensor product with $n$ $|a\rangle$ \\
\textbf{Transpose of Matrix $\textbf{\textit{A}}$} & $\textbf{\textit{A}}^\top$ & $\mathbf{A}^\top$ & Reflect over main diagonal \\
\textbf{Complex Conjugate of Matrix $\textbf{\textit{A}}$} & $\bar{\textbf{\textit{A}}}$ & $\mathbf{A}^*$ & Element-wise complex conjugate \\
\textbf{Conjugate Transpose of Matrix $\textbf{\textit{A}}$} & $\textbf{\textit{A}}^\dagger = (\textbf{\textit{A}}^*)^\top$ & $\mathbf{A}^\dagger$ &  \\
\textbf{Adjugate of Matrix $\textbf{\textit{A}}$} & $\textbf{\textit{A}}^\#$ & $\text{adj}(\mathbf{A})$ &  \\
\textbf{Determinant of Matrix $\textbf{\textit{A}}$} & $\det(\textbf{\textit{A}})$ & $\det(\mathbf{A})$ & Scalar value \\
\textbf{Trace of Matrix $\textbf{\textit{A}}$} & $\text{tr}(\textbf{\textit{A}})$ & $\operatorname{Tr}(\mathbf{A})$ & Sum of diagonal elements \\
\textbf{Identity Matrix} & $\textbf{\textit{I}}_n$ & $\mathbb{I}$ &  \\
\textbf{Zero Vector/Matrix} & $\textbf{\textit{0}}$ & $\mathbf{0}$ &  \\
\textbf{Expectation Value} & $\mathbb{E}[X]$ & $\langle X \rangle = \langle \psi | X | \psi \rangle$ & $|\psi\rangle$ is the state vector \\
\textbf{Variance} & $\text{Var}(X) = \mathbb{E}[(X - \mathbb{E}[X])^2]$ & $\langle (X - \langle X \rangle)^2 \rangle$ & Measure of spread \\
\textbf{Covariance} & $\text{Cov}(X,Y) = \mathbb{E}[XY] - \mathbb{E}[X] \mathbb{E}[Y]$ & $\langle XY \rangle - \langle X \rangle \langle Y \rangle$ &  \\
\textbf{Hermitian Operator} & $\textbf{\textit{A}} = \textbf{\textit{A}}^\dagger$ & $A = A^\dagger$ & Self-adjoint operator \\
\textbf{Unitary Operator} & $\textbf{\textit{U}} \textbf{\textit{U}}^\dagger = \textbf{\textit{I}}_n$ & $U U^\dagger = \mathbb{I}$ & Preserves norms \\
\textbf{Projection Operator} & $\mathbf{P} = \textbf{\textit{a}} \textbf{\textit{a}}'$ & $P = |a\rangle \langle a|$ & $P^2 = P$ \\
\textbf{Commutator of Operators $A$ and $B$} & $[A,B] = AB - BA$ & $[A, B] = AB - BA$ & Measures non-commutativity \\
\textbf{Anticommutator of Operators $A$ and $B$} & $\{A, B\} = AB + BA$ & $\{A, B\} = AB + BA$ &  \\
\textbf{Kronecker Delta} & $\delta_{ij}$ & $\delta_{ij}$ & $\delta_{ij} = \begin{cases} 1 & \text{if } i = j \\ 0 & \text{if } i \neq j \end{cases}$ \\
\textbf{Dirac Delta Function} & $\delta(x - x')$ & $\delta(x - x')$ & Generalized function \\
\textbf{Exponential of Operator} & $\exp(\textbf{\textit{A}})$ & $e^{A}$ & Defined via power series \\
\textbf{Fourier Transform of Function $f(x)$} & $\mathcal{F}\{f(x)\} = \int_{-\infty}^\infty e^{-i\omega x} f(x) \, dx$ & Same as econometric notation &  \\
\textbf{Tensor Product of Matrices $\textbf{\textit{A}}$ and $\textbf{\textit{B}}$} & $\textbf{\textit{A}} \otimes \textbf{\textit{B}}$ & $A \otimes B$ &  \\
\textbf{Spectral Decomposition} & $\textbf{\textit{A}} = \textbf{\textit{Q}} \mathbf{\Lambda} \textbf{\textit{Q}}^\dagger$ & $A = \sum_i \lambda_i |i\rangle \langle i|$ & $\lambda_i$ are eigenvalues \\
\textbf{Eigenvalue Equation} & $\textbf{\textit{A}} \textbf{\textit{v}} = \lambda \textbf{\textit{v}}$ & $A |v\rangle = \lambda |v\rangle$ &  \\
\textbf{Density Matrix} & N/A & $\rho = \sum_i p_i |i\rangle \langle i|$ & Represents mixed states \\
\textbf{Born Rule} & $P(x) = |\psi(x)|^2$ & $P(a) = |\langle a|\psi\rangle|^2$ & Probability of outcome $a$ \\
\textbf{Heisenberg Uncertainty Principle} & N/A & $\Delta X \Delta P \geq \frac{\hbar}{2}$ & Fundamental limit \\
\textbf{Pauli Matrices} & N/A & 
\(
\left\{
\begin{array}{l}
X = \begin{bmatrix} 0 & 1 \\ 1 & 0 \end{bmatrix}, \\
Y = \begin{bmatrix} 0 & -i \\ i & 0 \end{bmatrix}, \\
Z = \begin{bmatrix} 1 & 0 \\ 0 & -1 \end{bmatrix}
\end{array}
\right.
\) & SU(2) Rotation Elements\\
\textbf{Spin Operators} & N/A & $S_\alpha = \frac{\hbar}{2} \alpha$ & $\alpha = \{X, Y, Z\}$ \\
\textbf{Rotation Operator} & N/A & $U(\theta, \hat{n}) = e^{-i \theta \hat{n} \cdot \mathbf{S}/\hbar}$ & Rotates spin state \\
\hline
\end{tabular}
}
\label{table:notation_table_adjusted}
\end{table*}

\vspace{0.3cm}
\textit{ Postulate} 1: quantum states. --  The complete information about a quantum system is encoded in its state vector $|\psi\rangle$, which is a ray of the system Hilbert space $\mathcal{H}$. 
The quantum bit~(or qubit for short) is the smallest nontrivial unit in quantum computing, similar to the bit in classical computing.
It is physically implemented by a two-level quantum system, such as the spin of an electron.
Its mathematical form can be represented by a state vector in a Hilbert space $\mathcal{H}$, such that the single qubit state is defined as
\begin{equation}\label{psi_vector}
    |\psi\rangle = \alpha|0\rangle + \beta|1\rangle = \begin{bmatrix}
        \alpha \\ \beta
        \end{bmatrix},
\end{equation}
where $|\alpha|^2+|\beta|^2=1$ and $\alpha,\beta\in\mathcal{C}$.
Note that $|0\rangle$ and $|1\rangle$ are the basis of a two-dimension Hilbert space. 
The vector representation of these basis as
\begin{equation}
|0\rangle=\begin{bmatrix}
    1 \\ 0
    \end{bmatrix},
|1\rangle = \begin{bmatrix}
    0 \\ 1
    \end{bmatrix}
\end{equation}
And, 
\begin{equation}
\langle 0|=\begin{bmatrix}
    1, 0
    \end{bmatrix},
\langle 1| = \begin{bmatrix}
    0, 1
    \end{bmatrix}
\end{equation}
It is easy to see how qubits are fundamentally different from classical bits from this definition itself. By choosing $\alpha$ and $\beta$ arbitrarily, the state vector $|\psi\rangle$ can be expressed in an arbitrary but unique linear combination of $|0\rangle$ and $|1\rangle$ simultaneously, while classical bits can only be in one of these states. This characteristic of quantum systems is called  \textit{superposition}.

\vspace{0.3cm}
\emph{Postulate} 2: composition of quantum systems. -- When we consider a joint quantum system consisting of multiple qubits, the state of this system is described by the tensor product Hilbert space of each qubit $\mathcal{H}_i$. For example, if two qubits are described by their Hilbert spaces $\mathcal{H}_1$ and $\mathcal{H}_2$ respectively, the joint system comprising both these systems as subsystems is described by the tensor-product Hilbert space $\mathcal{H}_1 \otimes \mathcal{H}_2$. Thus, the state of two qubits is
    \begin{equation}
        \begin{aligned}
            |\psi\rangle &= |\psi_1\rangle\otimes|\psi_2\rangle\\
            &=\alpha_{00}|00\rangle + \alpha_{01}|01\rangle+\alpha_{10}|10\rangle+\alpha_{11}|11\rangle,
        \end{aligned}
    \end{equation}
where we ignore the $\otimes$ between qubits for notational simplicity~(e.g. $|0\rangle\otimes|0\rangle = |00\rangle$).
In the same way, for $n$-qubits state, its formulation is 
    \begin{equation}
        |\psi\rangle=\sum_{x1,x2,\cdots,x_n}^{0,1}\alpha_{x_1,x_2,\cdots,x_n}|x_1x_2\cdots x_n\rangle.
    \end{equation}
The dimension of the $n$-qubit system is $2^n$, and hence the tensor product space is a $2^n$-dimensional complex vector space $\mathcal{C}^{2^n}$. The dimension of the $n-$qubit system increases exponentially in the number $n$ of the qubits. Quantum entanglement is another important property of quantum states that is different from the classical states. It occurs when a joint quantum system can not be represented as a tensor product of the subsystems that make up it. Mathematically, if two quantum systems $A$ and $B$ are entangled, the joint system $AB$ might be represented as
    \begin{equation}
        |\psi\rangle_{AB}=\alpha|0\rangle_A|0\rangle_B+\beta|1\rangle_A|1\rangle_B,
    \end{equation}
    
where $|\alpha|^2+|\beta|^2=1$.
One can not then find any representation of $|\psi\rangle_{A}$ and $|\psi\rangle_{B}$ to satisfy $|\psi\rangle_{AB}=|\psi\rangle_{A}\otimes|\psi\rangle_{B}$. This interconnectedness leads to correlations between the qubits that are stronger than any classical correlation. These characteristics of quantum states - viz., \textit{superposition}, \textit{entanglement}, and \textit{exponentially growing state space}, give quantum computing the potential to solve problems efficiently. Another way to describe the quantum state is the density operator of the density matrix, which is mathematically equivalent to the state vector approach, but it provides a much more convenient language for thinking about some commonly encountered scenarios in quantum mechanics. The density operator of a pure $n$-qubit state is defined as:
\begin{equation}\label{eq:pure_state}
        \begin{split}
            \rho&=|\psi\rangle\langle\psi|\\
            &=\sum_{x_1,\cdots,x_n,x_1',\cdots,x_n'}^{\{0,1\}}\alpha_{x_1,\cdots,x_n}\alpha_{x_1',\cdots,x_n'}^*|x_1x_2\cdots x_n\rangle\langle x_1'x_2'\cdots x_n'|.
        \end{split}
\end{equation}
Moreover, suppose a quantum system is in one of several states $|\psi_i\rangle$, where $i$ is an index, with respective probabilities $p_i$. 
    The pairs $\{p_i,|\psi_i\rangle\}$ is named an ensemble of pure states.
    The density operator for the system is defined by the equation
    \begin{equation}\label{eq:mix_state}
        \rho=\sum_ip_i|\psi_i\rangle\langle\psi_i|,
    \end{equation}
where the normalization condition for probabilities implies $\sum_ip_i=1$.
Mathematically, we can recognize the pure states of Eq.~(\ref{eq:pure_state}) from the mixed states in Eq.~(\ref{eq:mix_state}) by the trace on the second order as $\mathrm{tr}(\rho^2)=1$ for pure states and $\mathrm{tr}(\rho^2)<1$ for mixed states. 

\vspace{0.3cm}
\textit{Postulate} 3: quantum dynamics. --
    The time evolution of a closed quantum system describes how the state of the system evolves over time, which is the most important way to deal with the information.
    The \textit{closed} means the evolution of systems that don't interact with the rest of the world.
    The time evolution is governed by the Schr{\"o}dinger equation, which is a fundamental partial differential equation in quantum mechanics.

    The time-dependent Schr{\"o}dinger equation is given by 
    \begin{equation}
        i\hbar \frac{d}{dt}|\psi(t)\rangle=H|\psi(t)\rangle,
    \end{equation}
    where $i$ is the imaginary unit, $\hbar$ is the reduced Planck's constant, $|\psi(t)\rangle$ is the state vector at time $t$, and $H$ is the Hamiltonian operator, which represents the total energy of the system.
    The solution of the Schr{\"o}dinger equation for a time-independent Hamiltonian $H$ is given by
    \begin{equation}
        |\psi(t)\rangle=e^{-\frac{i}{\hbar}Ht}|\psi(0)\rangle,
    \end{equation}
where $|\psi(0)\rangle$ is the initial state of the system at time $0$, and $e^{-\frac{i}{\hbar}Ht}$ is the time evolution operator.In practice, it is common to absorb the factor $\hbar$ into $H$, effectively setting $\hbar=1$. If the Hamiltonian $H$ depends on time, the situation is more complex, but we don't use it in this paper. It is not difficult to see that the time evolution operator $e^{-\frac{i}{\hbar}Ht}$ is a unitary matrix, so this time evolution is a unitary transformation, which preserves the norm of the quantum states, meaning the probability conservation. For simplicity, we usually use $U(t)$ to represent $e^{-iHt}$ as 
    \begin{equation}
        |\psi(t)\rangle=U(t)|\psi(0)\rangle.
    \end{equation}
In the context of digital quantum computing, time evolutions or quantum operations are also named quantum logical gates, similar to the logical gates in classical digital computing. Similarly, the time evolution implemented on the density operator can be easily defined as 
    \begin{equation}
        \rho(t)=U(t)\rho(0)U^\dagger(t),
    \end{equation}
where $U^\dagger(t)$ is the transpose conjugate of $U(t)$ as $U^\dagger(t)=e^{iHt}$.\\

\vspace{0.3cm}
\textit{ Postulate} 4: quantum measurement. --
    Above, we described the evolution of a \emph{closed} quantum system. However, to gain any information about the system, one has to necessarily consider an  experimentalist as an external physical system observing it. Unlike classical mechanics, where in principle, such observations can be made without changing the observed system itself, quantum mechanics distinguishes itself by postulating that the quantum system undergoes irreversible change due to such a measurement. The general quantum measurement postulate is described by a collection measurement operator $\{M_m\}$. The index $m$ refers to the measurement outcomes that may occur in the experiment with the probability $p_m$ given by the Born rule $p_m = \langle\psi| M^\dagger_mM_m |\psi\rangle$, where $|\psi\rangle$ is the state being measured.  To ensure the probabilities sum up to unity, the measurement operators satisfy the completeness equation
    \begin{equation}
        \sum_m M_m^\dagger M_m=I
    \end{equation}

In most cases of quantum computing, we care about the quantum measurement of an observable.  An observable, i.e., a dynamical variable like position or momentum, is described by a self-adjoint operator $A$ (namely $A=A^\dagger$), acting on the system Hilbert space $\mathcal{H}$. Via the spectral theorem, any observable has a spectral decomposition 
    \begin{equation}
        A=\sum_m a_m \ket{a_m}\bra{a_m},
    \end{equation}
where $a_m$'s are the eigenvalue and $\ket{a_m}$'s are the eigenvectors of $A$. Note that $\bra{a_m}$ is the dual of the vector $|a_m\rangle$, and is an element of the so-called dual space of $\mathcal{H}$ connected to $\mathcal{H}$ via the Riesz representation theorem. For a general quantum state $\ket{\psi}$, given in Eq.~(\ref{psi_vector}), the dual state is given by $\bra{\psi}=[\alpha^*, \beta^*]$, see Table~\ref{table:notation_table_adjusted}.
According to quantum physics, the outcome of measuring the operator $A$ on a quantum system described by state $\ket{\psi}$ probabilistically results in one of the eigenvalues $a_m$. The probability of each outcome is determined by the projector $M_m{=}\ket{a_m}\bra{a_m}$ as $p_m = \langle\psi| M^\dagger_mM_m |\psi\rangle=|\langle a_m|\psi\rangle|^2$. Hence, the expectation value of the measurement is given by 
\begin{equation}
   \langle A \rangle=\sum_m p_m a_m = \langle \psi| A|\psi \rangle
\end{equation}

Note here that the Hilbert space is a linear vector space, thus for any two states $|\psi_1\rangle$ and $|\psi_2\rangle$, the linear combination $\alpha |\psi_1\rangle + \beta |\psi_2\rangle$ is also a quantum state. Thus, in contrast to classical digital computation, where each bit (either 0 or 1) encodes one unit of classical information, theoretically there are an infinite number of superpositions are accessible to a single ``qubit" (a two-level quantum system). From a computational standpoint, the superposition property of quantum mechanics open up the possibility of encoding multiple gates in a single step. Indeed, famous quantum algorithms like Shor's algorithm \cite{shor1999polynomial} for factoring, or Grover's algorithm~\cite{grover1996fast} for search, use this inherent parallelism of quantum mechanics to solve classical computational problems far more efficiently than the best possible classical algorithms known. However, in practice, these quantum properties are extremely fragile to noise, and reliable implementation of error-correction schemes are still in their infancy~\cite{acharya2024quantum}. 

\section{Numerical Simulation Method}\label{Numerical}
{\color{black}To simulate QRC on a classical computer, we explicitly model the system-state evolution. Since our QRC contains only 10 qubits, we represent the quantum state using a density matrix$\rho$. The Hamiltonian dynamics are implemented via exact diagonalization, and the time-evolution operator is applied using a product-formula $e^{-i\tau H}\rho e^{i\tau H}$, which is a standard method for numerically simulating quantum dynamics with small quantum systems.

In the partial-trace step, we obtain the reduced density matrix of the hidden state $\rho_h$ by tracing out subsystem $I$ from the joint state $\rho_{Ih}$, i.e.,
\[
\rho_{h} = \mathrm{Tr}_{I}\!\left(\rho_{Ih}\right).
\]
Finally, the measurement outcome on each qubit is computed as the expectation value of the Pauli-$Z$ operator,
\[
\langle Z_i \rangle = \mathrm{Tr}\!\left(\rho_{Ih} Z_i\right).
\]}

\section{Exponential concentration of Quantum Measurement}\label{Exponential}
\begin{figure*}
    \centering
    \includegraphics[width=1\linewidth]{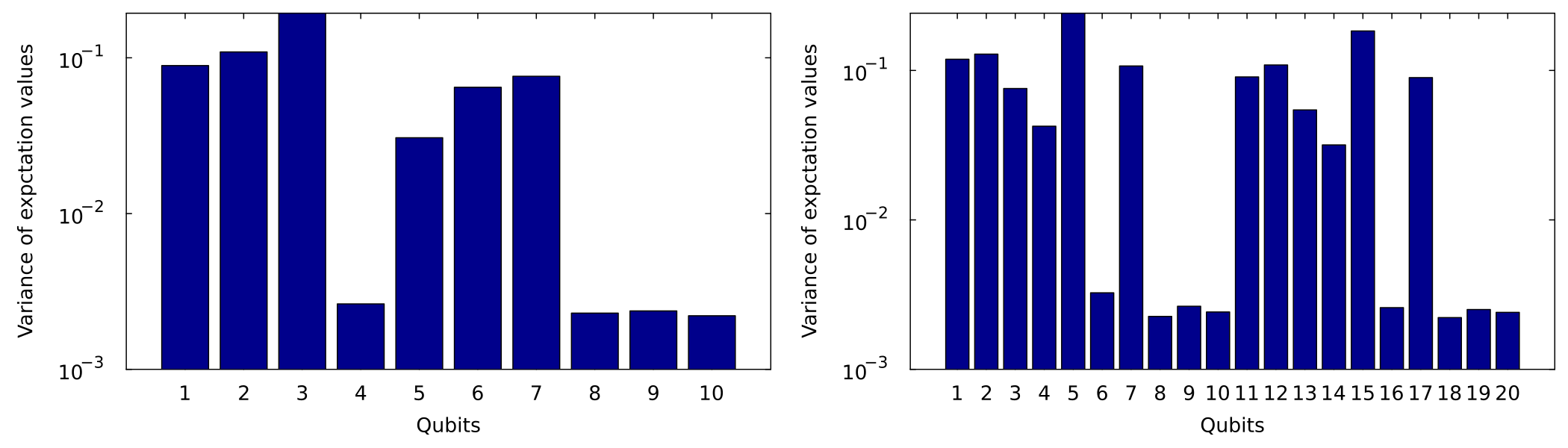}
    \caption{Variance of the measured expectation values across different qubits (left: QR1 model; right: QR2 model).}
    \label{fig:variance}
\end{figure*}

{\color{black} In this article\cite{xiongFundamentalAspectsQuantum2025}, the authors discuss exponential concentration: as the circuit depth increases, the system size increases, entanglement increases, or the noise strength increases, the measurement expectation values converge exponentially to a variable-independent value, such that the number of measurement shots growths exponentially to identify different expectation values. 
In our paper, however, we consider a specific example—namely, the application of QRC to predicting RV. We use a fixed system size (the number of qubits is 10) and circuit depth (3 layers). 
By calculating the variance of the expectation value on each qubit, we find that it does not exhibit exponential concentration (see Fig.~\ref{fig:variance}), thereby demonstrating the feasibility of QRC for this problem.}

\section{The hyperparameters of Classical Reservoir computing and LSTM}\label{Hyperparameters}

{\color{black} For the LSTM(x), we used stack LSTM model with 2 layers. The optimizer is ADAM, with 0.001 learning rate. The number of iteration is 100 epochs. And the batch size is 64. After these hyperparameters are chosen. We train these LSTM with different hidden size from 10 to 200 with 10 interval. We get for the LSTM model, the hidden size is 60 achieving best result, and LSTMX model, 50 achieving bset performance~(see Fig.~\ref{fig:MSEandHidden} right). 
\begin{figure}
    \centering
    \includegraphics[width=1\linewidth]{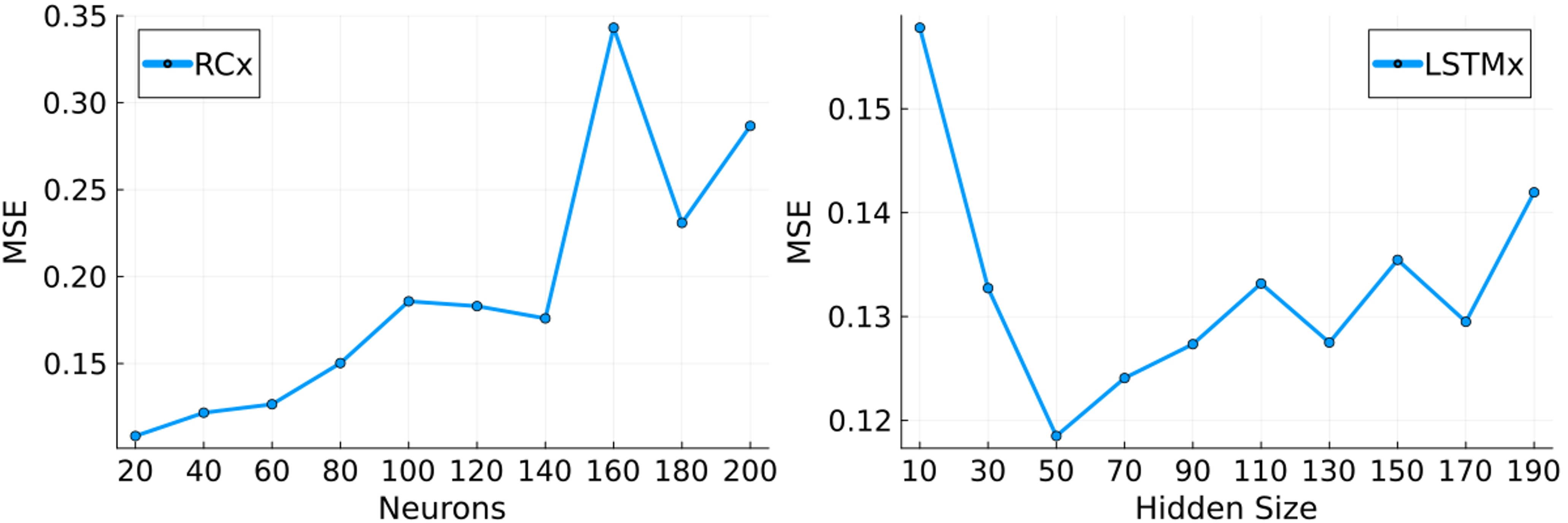}
    \caption{The MSE of RCx~(left) and LSTMx~(right) as the  hidden state size increases.}
    \label{fig:MSEandHidden}
\end{figure}
For Classical reservoir computing, RC(X), Apart from the number of neurons in reservoir, the other key hyperparameters are the leak rate~(lr), controlling the time constant of the reservoir, the spectral radius of W~(sr), the maximum absolute eigenvalue of the reservoir, and the input scaling~(is), the coefficient applied on $W_{in}$. We set them as $lr=0.6$, $sr=0.9$, $is=0.1$ by trying a lot of configurations. This~\href{https://reservoirpy.readthedocs.io/en/latest/user_guide/hyper.html}{document} provides a detailed description of these hyperparameters. Based on these hyperparameters, We train RC(X) models with the number of neurons varying from 10 to 200 with 10 interval. 
We get for the RC model, the number of neurons is 50 achieving best result, and RCx model, 20 achieving the best performance~(see Fig.~\ref{fig:MSEandHidden} left). 
As  shown in Fig.~\ref{fig:MSEandHidden} depicted, increasing the number of hidden nodes~(neurons) does not always reduce errors. So we selected the best model by comparing the performance of different hidden nodes~(neurons).} \\

\end{document}